\documentclass[aps, prd, reprint,10pt, notitlepage, a4paper,floats, floaqutfix,amsmath, amssymb, amsfonts,superscriptaddress,showpacs, showkeys,nofootinbib,longbibliography]{revtex4-1}

\pdfoutput=1

\usepackage{epsfig}
\usepackage{graphics}
\usepackage{graphicx}
\usepackage{amsmath,amssymb,mathrsfs}
\usepackage{amsfonts}
\usepackage[usenames,dvipsnames]{xcolor}
\usepackage{wasysym}
\usepackage{times}
\usepackage{mathptmx}
\usepackage{gensymb}
\usepackage{appendix}
\usepackage{listings}
\usepackage{url}
\usepackage[normalem]{ulem}
\usepackage{alltt}
\usepackage[colorlinks]{hyperref}
\usepackage{cleveref}
\usepackage{makecell}

\usepackage{enumitem}
\setlist{nosep}
\usepackage{color}
\usepackage{calc}
\usepackage{tensor}
\usepackage{bm}
\usepackage{times}
\usepackage{multirow}
\usepackage[varg]{txfonts}
\usepackage{float}
\usepackage{dcolumn}
\usepackage[nolist,nohyperlinks]{acronym}
\usepackage{xspace}
\usepackage[english]{babel}
\usepackage[abs]{overpic}
\usepackage{pict2e}
\usepackage[caption=false]{subfig} 
\allowdisplaybreaks[1]
\usepackage[utf8]{inputenc}
\usepackage{gensymb}
\usepackage{bm}
\usepackage{stackengine}
\usepackage{boldline,multirow}
\usepackage{braket}
\usepackage{longtable}

\usepackage{tabularx}
\usepackage{rotating}

\DeclareMathAlphabet{\mathcalstd}{OMS}{cmsy}{m}{n}
\DeclareMathAlphabet{\mathpzc}{OT1}{pzc}{m}{it}

\newcommand{\AEI}{Max Planck Institute for Gravitational Physics (Albert Einstein Institute), Am M{\"u}hlenberg 1, Potsdam, 14476, Germany}
\newcommand{\Maryland}{Department of Physics, University of Maryland, College Park, MD 20742, USA}

\definecolor{dodgerblue}{HTML}{1E90FF}
\definecolor{viennared}{HTML}{DA0A14}

\hypersetup{citecolor=dodgerblue}

\def\mr{\mathrm}

\acrodef{PN}{post-Newtonian}
\acrodef{EOB}{effective-one-body}
\acrodef{NR}{numerical relativity}
\acrodef{GW}{gravitational wave}
\acrodef{BBH}{binary black hole}
\acrodef{BH}{black hole}
\acrodef{BNS}{binary neutron star}
\acrodef{NSBH}{neutron star-black hole}
\acrodef{SNR}{signal-to-noise ratio}
\acrodef{aLIGO}{Advanced LIGO}
\acrodef{AdV}{Advanced Virgo}

\begin{document}




\title{Bayesian inference of binary black holes with \\ inspiral-merger-ringdown waveforms using two eccentric parameters}


\author{Antoni Ramos-Buades}
\affiliation{\AEI}
\author{Alessandra Buonanno}
\affiliation{\AEI}
\affiliation{\Maryland}
\author{Jonathan Gair}
\affiliation{\AEI}
\date{\today}

\begin{abstract}
  Orbital eccentricity is a crucial physical effect to unveil the
  origin of compact-object binaries detected by ground- and
  spaced-based gravitational-wave (GW) observatories.  Here, we perform
  for the first time a Bayesian inference study of  
  inspiral-merger-ringdown eccentric waveforms for binary black holes
  with non-precessing spins using two (instead of one) eccentric parameters:
  eccentricity and relativistic anomaly. We employ for our study 
the multipolar effective-one-body (EOB) waveform model \texttt{SEOBNRv4EHM}, 
and use initial conditions such that the eccentric parameters are
  specified at an orbit-averaged frequency. We show that this new
  parametrization of the initial conditions leads to a more efficient sampling of the parameter
  space. We also assess the impact of the relativistic-anomaly
  parameter by performing mock-signal injections, and we show that
  neglecting such a parameter can lead to significant biases in
  several binary parameters. We validate our model with mock-signal
  injections based on numerical-relativity waveforms, and we
  demonstrate the ability of the model to accurately recover the
  injected parameters.  Finally, using standard stochastic samplers 
employed by the LIGO-Virgo-KAGRA  Collaboration, we analyze a set of real GW signals
 observed by the LIGO-Virgo detectors during the first and third runs. We do not 
find clear evidence of eccentricity in the signals analyzed, more specifically we measure
 $e^{\text{GW150914}}_{\text{gw, 10Hz}}=
  0.08^{+0.09}_{-0.06}$, $e^{\text{GW151226}}_{\text{gw, 20Hz}}=
  {0.04}^{+0.05}_{-0.04} $, and $e^{\text{GW190521}}_{\text{gw, 5.5Hz}}=
  0.15^{+0.12}_{-0.12}$. 
\end{abstract}

\maketitle


\section{Introduction}
\label{sec:intro}

In the second part of this decade, the LIGO, Virgo and KAGRA (LVK) ground-based detectors
\cite{LIGOScientific:2018mvr,LIGOScientific:2020ibl,LIGOScientific:2021usb,LIGOScientific:2021djp,Nitz:2021uxj,Olsen:2022pin,KAGRA:2018plz,KAGRA:2020tym} will be reaching design sensitivity, promising several hundreds 
of detection per year of coalescing binary black holes. Most of these binaries are expected to be formed
via isolated binary evolution \cite{Bethe:1998bn,Belczynski:2001uc,
  Dominik:2013tma, Belczynski:2014iua,
  mennekens2014massive,spera2015mass, Belczynski:2016obo,
  Eldridge:2016ymr,
  Marchant:2016wow,Mapelli:2017hqk,Mapelli:2018wys,Stevenson:2017tfq,Giacobbo:2018etu,Kruckow:2018slo,Kruckow:2018slo},
and circularize due to GW emission~\cite{Peters:1964zz} by the time
they enter the detector frequency band. Nonetheless, a small fraction
of these binaries may have non-negligible orbital eccentricity in the
frequency band of ground-based detectors, if they form through
dynamical captures and interactions in dense stellar environments,
such as globular clusters~\cite{PortegiesZwart:1999nm,
  Miller:2001ez,Miller:2002pg,Gultekin:2004pm,Gultekin:2005fd,OLeary:2005vqo,
  Sadowski:2007dz, downing2010compact,downing2011compact,
  Samsing:2013kua,Rodriguez:2015oxa,Askar:2016jwt,Rodriguez:2016kxx,
  Rodriguez:2016avt,
  Samsing:2017rat,Samsing:2017xmd,Rodriguez:2017pec,Rodriguez:2018rmd,
  Fragione:2018vty, Zevin:2018kzq,Gondan:2020svr} or galactic
nuclei~\cite{OLeary:2008myb,Antonini:2012ad,Tsang:2013mca,Antonini:2016gqe,Petrovich:2017otm,Stone:2016wzz,Stone:2016ryd,Rasskazov:2019gjw}. Hence, 
measuring orbital eccentricity in the GW signal from merging binaries
provides crucial information about the origin, evolution and the properties of
the population of such binaries
\cite{Mandel:2009nx,LIGOScientific:2016vpg, Farr:2017uvj,
  LIGOScientific:2018jsj, Zevin:2021rtf,LIGOScientific:2021psn}.

With upgrades of existing ground-based detectors (e.g., 
$A^+$  and Virgo$^+$ \cite{KAGRA:2013rdx,LIGOInstrumentWhitePaper,VirgoInstrumentWhitePaper})  
, and future next-generation detectors on the earth, such as the Einstein Telescope and 
Cosmic Explorer
\cite{Punturo:2010zz,LIGOScientific:2016wof,Reitze:2019dyk,
  Reitze:2019iox}, and in space, such as LISA
and TianQin \cite{amaroseoane2017laser,TianQin:2015yph}, the fraction
of GW events with non-negligible orbital eccentricity is expected to
significantly increase~\cite{Sesana:2010qb,Breivik:2016ddj,Samsing:2018isx,Cardoso:2020iji}.
All these projections have motivated, in the last years, the development of 
waveform models that include the effect of orbital eccentricity
\cite{Romero-Shaw:2019itr,Nitz:2019spj,Romero-Shaw:2020thy,Gayathri:2020coq,Wu:2020zwr,Favata:2021vhw,
  OShea:2021ugg,
  Romero-Shaw:2021ual,Gamba:2021gap,Bonino:2022hkj,Iglesias:2022xfc}. Those waveforms 
have been used to analyze GW signals observed by LIGO and Virgo. Some
of these studies have found evidence of eccentricity, for
instance, for GW190521, where Ref.~\cite{Gayathri:2020coq} showed
evidence for a highly eccentric precessing-spin binary,
Ref.~\cite{Romero-Shaw:2021ual} showed evidence for an eccentric
non-precessing binary, and Ref.~\cite{Gamba:2021gap} found evidence
for a nonspinning dynamical capture merger. However, each of these
studies comes with their own limitations. For example,
Ref.~\cite{Gayathri:2020coq} used a sparse grid of numerical
relativity (NR) waveforms for inference,
Ref.~\cite{Romero-Shaw:2021ual} applied reweighing techniques for
inference \cite{Romero-Shaw:2019itr} combined with an
inspiral-merger-ringdown (IMR) model based on the effective-one-body
(EOB) formalism~\cite{Buonanno:1998gg,Buonanno:2000ef,Damour:2000we,Damour:2001tu,Buonanno:2005xu},
\texttt{SEOBNRE} \cite{PhysRevD.86.024011,Cao:2017ndf,Liu:2021pkr} with only one
eccentricity parameter, although the impact of the latter in the
eccentricity posterior might be limited \cite{Clarke:2022fma}. While
Ref.~\cite{Gamba:2021gap} restricted the study  to nonspinning dynamical captures
using the waveform model \texttt{TEOBResumS-Dali}~\cite{Chiaramello:2020ehz,Nagar:2021gss}.

In this paper, we perform a Bayesian inference study using the waveform model \texttt{SEOBNRv4EHM}~
\cite{Khalil:2021txt,Ramos-Buades:2021adz}, which describes eccentric effects
using two eccentric parameters: the initial eccentricity and
the relativistic anomaly. \texttt{SEOBNRv4EHM} is built upon the
quasi-circular \texttt{SEOBNRv4HM} model~\cite{Cotesta:2018fcv} for
binary black holes (BBHs) with non-precessing spins and includes
eccentric corrections up to 2PN order \cite{Khalil:2021txt} in the
$(l,|m|)=(2,2),(2,1),(3,3),(4,4),(5,5)$ multipoles. When restricting
to the $(l,|m|)=(2,2)$ modes, we refer to the model as
\texttt{SEOBNRv4E}. We modify the initial conditions of the original
version of the \texttt{SEOBNRv4EHM} model, which was presented in
Ref.~\cite{Ramos-Buades:2021adz}, so that now they are specified at an orbit-averaged
orbital frequency instead of an instantaneous orbital frequency. We 
show that this new parametrization simplifies the stochastic sampling
across parameter space. Furthermore, in order to increase the efficiency 
in evolving the dynamics, we combine reduced tolerances of the Runge-Kutta integrator 
during the inspiral with the optimized Hamiltonian and integrator from Refs. \cite{Devine:2016ovp,Knowles:2018hqq}, and assess the impact of these
 modifications in the accuracy of the waveform model across parameter space. We find
that for most of the parameter space the loss of accuracy is $<1\%$ in
unfaithfulness, indicating that the more efficient version of the
\texttt{SEOBNRv4EHM} is still highly accurate for our purposes.

With the more efficient model, we use a highly parallelizable nested
sampler \texttt{parallel Bilby} \cite{Smith:2019ucc} to assess firstly
the impact of the different initial conditions, which available in the 
\texttt{SEOBNRv4EHM} model. We find that the initial conditions
specified at an orbit-averaged frequency perform as accurately as the
ones specified at an instantaneous orbital frequency, but with a more
efficient sampling of the parameter space, which translates into
shorter wall clock times for the parameter-estimation
runs. Furthermore, we show, for the first time, the impact of the radial
phase parameter (i.e., the relativistic anomaly), using an inspiral-merger-ringdown 
model and a full parameter-estimation code (see, e.g., Ref. \cite{Clarke:2022fma} for a 
study of the argument of periapsis based on the overlaps).

We also validate the model \texttt{SEOBNRv4EHM} by performing zero-noise
injections of public eccentric NR waveforms from the Simulating
eXtreme Spacetimes (SXS) catalog
\cite{Boyle:2019kee,Hinderer:2017jcs}. In particular we select a set
of three equal-mass, nonspinning eccentric simulations with initial
eccentricities measured from the orbital frequency at first periastron
passage of $0.07$, $0.13$ and $ 0.25$, respectively. In order to
compare the eccentricity from NR waveforms and the
\texttt{SEOBNRv4EHM} model, we choose a common definition of
eccentricity based on the GW signal, $e_{\text{gw}}$, as introduced in
Ref.~\cite{Ramos-Buades:2022lgf}, and measure it from the waveform
using the \texttt{gw\_eccentricity} package \cite{Shaikh:2023ypz}. We
find that \texttt{SEOBNRv4EHM} is able to accurately recover the
injected parameters of the NR signals for all the eccentricity values
considered.

Finally, we analyze some GW events observed by the LVK collaboration
with \texttt{SEOBNRv4EHM}. In particular, we analyze GW150914
\cite{LIGOScientific:2016aoc}, GW151226 \cite{LIGOScientific:2016sjg}
and GW190521 \cite{LIGOScientific:2020iuh,LIGOScientific:2020ufj}. The
choice is based on the fact that the first GW signal is still one of the loudest  
GW events so far, and has been found as a non-eccentric binary by many studies in the
literature \cite{LIGOScientific:2016ebw,Romero-Shaw:2019itr,Bonino:2022hkj,Iglesias:2022xfc}. 
On the other hand, GW151226 and GW190521 have been found to have signatures of
eccentricity by some studies in the literature
\cite{Gayathri:2020coq,OShea:2021ugg,
  Romero-Shaw:2021ual,Gamba:2021gap,Iglesias:2022xfc}. We measure the
following values of eccentricity for these GW events,
$e^{\text{GW150914}}_{\text{gw, 10Hz}}= 0.08^{+0.09}_{-0.06}$,
$e^{\text{GW151226}}_{\text{gw, 20Hz}}= {0.04}^{+0.05}_{-0.04} $, and
$e^{\text{GW190521}}_{\text{gw, 5.5Hz}}= 0.15^{+0.12}_{-0.12}$, and
therefore, find no clear evidence of eccentricity in these signals.

This paper is organized as follows. In Sec. \ref{sec:EOBmodel}, we
provide an overview of the eccentric model \texttt{SEOBNRv4EHM}. In
Sec. \ref{sec:eccIC}, we introduce new initial conditions specified at
an orbit-averaged orbital frequency, and in
Sec. \ref{sec:ODEloosening}, we introduce an \textit{optimized} version
of the model, 
\texttt{SEOBNRv4EHM\_opt}, and assess its performance. We
present the methodology for parameter estimation in
Sec. \ref{sec:PEsetup}. In Sec. \ref{sec:QClimit} we demonstrate the
accuracy of \texttt{SEOBNRv4EHM\_opt} in the quasi-circular limit, in
Sec. \ref{sec:ModelInj} we show the importance of the radial phase
parameter with mock-signal injections, and in Sec. \ref{sec:NRInj} we
asses the accuracy of the model against eccentric NR waveforms from
the SXS catalog using zero noise. In Sec.~\ref{sec:GWevents}, we
analyze GW events detected by the LVK Collaboration. In
Sec. \ref{sec:conclusions}, we summarize our main conclusions and
discuss future work. Finally, in Appendix \ref{sec:AppendixA} we
provide details about the derivation of the expressions used in the
initial conditions specified at an orbit-averaged frequency.

\section*{Notation}\label{sec:notation}
In this paper, we use geometric units, setting $G=c=1$ unless otherwise specified.

We consider a binary with masses $m_{1,2}$, with $m_1 \geq m_2$, and spins $\bm{S}_{1,2}$. We define the following combination of masses 
\begin{equation}
\begin{gathered}
M \equiv m_1 + m_2, \quad \mu \equiv \frac{m_1m_2}{M}, \quad \nu \equiv \frac{\mu}{M},  \\
\delta \equiv\frac{m_1 - m_2}{M},  \quad q \equiv \frac{m_1}{m_2}.
\end{gathered}
\end{equation}
A relevant combination of masses for GW data analysis is the \textit{chirp mass} defined as \cite{Sathyaprakash:2009xs}
\begin{equation}
\mathcal{M} = \nu^{3/5} M.
\label{eq:chirpMass}
\end{equation}
For non-precessing configurations the spin components are aligned/anti-aligned with the orbital angular momentum, $\bm{L}$, with direction $\hat{\bm L}$. In this case the dimensionless spin components can be defined as 
\begin{equation}
\chi_i=\bm{S}_i \cdot \hat{\bm L}/m^2_i,
\end{equation}
where $\mr i = 1,2$. It is convenient to define the effective-spin parameter $\chi_{\rm eff}$ \cite{Damour:2001tu,Racine:2008qv,Santamaria:2010yb},
\begin{equation}
\chi_{\rm eff} = \frac{1}{M}(m_1 \chi_1 + m_2 \chi_2),
\label{eq:chi_eff}
\end{equation}
and the spin combinations,
\begin{equation}
 \chi _S = \frac{1}{2}(\chi_1 +\chi_2), \quad \chi _A = \frac{1}{2}(\chi_1 -\chi_2).
\label{eq:chi_SA}
\end{equation} 

\section{Eccentric effective-one-body waveform model with non-precessing spins}
\label{sec:EOBmodel}

In this section we describe the waveform model, \texttt{SEOBNRv4EHM},
used throughout the paper. We provide an overview of the model,
develop a new procedure to specify the initial conditions at an
orbit-averaged frequency, and introduce modifications to increase the
computational efficiency of the model.

\subsection{Overview}
\label{sec:modelOverview}

The \texttt{SEOBNRv4EHM} model was presented in
Ref. \cite{Ramos-Buades:2021adz}, and we refer therein for further
details. \texttt{SEOBNRv4EHM} is constructed within the EOB
formalism~\cite{Buonanno:1998gg,Buonanno:2000ef,Damour:2000we,Damour:2001tu,Buonanno:2005xu},
and thus it is constituted by three main building blocks:
\\
\begin{itemize}
\item \textit{{\rm EOB} Hamiltonian}: The EOB conservative dynamics is determined by the EOB Hamiltonian constructed from the effective Hamiltonian, $H_{\text{eff}}$, as described in Refs.~\cite{Barausse:2011ys,Taracchini:2013rva}, augmented with the 
parameters $(K,d_{\text{SO}}, d_{\text{SS}}, \Delta t^{22}_{\text{peak}})$ calibrated to NR waveforms from Ref.~\cite{Bohe:2016gbl}, through the energy map~\cite{Buonanno:1998gg}
\begin{equation}
H_{\text{EOB}}=M \sqrt{1+ 2 \nu \left(\frac{H_{\text{eff}}}{\mu}-1 \right) }\,. 
\label{eq:eq0}
\end{equation}
For spins anti-aligned/aligned with the orbital angular momentum the motion is restricted to a plane. As a consequence, the dynamical variables in the Hamiltonian are the (dimensionless) radial separation $r \equiv R/M$, the orbital phase $\phi$, and their (dimensionless) conjugate momenta $p_r \equiv P_r/\mu$ and $p_\phi \equiv P_\phi / \mu$.
\\
\item \textit{Radiation-reaction force}: The dissipative effects in the EOB dynamics are described by a radiation-reaction (RR) force 
$\mathcal{F}$, which enters the Hamilton equations of motion, as~\cite{Pan:2011gk,Pan:2013rra}
\begin{align}
\qquad & \dot{r} = \xi(r) \frac{\partial \hat{H}_\text{EOB}}{\partial p_{r_*}}(r,p_{r_*},p_\phi), \quad \dot{\phi} = \frac{\partial \hat{H}_\text{EOB}}{\partial p_\phi}(r,p_{r_*},p_\phi), \nonumber\\
\qquad & \dot{p}_{r_*} = - \xi(r) \frac{\partial \hat{H}_\text{EOB}}{\partial r}(r,p_{r_*},p_\phi) + \hat{\mathcal{F}}_r, \quad \dot{p}_\phi = \hat{\mathcal{F}}_\phi,
\label{eq:eq1}
\end{align}
where the dot represents the time derivative $d/d\hat{t}$, with respect to the dimensionless time $\hat{t} \equiv T/M$, $\hat{H}_\text{EOB}\equiv H_\text{EOB}/\mu$, and $\hat{\mathcal{F}}_\phi \equiv \mathcal{F}_\phi / M$. 

The equations are expressed in terms of $p_{r_*} \equiv p_r\,\xi(r)$, which is the conjugate momentum to the tortoise-coordinate $r_*$, and $\xi(r)\equiv dr/dr_*$ can be expressed in terms of the potentials of the effective Hamiltonian~\cite{Pan:2011gk}.
The components of the RR force are computed using the following relations~\cite{Buonanno:2000ef,Buonanno:2005xu}
\begin{equation}
\hat{\mathcal{F}}_\phi = - \frac{\Phi_E}{\omega}, \quad  \hat{\mathcal{F}}_r=\hat{\mathcal{F}}_\phi \frac{p_r}{p_\phi}, \\
\label{eq:eq2}
\end{equation}
where $\omega = \dot{\phi}$ is the (dimensionless) orbital frequency, and $\Phi_E$ is the energy flux for quasi-circular orbits written as a sum over waveform modes using~\cite{Damour:2008gu,Pan:2010hz}
\begin{equation}
\Phi_E = \frac{\omega^2}{16\pi} \sum_{l=2}^{8} \sum_{m=-l}^{l} m^2 \left| \frac{d_L}{M} h_{lm}\right|^2,
\label{eq:eq2.1}
\end{equation}
where $d_L$ is the luminosity distance between the binary system and the observer. 

\item \textit{Waveform multipoles}: The GW multipoles are composed by two main parts: the inspiral-plunge multipoles $h_{lm}^\text{insp-plunge}$, and the merger-ringdown $h_{lm}^\text{merger-RD}$ modes,
\begin{align}
\label{eq:complete_mode}
h_{\ell m}(t) = \begin{cases}
h_{\ell m}^{\mathrm{insp-plunge}}(t), &t < t_{\textrm{match}}^{\ell m}\\
h_{\ell m}^{\mathrm{merger-RD}}(t), &t > t_{\textrm{match}}^{\ell m},\\
\end{cases}
\end{align}
where $t_{\textrm{match}}^{\ell m}$ is defined from the peak of the
amplitude of the $(2,2)$-mode (see Ref. \cite{Cotesta:2018fcv} for
details).  The merger-ringdown modes of \texttt{SEOBNRv4EHM} are the
same as in the underlying \texttt{SEOBNRv4HM} model
\cite{Bohe:2016gbl,Cotesta:2018fcv}, and thus we assume that the
effects of eccentricity at merger-ringdown are negligible. The 
inspiral-plunge multipoles are constructed upon the NR-calibrated
quasi-circular \texttt{SEOBNRv4HM} multipoles \cite{Cotesta:2018fcv}
by incorporating the 2PN eccentric corrections, including spin-orbit
and spin-spin effects, as derived in Ref.~\cite{Khalil:2021txt}, and 
are augmented by an orbit-averaged procedure to compute the 
nonquasi-circular (NQC) terms (see Sec. II B from Ref. \cite{Ramos-Buades:2021adz} for details).
\end{itemize}

\subsection{New eccentric initial conditions}
\label{sec:eccIC}

The initial conditions of \texttt{SEOBNRv4EHM} are expressed in terms of the eccentricity, $e$, and the relativistic anomaly $\zeta$ defined in the Keplerian parametrization of the orbit \cite{Ramos-Buades:2021adz}
\begin{equation}
r = \frac{1}{u_p (1 + e \cos \zeta)} \,,
\end{equation}
where $u_p$ is the inverse semilatus rectum.
Given the  masses, spins, the initial instantaneous orbital frequency $\omega_0$, initial eccentricity $e_0$ and relativistic anomaly $\zeta_0$, the initial conditions for $r_0$ and ${p_\phi}_0$, in absence of radiation reaction, can be obtained by solving the equations \cite{Ramos-Buades:2021adz}
\begin{equation}
\left[\frac{\partial \hat{H}_\text{EOB}}{\partial r}\right]_0 = - \left[\dot p_r(p_\phi,e,\zeta)\right]_0, \qquad \left[\frac{\partial \hat{H}_\text{EOB}}{\partial p_\phi}\right]_0 = \omega_0,
\label{eq:eqR0omega0}
\end{equation}
with $p_r(p_\phi,e,\zeta)$ and $\dot p_r(p_\phi,e,\zeta)$ given by the 2PN-order expressions in Eqs. (C3) and (C4) of the Appendix C in Ref. \cite{Ramos-Buades:2021adz}.

The initial condition for $p_{r_0}$ can be computed using the solution for $r_0$ and ${p_\phi}_0$, and numerically solving
\begin{equation}
\left[\frac{\partial \hat{H}_\text{EOB}}{\partial p_r}\right]_0 = \left[\dot{r}^{(0)} + \dot{r}^{(1)}\right]_0,
\end{equation}
where $\dot{r}^{(0)}$ is the 2PN-order expression for $\dot{r}$ at zeroth order in the RR effects (see Eq. (C5) in Ref. \cite{Ramos-Buades:2021adz}), while $\dot{r}^{(1)}$ is the first-order term in the RR part of $\dot{r}$, for which we use the quasi-circular expression derived in Ref.~\cite{Buonanno:2005xu}
\begin{equation}
\dot{r}^{(1)} = - \frac{\Phi_E^\text{qc}}{\omega} \frac{\partial^2 \hat{H}_\text{EOB} / \partial r\partial p_\phi}{\partial^2 \hat{H}_\text{EOB} / \partial r^2},
\end{equation}
being $\Phi_E^\text{qc}$ the quasi-circular energy flux given in Eq.~\eqref{eq:eq2.1}. Finally, the initial value  $p_{r_0}$ is converted into the tortoise-coordinate conjugate momentum ${p_{r_*}}_0$, using the relations in Sec. II A of Ref.~\cite{Ramos-Buades:2021adz}, so that together with $r_0$ and ${p_\phi}_0$, it can be introduced in Eqs. \eqref{eq:eq1} to  evolve the EOB equations of motion. 

\begin{figure}[!]
\begin{center}
\includegraphics[width=1.\columnwidth]{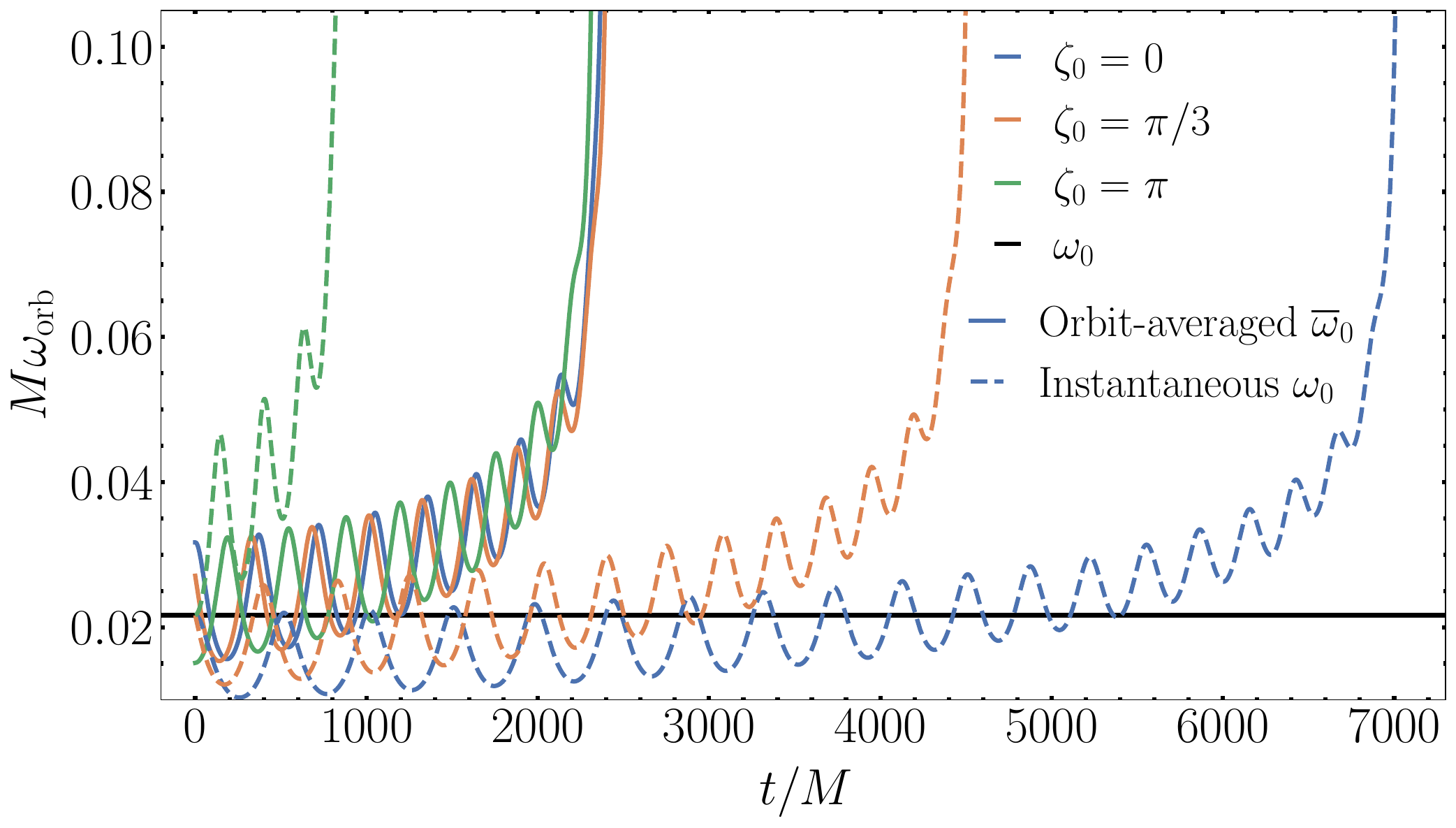}
\includegraphics[width=1.\columnwidth]{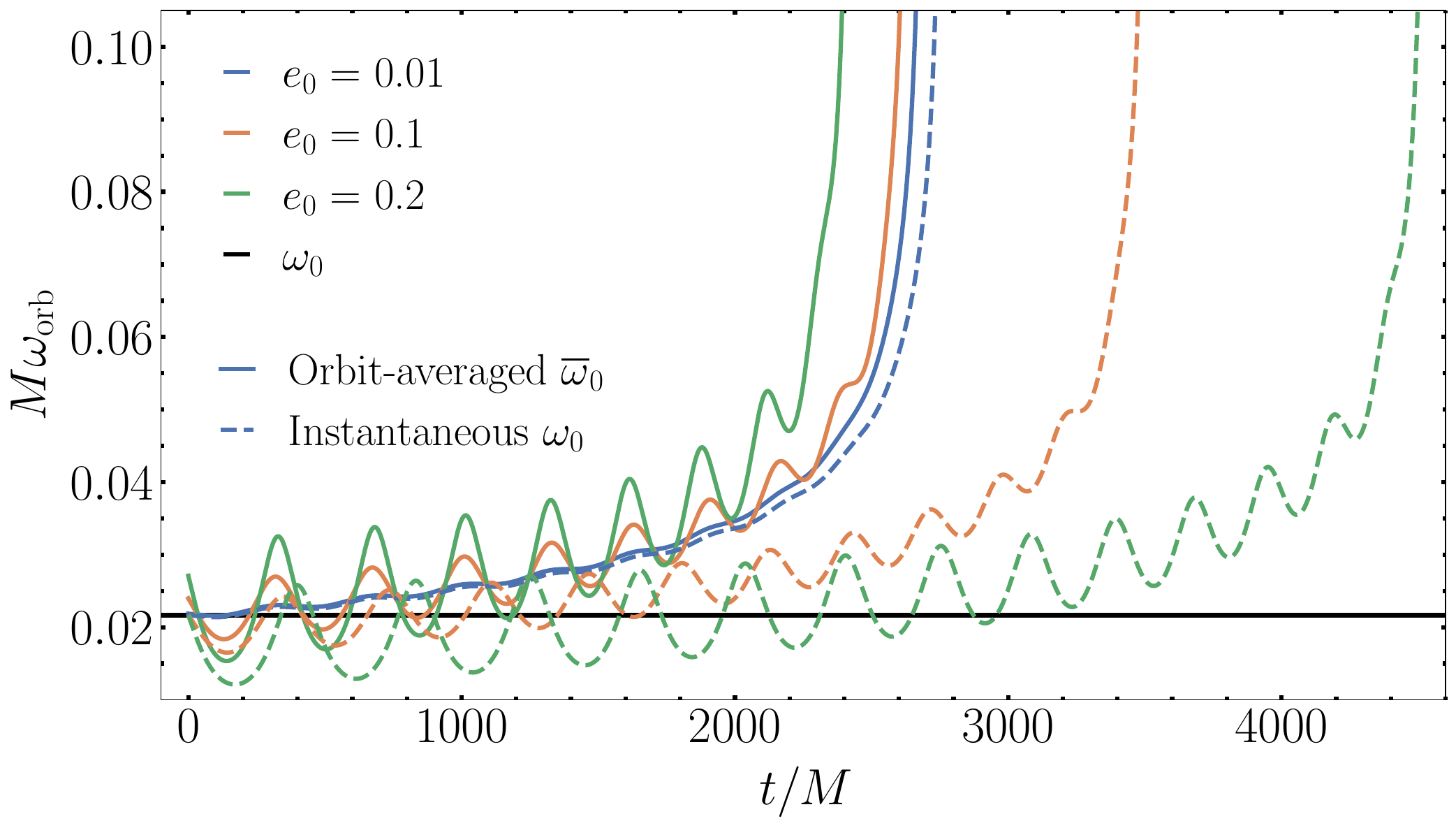}
\caption{\textit{Top panel:} Orbital frequency evolution of \texttt{SEOBNRv4EHM} for a configuration with mass ratio $q=3$, dimensionless spins $\chi_1=0.3$, $\chi_2=0.5$, $e_0=0.2$, total mass $70 M_\odot$ at a starting frequency of $20$Hz, and three different values of initial relativistic anomaly $\zeta_0 = [0,\pi/3,\pi]$. The dashed curves correspond to the specification of $(e_0, \zeta_0)$ at an instantaneous orbital frequency, $\omega_0$, while the solid lines correspond to the initial conditions specified at an orbit-averaged orbital frequency $\overline{\omega}_0$. \textit{Bottom panel:} Same configuration as in the upper panel, but fixing $\zeta_0 = \pi/3$, and using three different values of initial eccentricity, $e_0=[0.01,0.1,0.2]$, for both types of initial conditions. The horizontal black solid line in both panels indicates a dimensionless frequency of $M\omega_0 = 0.0216$, corresponding to $20$Hz at $70 M_\odot$.  
}
\label{fig:OrbitalFreqExample}
\end{center}
\end{figure}

The specification of eccentricity and relativistic anomaly at an instantaneous orbital frequency, which enters the right-hand side (RHS) of Eq. \eqref{eq:eqR0omega0}, implies significant variations of the length of the EOB dynamics (and thus of the waveform length), as one of the eccentric parameters is modified. This effect is displayed in Fig.~\ref{fig:OrbitalFreqExample} for the orbital frequency evolution for a configuration with mass ratio $q=3$, dimensionless spins $\chi_1=0.3$, $\chi_2=0.5$, and total mass $70 M_\odot$ at a starting frequency of $20$Hz.
In particular, we show in the upper panel of Fig.~\ref{fig:OrbitalFreqExample}, the instantaneous 
orbital frequency at a fixed
initial eccentricity, $e_0=0.2$, for three values of the initial
relativistic anomaly, $\zeta_0 = [0,\pi/3,\pi]$ (dashed lines). Imposing 
that the initial eccentricity and relativistic
anomaly are specified at an instantaneous orbital frequency causes
necessarily that the length of the evolution is substantially different when the 
initial relativistic anomaly is at periastron ($\zeta_0=0$) or apastron
($\zeta_0=\pi$). In the bottom panel of Fig.~\ref{fig:OrbitalFreqExample}, we display 
the instantaneous orbital frequency for three values of
initial eccentricity, $e_0=[0.01,0.1,0.2]$, at a fixed value of the
initial relativistic anomaly ($\zeta_0=\pi/3$). The
evolutions (dashed lines) show that the higher the eccentricity the
longer the evolution due to the larger oscillations in the
instantaneous orbital frequency, and therefore, the chosen value of
$\omega_0$ is crossed at earlier times in the inspiral.

We note that specifying the initial eccentricity and relativistic
anomaly at an instantaneous orbital frequency is a particular
parametrization of elliptical orbits. Other possible
parameterizations exist in the literature. For instance, in
post-Newtonian (PN) models based on the quasi-Keplerian
parametrization~\cite{Gopakumar:1997bs, Gopakumar:2001dy,
  Damour:2004bz, Konigsdorffer:2006zt, Arun:2007rg,Arun:2007sg,
  Arun:2009mc,
  Memmesheimer:2004cv,Yunes:2009yz,Huerta:2014eca,Mishra:2015bqa,
  Loutrel:2017fgu,Klein:2018ybm,Moore:2018kvz,Moore:2019xkm,Tanay:2019knc,Tiwari:2020hsu},
the initial conditions for the evolution of a binary in elliptical
orbits are specified at an initial orbit-averaged orbital
frequency. In these evolutions the initial parameters are the
orbit-averaged orbital frequency, eccentricity and radial phase
(typically the mean anomaly). One of the consequences of this
parametrization is that at a fixed value of eccentricity, changes in
the radial phase correspond to different positions in the same orbit.
While an increase (decrease) of the value of eccentricity at fixed
value of the radial phase causes a decrease (increase) of the length
of the evolution.

\begin{figure}[h!]
\begin{center}
\includegraphics[width=1.\columnwidth]{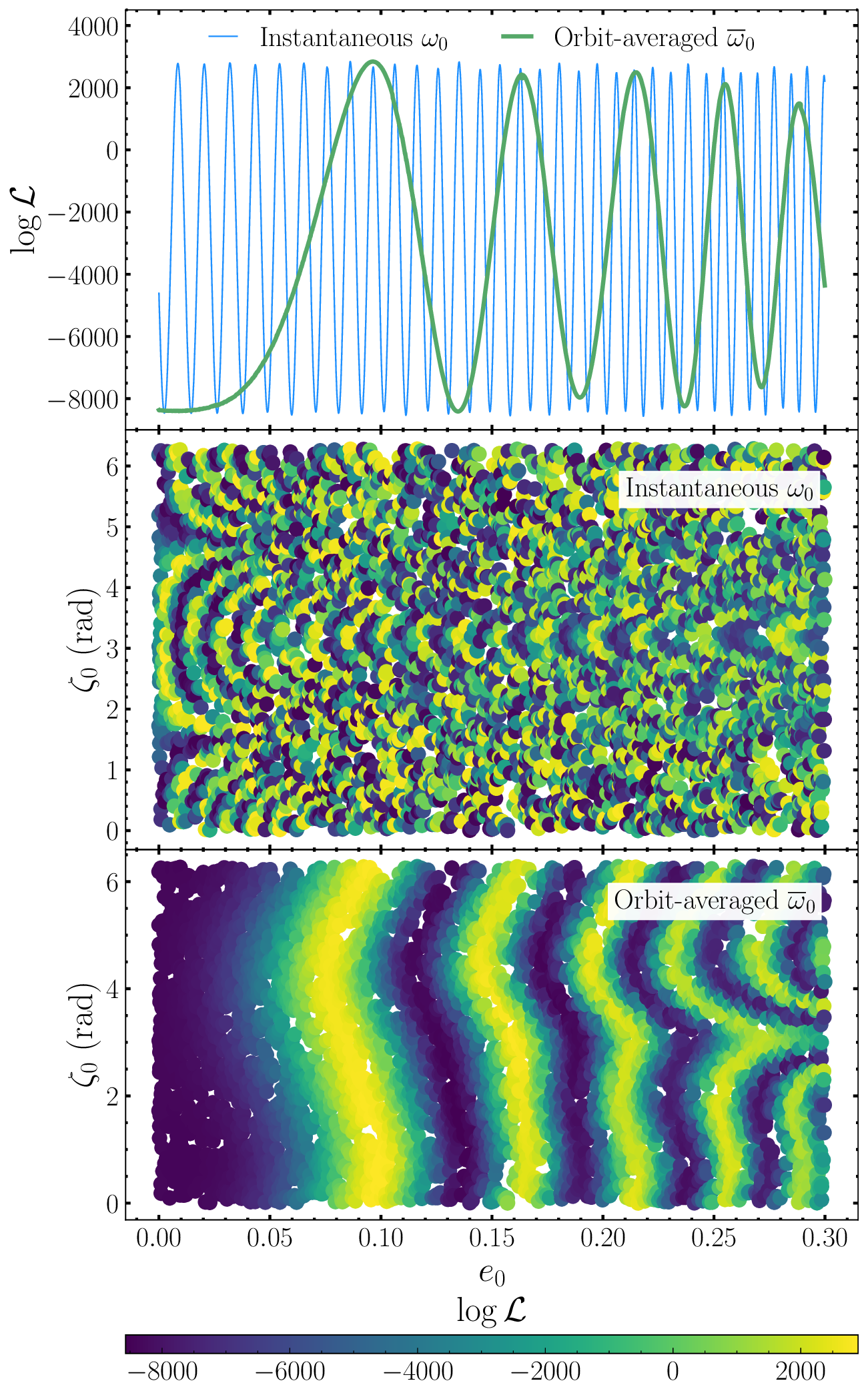} 
\caption{Zero-noise injection of a \texttt{SEOBNRv4E} waveform with initial eccentricity $e_0=0.1$, total mass $65 M_\odot$ and spins $\chi_1=0.3$, $\chi_2=0$ at a starting frequency of 20Hz, and recovering with  \texttt{SEOBNRv4E} with all the parameters fixed to injected values except for the eccentric ones ($e_0$, $\zeta_0$). In all the plots we consider $5000$ points randomly distributed in the parameter space $\zeta_0 \in [0,2\pi]$ and $e_0\in[0,0.3]$. In the top panel, the relativistic anomaly is fixed to $\zeta_0=0$, and only $e_0$ is sampled. The blue curve corresponds to using initial conditions based on the instantaneous orbital frequency, $\omega_0$, while the green curve corresponds to using the orbit-averaged orbital frequency, $\overline{\omega}_0$. The mid and bottom plots correspond to an injected  value of $\zeta_0=1$, and the use of $\omega_0$ and $\overline{\omega}_0$ in the initial conditions, respectively. In both mid and bottom panels, $e_0$ and $\zeta_0$ are sampled, and each point in parameter space is color-coded by its log-likelihood value.  
}
\label{fig:logLExamples}
\end{center}
\end{figure}

An additional motivation to explore a new parameterization of the
initial conditions for \texttt{SEOBNRv4EHM} is the application of the
model to Bayesian inference studies. Particularly, when performing
parameter estimation with \texttt{SEOBNRv4EHM}, the usage of the
initial conditions based on the instantaneous orbital frequency
produce an increase in the structure of the log-likelihood surface as
can be observed in Fig. \ref{fig:logLExamples}. There, we show the
log-likelihood computed for a zero-noise injection of a
\texttt{SEOBNRv4E} waveform with initial eccentricity $e_0=0.1$, total
mass $65M_\odot$, and dimensionless spins $\chi_1=0.3$, $\chi_2=0$ at
a starting frequency of 20Hz, and recovering with \texttt{SEOBNRv4E}
with all the parameters fixed to injected values except for the
initial eccentricity and relativistic anomaly. We consider 5000 random
points in the parameter space $\zeta_0 \in [0,2\pi]$ and
$e_0\in[0,0.3]$, and use the parameter estimation code \texttt{Bilby}
\cite{Ashton:2018jfp,Romero-Shaw:2020owr} to compute the
likelihood. In the uppermost panel we fix the value of the initial
relativistic anomaly to periastron ($\zeta_0=0$) and sample only in
eccentricity, while in the mid panel we fix $\zeta_0=1$ and sample
both in initial eccentricity and relativistic anomaly. In both panels
the specification of the initial conditions at an instantaneous
orbital frequency leads to a complex structure in the log-likelihood
values, which can pose a challenge for stochastic samplers as shown
in Sec. \ref{sec:ModelInj}.

Therefore, we introduce a new parametrization of the EOB initial conditions where the initial eccentricity and relativistic anomaly are specified at an orbit-averaged orbital frequency, $\overline{\omega}_0$. The orbit-averaged orbital frequency can be defined as
\begin{equation}
\overline{\omega}_0 = \frac{1}{T_r} \oint \omega(t) \, dt,  
\label{eq:eq3}
\end{equation}
where $T_r$ is the radial period. The integral in Eq. \eqref{eq:eq3} can be computed using the 2PN expressions in the Keplerian parametrization from Ref.~\cite{Khalil:2021txt}, and a detailed derivation can be found in Appendix \ref{sec:AppendixA}. 

The expression for $\overline{\omega}_0$ can be inverted so that the instantaneous orbital frequency is expressed in terms of the orbit-averaged frequency, eccentricity, relativistic anomaly, mass ratio and spins. Therefore, the instantaneous initial orbital frequency, $\omega_0$, entering the RHS of Eq. \eqref{eq:eqR0omega0}, can be expressed as

\begin{align}
\label{eq:omegaInstAvg}
\omega_0 &= \frac{\overline{\omega }_0 (e \cos \zeta +1)^2}{\left(1-e^2\right)^{3/2}}
-\frac{e \overline{\omega }_0^{5/3} (3 e+2 \cos \zeta ) (e \cos \zeta +1)^2}{\left(1-e^2\right)^{5/2}} \nonumber\\
&\quad
-\frac{\overline{\omega }_0^2e (e+\cos \zeta ) }{\left(e^2-1\right)^3}\left(1+e  \cos \zeta  \right)^2 \left[2 \delta  \chi _A-(\nu -2) \chi _S\right]\nonumber\\
&\quad
-\frac{\overline{\omega }_0^{7/3} (e \cos \zeta +1)^2}{12\left(1-e^2\right)^{7/2}}
\bigg\lbrace
12 e^4 (\nu -6)+8 e \left(e^2 (\nu -15) \right. \nonumber\\
&\qquad
\left. -\nu +6\right) \cos \zeta -3 e^2 \left[2 \left(6 \sqrt{1-e^2}+7\right) \nu  \right. \nonumber\\
&\qquad
\left. -30 \sqrt{1-e^2}+17\right] +18 \left(\sqrt{1-e^2}-1\right) (2 \nu -5)
\bigg\rbrace
\nonumber\\
&\quad
+ \frac{e \overline{\omega }_0^{7/3} (e \cos \zeta +1)^2}{2 \left(1-e^2\right)^{7/2}}
\bigg\lbrace
2 \delta  \chi _A \chi _S \left[e (2 \nu -1) \cos (2 \zeta ) \right. \nonumber\\
&\qquad \left. +2 e (\nu -1)+(8 \nu -4) \cos \zeta \right] +\chi _S^2 \left[-e \left( 2 -4 \nu + 4 \nu ^2\right. \right.  \nonumber\\
&\qquad \left.\left. +(1-2 \nu )^2 \cos (2 \zeta )  \right)-4 (1-2 \nu )^2 \cos \zeta \right] \nonumber\\
&\qquad
+(4 \nu -1) \chi _A^2 \left[e (\cos (2 \zeta )+2)+4 \cos \zeta \right]
\bigg\rbrace +\mathcal{O}\left(\overline{\omega}_0^{8/3}\right).
\end{align}

Given masses, spins, initial eccentricity and relativistic anomaly at a particular orbit-averaged frequency $\overline{\omega}_0$, we use Eq. \eqref{eq:omegaInstAvg} to compute the corresponding instantaneous orbital frequency $\omega_0$, which enters the RHS of Eq. \eqref{eq:eqR0omega0}. The rest of the procedure to compute the EOB initial conditions is not modified. The impact of this new parameterization in the EOB dynamics, can be observed in Fig. \ref{fig:OrbitalFreqExample}, where the solid lines correspond to the orbital frequency evolution of \texttt{SEOBNRv4EHM} using initial conditions specified at an orbit-averaged frequency of 20Hz. The main effect of this new parameterization is the almost constant merger time when varying the initial relativistic anomaly at a fixed value of initial eccentricity (top panel), and the reduction of the length of the evolution with increasing $e_0$ at a fixed value of $\zeta_0$ (bottom panel) due to the larger emission of radiation at periastron passages. This behavior resembles the one in PN models based on the quasi-Keplerian parametrization \cite{Memmesheimer:2004cv, Huerta:2014eca, Klein:2018ybm, Tanay:2019knc,Tiwari:2020hsu}, except for the fact that the EOB initial conditions are not adiabatic as in PN, and thus causing small variations of the merger time with different values of $\zeta_0$ at a fixed value of $e_0$ (see top panel of Fig. \ref{fig:OrbitalFreqExample}).

Finally, sampling in $e_0$ and $\zeta_0$ at $\overline{\omega}_0$, produces notably simpler log-likelihood structures as shown in Fig. \ref{fig:logLExamples}. In the top panel, where the sampling is performed only in $e_0$, the number of oscillations in the log-likelihood curves is reduced with respect to the initial conditions based on the instantaneous orbital frequency. The results of the orbit-average initial conditions create a pattern with an easy to identify maximum in log-likelihood (green curve) at the injected value of $e_0=0.1$. While in the bottom panel, where the sampling in $e_0$ and $\zeta_0$ is performed, it emerges a clearly defined pattern, and the values of $e_0$ and $\zeta_0$ at the maximum log-likelihood point for the orbit-averaged initial conditions are $e^{\text{orb-avg}}_{\log \mathcal{L}_{\text{max}}}= 0.099$, $\zeta^{\text{orb-avg}}_{\log \mathcal{L}_{\text{max}}}=0.77$, while for the instantaneous initial conditions are $e^{\text{orb-avg}}_{\log \mathcal{L}_{\text{max}}}= 0.086$, $\zeta^{\text{orb-avg}}_{\log \mathcal{L}_{\text{max}}}=4.829$. An accurate recovery of the injected values requires more than 5000 points in the 2D parameter, however, the results already indicate that even with a low number of points the initial conditions based on the orbit-averaged orbital frequency are closer to the injected value, and may need less points than the recovery using the initial conditions based on the instantaneous orbital frequency. Thus, the initial conditions at an orbit-averaged frequency may be more adequate for data analysis applications such as parameter estimation, and in Sec.  \ref{sec:ModelInj} we further explore the consequences of these different initial conditions with stochastic sampling techniques.

\subsection{Computational performance}
\label{sec:ODEloosening}
One of the main applications of waveform models is the inference of the source parameters using Bayesian inference  methods. These methods typically require of the order of $\sim 10^7-10^8$ or more waveform evaluations. 
The \texttt{SEOBNRv4EHM} model is built upon the \texttt{SEOBNRv4HM} model \cite{Cotesta:2018fcv}, and thus it inherits the low computational efficiency of the previous generation of \texttt{SEOBNR} models\footnote{The computational efficiency of the \texttt{SEOBNR} models has been recently significantly improved by the new generation of \texttt{SEOBNRv5} models \cite{Khalil:2023kep,Mihaylov:2023bkc,Pompili:2023tna,Ramos-Buades:2023ehm,vandeMeent:2023ols}.}. Several techniques exist to increase the computational efficiency of EOB waveforms, such as reduced-order or surrogate models~\cite{Field:2013cfa,Purrer:2014fza,Purrer:2015tud,Lackey:2016krb,Lackey:2018zvw,Cotesta:2020qhw,Gadre:2022sed,Khan:2020fso,Thomas:2022rmc}, the post-adiabatic approximation \cite{Nagar:2018gnk,Mihaylov:2021bpf},  as well as methods targeting specifically parameter estimation, such as reduced order quadratures \cite{Canizares:2014fya,Smith:2016qas,Morisaki:2020oqk,Tissino:2022thn} or relative binning \cite{Zackay:2018qdy}. %

Here, we decide to increase the efficiency of \texttt{SEOBNRv4EHM} by reducing the absolute and relative tolerances of the 4th-order Runge-Kutta integrator from $10^{-10}$ and $10^{-9}$ to $10^{-8}$ and $10^{-8}$, respectively\footnote{The reduction of the tolerances is a similar approach to the one followed in Ref. \cite{OShea:2021ugg} to improve the efficiency of the \texttt{TEOBResumS-Dali} model \cite{Nagar:2021gss}.}.  Furthermore, as in the \texttt{SEOBNRv4EHM} model the Hamiltonian and the radiation-reaction force are the same as in the \texttt{SEOBNRv4HM}
model, we also use the optimized Hamiltonian and integrator from Refs. \cite{Devine:2016ovp,Knowles:2018hqq}. In order to ease notation we refer to
the model with reduced tolerances and optimizations as \texttt{SEOBNRv4EHM\_opt}, and \texttt{SEOBNRv4E\_opt} when referring to the model containing only the $(l,|m|)=(2,2)$ multipoles. 

The reduction of the ODE tolerances implies an increase in the efficiency of the model with waveform evaluation timescales of the order of $\mathcal{O}(100$ ms), while decreasing the accuracy of the model. In order to quantify the latter across parameter space, we compute the unfaithfulness between the \texttt{SEOBNRv4EHM} and \texttt{SEOBNRv4EHM\_opt} models for 4500 points in the parameter space $q\in [1,50]$, $\chi_{1,2} \in [-0.9.0.9]$, $e_0 \in [0,0.5]$, $\zeta_0 \in [0. 2 \pi]$ for a dimensionless starting frequency of $M \omega_0 = 0.023$. We define the inner product between two waveforms, $h_A$ and $h_B$ \cite{Sathyaprakash:1991mt,Finn:1992xs}
\begin{equation}
\langle h_A, h_B \rangle \equiv 4\ \textrm{Re}\int_{f_{\rm in}}^{f_{\rm max}} df\,\frac{\tilde{h}_A(f) \ \tilde{h}_B^*(f)}{S_n(f)},
\end{equation}
 where a tilde indicates Fourier transform, a star complex conjugation and $S_n(f)$   the power spectral density (PSD) of the detector noise. In this work, we employ for the PSD the LIGO’s “zero-detuned high-power” design sensitivity curve~\cite{Barsotti:2018}. Similarly, as in Ref. \cite{Ramos-Buades:2023ehm}, we use $f_{\rm in} = 10 {\rm Hz}$  and $f_{\rm max} = 2048 {\rm Hz}$. 
 
To assess the agreement between two waveforms --- for instance, the signal, $h_s$, and the template, $h_t$, observed by a detector, we define the faithfulness function \cite{Cotesta:2018fcv,Ossokine:2020kjp},
\begin{equation}
\label{eq:eq16}
\mathcal{F}(M_{\textrm{s}},\iota_{\textrm{s}},{\varphi_0}_{\textrm{s}}) =  \max_{t_c, {\varphi_0}_{t}} \left[\left . \frac{ \langle h_s|h_t \rangle}{\sqrt{  \langle h_s|h_s \rangle  \langle h_t|h_t \rangle}}\right \vert_{\substack{\iota_{\mathrm{s}} = \iota_{t} \\\boldsymbol{\lambda}_\mathrm{s}(t_{\mathrm{s}} = t_{0_\mathrm{s}}) = \boldsymbol{\lambda}_{t}(t_t = t_{0_\mathrm{t}})}} \right ],
\end{equation}
where  $\boldsymbol{\lambda} =\{m_{1,2}, \chi_{1,2},e_0, \zeta_0 \}$ denotes the set of intrinsic parameters of the binary. When comparing waveforms, we choose the same inclination angle for the signal and template $\iota_s = \iota_t = \pi/3$, while the coalescence time and azimuthal angles of the template, $(t_{0_t},\varphi_{0_t})$, are adjusted to maximize the faithfulness of the template.  The maximizations over the coalescence time $t_c$ and coalescence phase ${\varphi_0}_{t}$ are performed numerically. Similarly as in Sec. IV of Ref. \cite{Ramos-Buades:2023ehm} we set a grid of 8 points in the coalescence phase of the signal ${\varphi_0}_s \in [0,2\pi]$, and average over it to compute $\mathcal{F}$.  Finally, we introduce the unfaithfulness or mismatch as
\begin{equation}
\overline{\mathcal{M}}=1-\overline{\mathcal{F}}.
\label{eq:eq20}
\end{equation}

\begin{figure}[H]
\begin{center}
\includegraphics[width=1.\columnwidth]{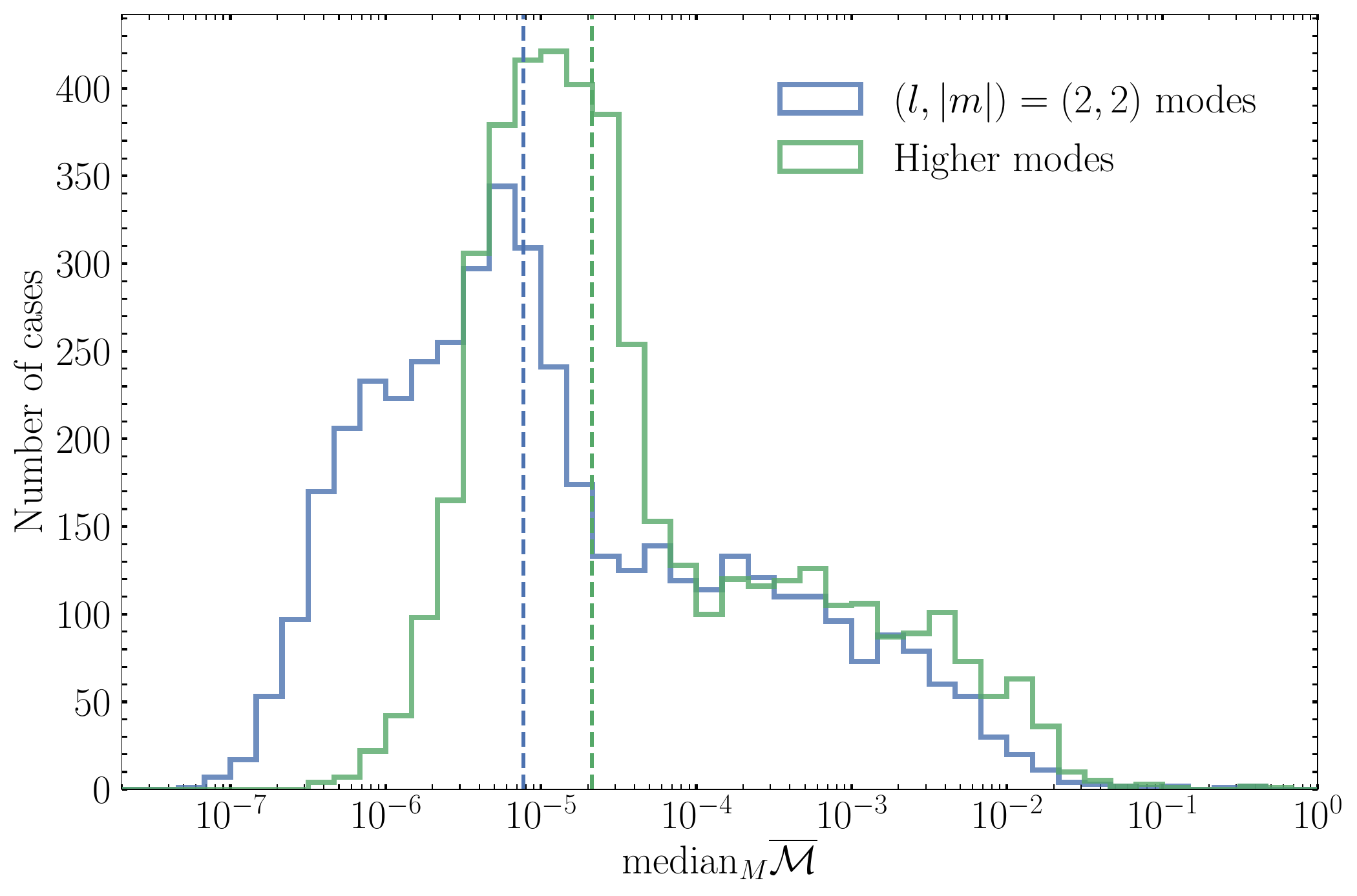}
\caption{Distribution of median unfaithfulness over the total mass range between $[20-300] M_\odot$ for an inclination $\iota=\pi/3$, between the $(l,|m|)=(2,2)$-modes models, \texttt{SEOBNRv4E}  and \texttt{SEOBNRv4E\_opt} (blue), as well as between the higher-order mode models, \texttt{SEOBNRv4EHM}  and \texttt{SEOBNRv4EHM\_opt} (green) for 4500 configurations  in the parameter space $q\in [1,50]$, $\chi_{1,2} \in [-0.9.0.9]$, $e_0 \in [0,0.5]$ and $\zeta_0 \in [0. 2 \pi]$ for a dimensionless starting frequency of $M \omega_0 = 0.023$. The vertical dashed lines indicate the median values of the distribution.}
\label{fig:v4ehmMismatches}
\end{center}
\end{figure}

In Fig. \ref{fig:v4ehmMismatches} we show the distribution of median unfaithfulness over the total mass range $[20-300] M_\odot$ between the models containing only the  $(l,|m|)=(2,2)$ modes, \texttt{SEOBNRv4E}  and \texttt{SEOBNRv4E\_opt} models, as well as for the models including also the subdominant harmonics, i.e.,   $(l,|m|)=(2,2),(2,1),(3,3),(4,4),(5,5)$ multipoles. The results demonstrate a remarkable good agreement between the optimized (\texttt{SEOBNRv4EHM\_opt}  and \texttt{SEOBNRv4E\_opt}) and the original models (\texttt{SEOBNRv4EHM}  and \texttt{SEOBNRv4E}), with a median of unfaithfulness of $7.7\times 10^{-6}$ for the dominant mode models, and $2.1\times 10^{-5}$ for the models including higher order modes. 
As expected the differences between models with higher order modes are larger than for the dominant-mode  models, due to the fact that small changes in the termination of the dynamics caused by the modifications in \texttt{SEOBNRv4EHM\_opt} impact more significantly the higher multipoles. In particular, the reduced tolerances in the integration can lead to small differences in the non-quasicircular coefficients computed from the input values (see Ref. \cite{Bohe:2016gbl} for details), which affect more the higher modes due to their low power.
 Furthermore, we also observe a tail of cases with unfaithfulness larger than $1\%$ between the original and optimized models. These cases correspond to the more challenging parts of the parameter space considered with $e_0 >0.3-0.5$ and high spins ($\chi_{1,2}>0.8-0.9$),  where the models have known limitations, such as the orbit-averaged procedure (see Appendix B of Ref. \cite{Ramos-Buades:2021adz}), and the use of non-quasicircular corrections calibrated to quasi-circular binaries, which may increase the differences between the models.

\begin{figure}[H]
\begin{center}
\includegraphics[width=1.\columnwidth]{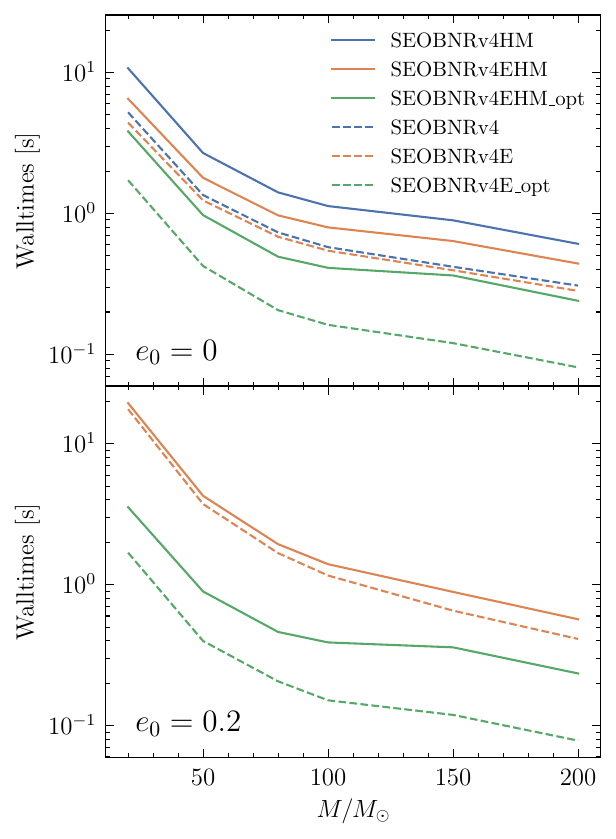}
\caption{Walltimes of the \texttt{SEOBNRv4HM}, \texttt{SEOBNRv4EHM}, \texttt{SEOBNRv4EHM\_opt} models (solid lines) for a configuration with mass ratio $q=3$, dimensionless spins $\chi_1=-0.5$, $\chi_2=0.3$, initial relativistic anomaly $\zeta_0=1 $, total mass range $M \in[20,200]M_\odot$, starting frequency $f_{\rm start}=10$Hz and two different initial eccentricities $e_0=0$ (top panel) and $e_0=0.2$ (bottom panel). The dashed lines correspond to the models with only the $(l,|m|)=(2,2)$ multipoles (\texttt{SEOBNRv4}, \texttt{SEOBNRv4E} and \texttt{SEOBNRv4E\_opt} ). The walltimes are computed using a sampling rate of $8192$Hz for all the masses considered. }
\label{fig:benchmark}
\end{center}
\end{figure}

Finally, we assess the computational efficiency of the
\texttt{SEOBNRv4EHM\_opt} model by timing the waveform generation and
comparing it to the original \texttt{SEOBNRv4EHM} model, as well as
the \texttt{SEOBNRv4HM} model in the quasi-circular limit. We consider
binary's configurations with mass ratio $q=3$, dimensionless spins
$\chi_1=-0.5$, $\chi_2=0.3$, initial relativistic anomaly $\zeta_0=1
$, total mass range $M \in[20,200]M_\odot$, starting frequency $f_{\rm
  start}=10$Hz and two different initial eccentricities $e_0
=0,0.2$. The results of the walltimes to generate the waveforms are
shown in Fig. \ref{fig:benchmark}, where we are including all the
modes up to $l=4$, and a fixed sampling rate of 8192Hz for all the
total masses considered~\footnote{The benchmarks of the waveform
  generation timing were performed on a computing node (dual-socket,
  128-cores per socket, SMT-enabled AMD EPYC (Milan) 7742 (1.5 GHz),
  with 4 GB RAM per core) of the \texttt{Hypatia} cluster at the Max
  Planck Institute for Gravitational Physics in Potsdam.}. The outcome
of the benchmark demonstrates the significant increase in speed of the
\texttt{SEOBNRv4EHM\_opt} model with respect to the
\texttt{SEOBNRv4EHM} and \texttt{SEOBNRv4HM} models.
For the configurations considered, we observe approximately a factor
2-5 improvement in speed. In the quasi-circular limit
\texttt{SEOBNRv4EHM} is on average, over the total mass considered, 
slightly ($\sim 1.5$ times) faster than \texttt{SEOBNRv4HM} due to a more
efficient implementation of some operations involved in the waveform
calculation of \texttt{SEOBNRv4EHM}, while \texttt{SEOBNRv4EHM\_opt}
is a factor $\sim 2.7$ faster than \texttt{SEOBNRv4EHM}.
Regarding models with only the $(l,|m|)=(2,2)$ modes, the hierarchy of curves is similar with the \texttt{SEOBNRv4E\_opt} being faster than \texttt{SEOBNRv4E} by a factor $\sim 3.1$. This increase in speed compared to models with higher-order modes is due to the lack of common operations performed for models with higher-order modes, which limit the speed of the latter. 
When considering an initial eccentricity $e_0=0.2$ in the bottom panel of Fig. \ref{fig:benchmark}, we find an average (over total masses) increase of speed of $\sim 7.7$ and $\sim 3.8$ for the \texttt{SEOBNRv4E\_opt} and \texttt{SEOBNRv4EHM\_opt} models. The speed increase is more significant at low total masses as the main cost of waveform generation comes from the evaluation of the EOB dynamics, which is computed less frequently in the optimized model due to the reduced integration tolerances. 

In summary, the \texttt{SEOBNRv4E\_opt} and \texttt{SEOBNRv4EHM\_opt}
models imply a significant acceleration in waveform evaluation with
respect to their original counterparts with a minor reduction in
accuracy, in the region of parameter space of interest for our study. 
Furthermore, the sampling rate used for the benchmarks here
is rather high\footnote{We choose such a sampling rate value in order
  to resolve all the modes up to l=4 for all the total mass range
  considered, $M \in[20,200]M_\odot$. } (8192Hz), and typical
applications for data analysis may use lower ones, which implies that
the walltimes for waveform evaluation may be further reduced. As a
consequence the optimized models reach speeds competitive for
parameter estimation, and we show in Sec. \ref{sec:PE} that they can
be used to perform parameter-estimation runs in the order of hours and
days.

\section{Bayesian inference study}\label{sec:PE}
 
The main application of the \texttt{SEOBNRv4EHM\_opt} waveform model
is the Bayesian inference of source parameters of GWs emitted by
BBHs. Thus, we introduce the methods and parameter-estimation codes
used to infer the binary parameters in Sec. \ref{sec:PEsetup}, we show
the accuracy of the model in the quasi-circular limit in
Sec. \ref{sec:QClimit}, we assess in Sec. \ref{sec:ModelInj} the
impact of the different initial conditions discussed in
Sec. \ref{sec:eccIC}, as well as the importance of the relativistic-
anomaly parameter. In Sec. \ref{sec:NRInj} we further investigate the
accuracy of the model by performing a series of synthetic NR signal
injections into zero detector noise. Finally, in
Sec. \ref{sec:GWevents}, we analyze three real GW events detected by
the LVK collaboration (GW150914, GW151226 and GW190521), and compare
with results from the literature.

\subsection{Methodology for parameter estimation}\label{sec:PEsetup}

For the parameter-estimation study we employ {\tt parallel Bilby}\footnote{In this paper we employ the {\tt parallel Bilby} code from the public repository  \url{https://git.ligo.org/lscsoft/parallel\_bilby} with the git hash \texttt{b56d25b87b3b33b33a91a8410ae3a6c2a5c92a2e}, which corresponds to the version \texttt{2.0.2}.} \cite{Smith:2019ucc}, a highly parallelized version of the Bayesian inference Python package {\tt Bilby} \cite{Ashton:2018jfp,Romero-Shaw:2020owr}, incorporating the nested sampler \texttt{dynesty} \cite{Speagle:2019ivv}. Based on previous experience with {\tt parallel Bilby} \cite{Ramos-Buades:2023ehm}, we use a number of auto-correlation times  $\mathrm{nact}=30$, number of live points $\mathrm{nlive}=2048$, and the remaining sampling parameters with their default values, unless otherwise specified. Furthermore, the runs are performed using distance marginalization as implemented in \texttt{Bilby}, and the phase marginalization is activated when using the $(l,|m|)=(2,2)$-mode models to further reduce the computational cost.

For the choice of priors, we follow broadly Refs. \cite{Veitch:2014wba,LIGOScientific:2018mvr,Ramos-Buades:2023ehm}. We choose a prior in inverse mass ratio, $1/q$, and chirp mass, $\mathcal{M}$, such that it is uniform in component masses. The priors in initial eccentricity, $e_0$, and relativistic anomaly, $\zeta_0 \in [0,2\pi ]$, are chosen to be uniform. In order to facilitate the comparison with precessing-spin results, the priors on the spin-components, $\chi_{1,2}$, are chosen such that they correspond to the projections of a uniform and isotropic spin distribution along the $\hat{z}$-direction \cite{Veitch:2014wba}. The luminosity distance prior is chosen to be proportional to $\propto d^2_L$, unless otherwise specified. The rest of the priors are set according to Appendix C of Ref.~\cite{LIGOScientific:2018mvr}. The specific values of the prior boundaries for the different parameters vary depending on the application, and we specify them in the subsequent sections.

\subsection{Quasi-circular limit}\label{sec:QClimit}
Eccentricity is a parameter, which defines the ellipticity of an orbit between two limits: the parabolic and the  circular case. In this section we consider the latter, and demonstrate that the  \texttt{SEOBNRv4EHM\_opt} model is able to accurately describe GWs from quasi-circular BBHs. In Sec. III of Ref. \cite{Ramos-Buades:2021adz} it is shown that \texttt{SEOBNRv4EHM} has a comparable accuracy to \texttt{SEOBNRv4HM} by computing the unfaithfulness against quasi-circular NR waveforms. 

\begin{figure}[H]
\begin{center}
\includegraphics[width=1.\columnwidth]{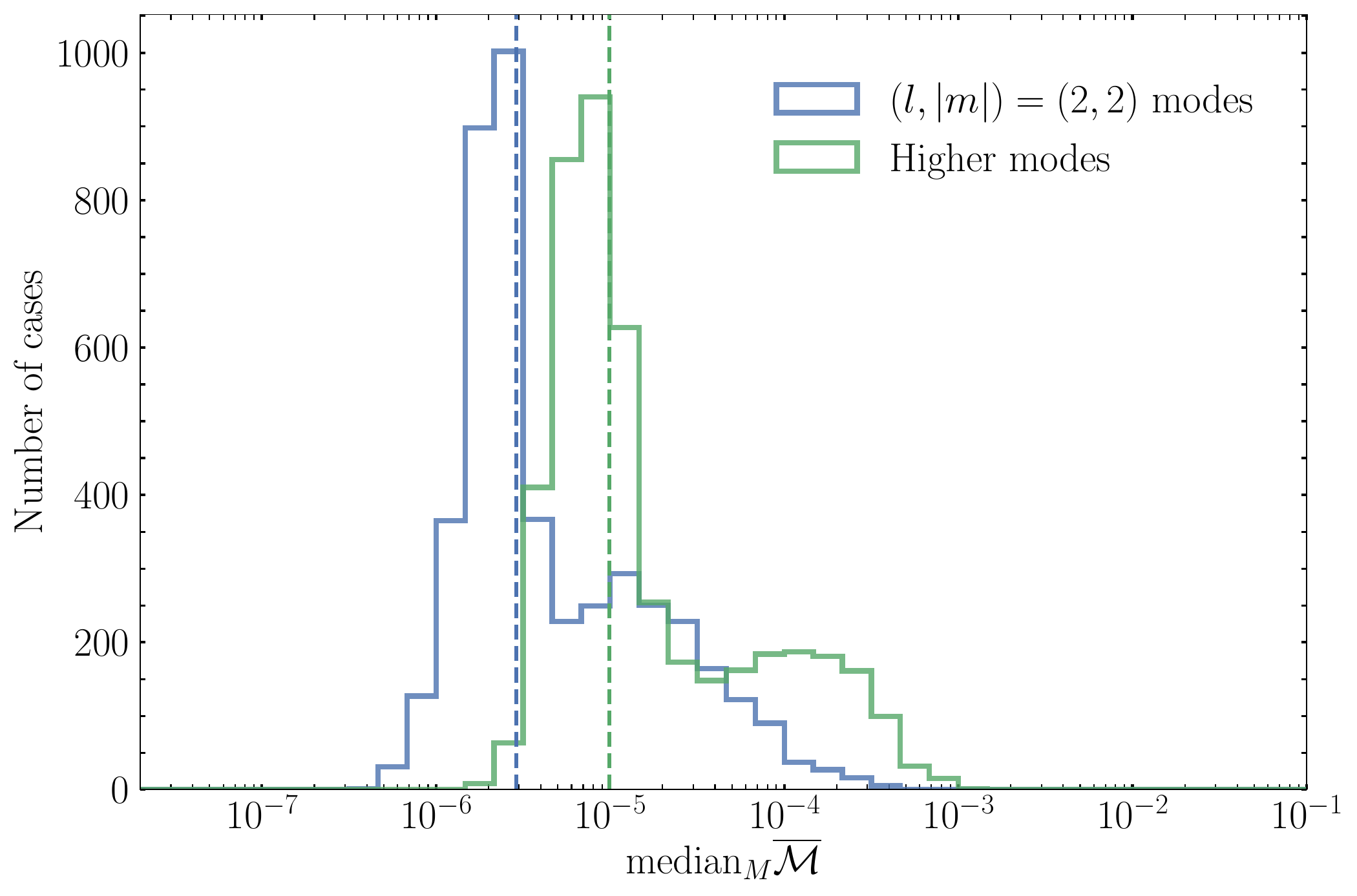}
\caption{Distribution of median unfaithfulness over the total mass range between $[20-300] M_\odot$ for an inclination $\iota=\pi/3$, between the $(l,|m|)=(2,2)$-modes models, \texttt{SEOBNRv4E\_opt}  and \texttt{SEOBNRv4} (blue), as well as between the higher-order mode models, \texttt{SEOBNRv4EHM\_opt}  and \texttt{SEOBNRv4HM} (green) for 4500 configurations in the parameter space $q\in [1,50]$, $\chi_{1,2} \in [-0.9.0.9]$ for a dimensionless starting frequency of $M \omega_0 = 0.023$. The vertical dashed lines indicate the median values of the distribution.}
\label{fig:qcMismatches}
\end{center}
\end{figure}

Here, we assess the accuracy of \texttt{SEOBNRv4EHM\_opt} in the zero-eccentricity limit by computing the unfaithfulness, as defined in Sec. \ref{sec:ODEloosening}, against the accurate quasi-circular \texttt{SEOBNRv4HM} model for 4500 random points distributed in the following parameter space:  $q\in [1,50]$, $\chi_{1,2} \in [-0.9.0.9]$, with an inclination angle of $\iota = \pi/3$, for a total mass range $[20-300] M_\odot$ and starting frequency of $M \omega_0 = 0.023$ in geometric units. In Fig. \ref{fig:qcMismatches} we show the distribution of the median unfaithfulness over the total mass range considered when comparing the $(l,|m|)=(2,2)$-mode models, \texttt{SEOBNRv4} and \texttt{SEOBNRv4E\_opt}, as well as the corresponding models including higher multipoles, \texttt{SEOBNRv4HM} and \texttt{SEOBNRv4EHM\_opt}. For \texttt{SEOBNRv4E\_opt} the median value of unfaithfulness is $8.1\times 10^{-6}$, while when including higher order modes it degrades to $3.8 \times 10^{-5}$. In both cases there are no configurations with unfaithfulness larger than $1\%$. Therefore, both the \texttt{SEOBNRv4E\_opt} and \texttt{SEOBNRv4EHM\_opt} models are faithful across parameter space to the \texttt{SEOBNRv4} and  \texttt{SEOBNRv4HM} models, respectively, considering that \texttt{SEOBNRv4} was calibrated to NR with unfaithfulness below $1\%$.

We further investigate the implications of these unfaithfulness results in parameter estimation by performing a mock-signal injection into zero-detector noise. With zero noise, and flat priors, the likelihood will peak at the true parameters when using the same model for injection and recovery. This makes it easier to see biases that are arising from model differences. 
We use the  \texttt{SEOBNRv4} model as a signal, and recover the injected parameters with the reduced order model (ROM) \texttt{SEOBNRv4\_ROM} \cite{Bohe:2016gbl} and the \texttt{SEOBNRv4E\_opt} model. For the latter we also sample in initial eccentricity and relativistic anomaly. We consider a configuration with mass ratio $q=4$, total mass $M=90.08 M_\odot$ and BH's dimensionless spins $\chi_1 = 0.5$ and $\chi_2 = -0.1$ defined at 20Hz.

\begin{figure*}[htpb!] 
\includegraphics[width=0.65\columnwidth]{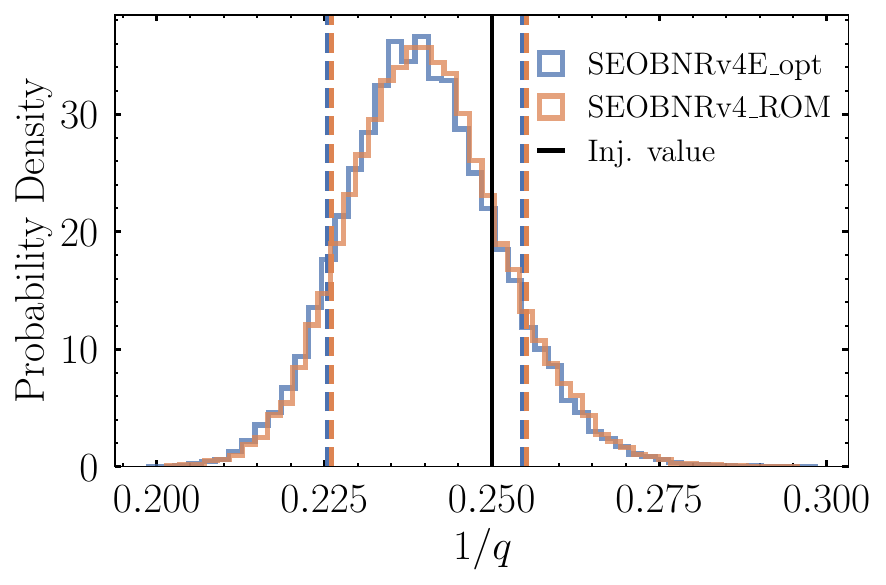} 
\includegraphics[width=0.65\columnwidth]{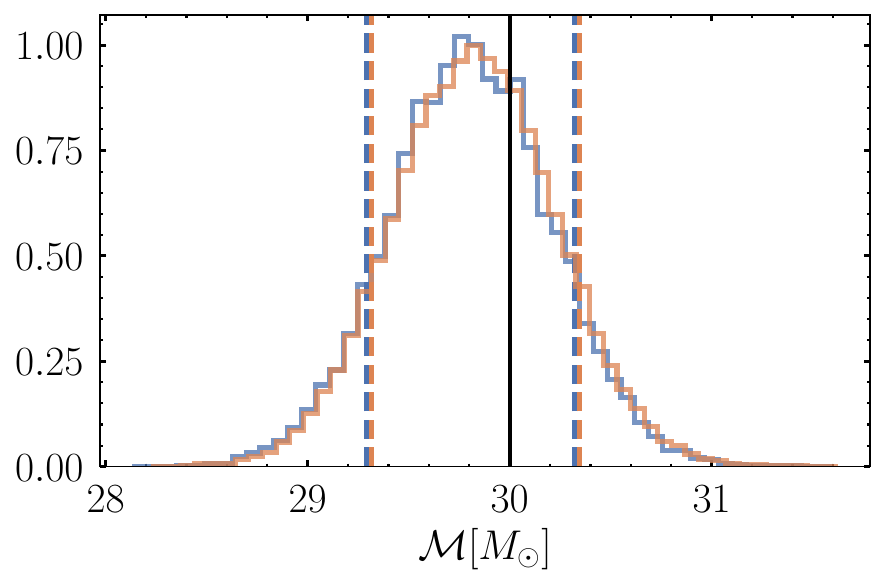} 
\includegraphics[width=0.65\columnwidth]{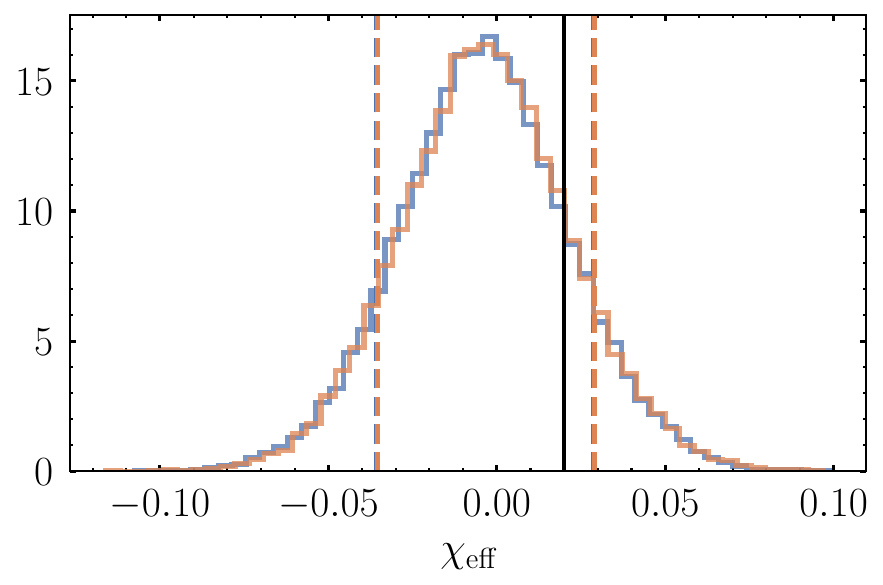} 
\includegraphics[width=0.65\columnwidth]{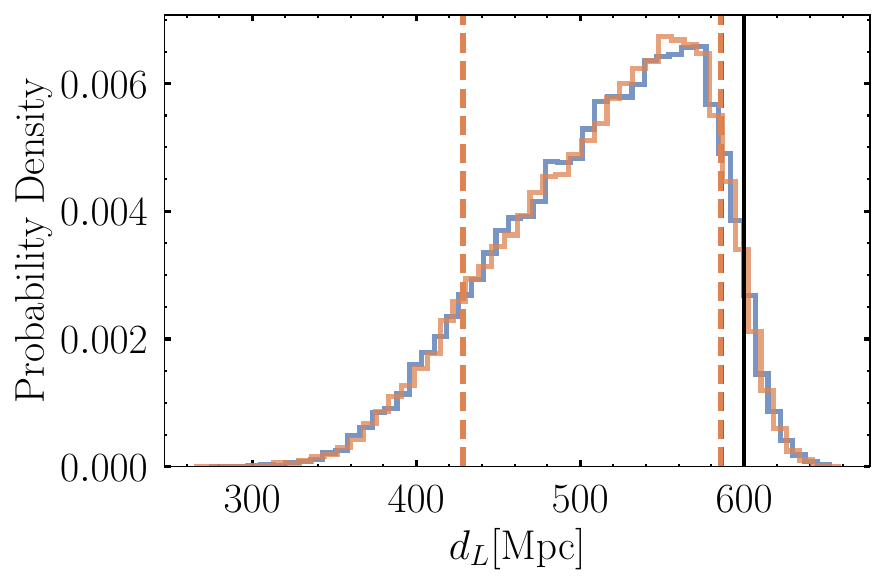} 
\includegraphics[width=0.65\columnwidth]{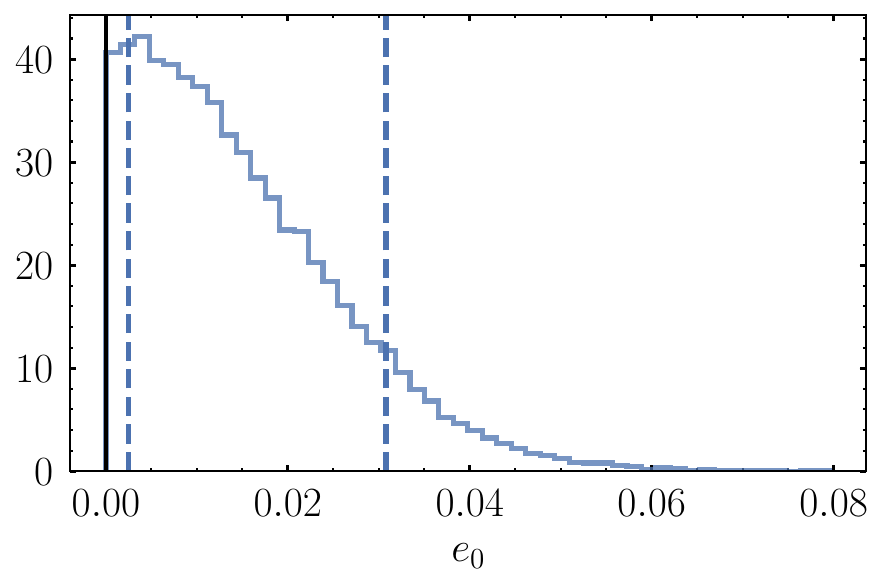} 
\includegraphics[width=0.65\columnwidth]{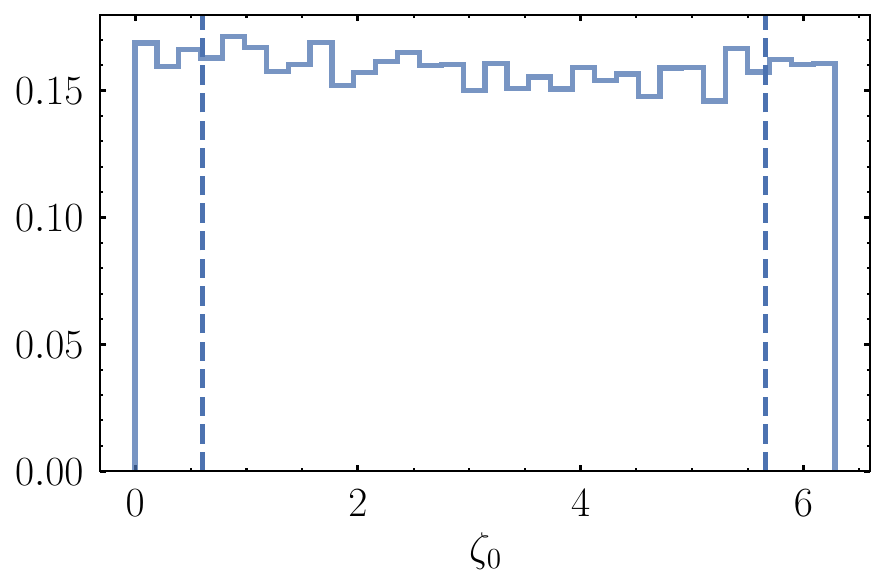}
 \caption{ Inverse mass ratio, chirp mass, effective spin parameter, luminosity distance, initial eccentricity and relativistic anomaly posterior distributions for a synthetic quasi-circular BBH signal injection using the \texttt{SEOBNRv4} model, and recovering the parameters with \texttt{SEOBNRv4E\_opt} (blue) and \texttt{SEOBNRv4\_ROM} (orange). The dashed vertical lines indicate the $90\%$ credible intervals. The solid vertical lines correspond to the injected parameters, which are also shown in Table \ref{tab:injection_settings}.}
\label{fig:qcInj}
\end{figure*}

For this injection we choose the inclination with respect to the line of sight of the BBH to be $\iota=0.1$ rad, coalescence and polarization phases are $\phi=0.6$ rad and $\psi=0.33$ rad, respectively. The luminosity distance to the source is chosen to be 600 Mpc, which produces a three-detector (LIGO Hanford, LIGO Livingston and Virgo) network-SNR of $\rho_{\mathrm{mf}}^N = 67.9$ when using the LIGO and Virgo PSD at design sensitivity \cite{Barsotti:2018}. 

We choose a uniform prior in inverse mass ratio and chirp mass, with ranges $1/q \in [0.05,1]$ and $\mathcal{M} \in [5,100] M_\odot$. The priors on the magnitudes of the dimensionless $z$-components of the spins are  $a_i \in [0,0.99]$.
For \texttt{SEOBNRv4E\_opt} we take a uniform prior in the initial eccentricity $e_0 \in [0,0.3]$, and uniform in the initial relativistic anomaly $\zeta_0 \in [0,2 \pi]$. 

The resulting posteriors for the inverse mass ratio, chirp mass, effective-spin parameter, luminosity distance, initial eccentricity and relativistic anomaly are shown in Fig. \ref{fig:qcInj}. We find remarkable agreement between the posteriors of \texttt{SEOBNRv4E\_opt}  and \texttt{SEOBNRv4E\_ROM} for all the parameters. The injected values and recovered parameters are displayed in Table \ref{tab:injection_settings}, where additional parameters are shown. Even for a relatively high SNR injection ($\rho^N_\mathrm{mf} \sim 68$) both models are able to accurately recover the injected parameters for the inverse mass ratio, chirp mass and effective-spin parameter within the $90 \%$ credible intervals, while the luminosity distance  presents a small bias due to the limited mode content of both models (only the $(l,|m|)=(2,2)$ multipoles), which creates a degeneracy between the inclination angle and the luminosity distance and thus, complicates the measurement of both quantities ~\cite{Littenberg:2012uj,Varma:2014jxa,Cotesta:2018fcv,Krishnendu:2021cyi}. 
This degeneracy and, thus the bias, can be removed by including higher multipoles in the waveform, but we do not use models with higher multipoles, as the focus of this injection study is the assessment of the agreement between the \texttt{SEOBNRv4E\_opt}  and \texttt{SEOBNRv4E\_ROM} models.

\setlength{\extrarowheight}{8pt}
\begin{table}[h!]
    \centering
    \begin{tabular}{ c  c  c  c   }
 \hline
 \hline
 Parameter & \makecell[cc]{Injected \\ value} & \makecell[cc]{SEOBNRv4E\_opt} & \makecell[cc]{SEOBNRv4\_ROM}  \\
 \hline
 \centering
$M/M_\odot$ &   $90.08$ &  ${90.96}^{+1.44}_{-1.51} $ &  ${90.95}^{+1.42}_{-1.5} $ \\
 $\mathcal{M}/M_\odot$ &   $30.00$ & ${29.81}^{+0.51}_{-0.51} $ &   ${29.83}^{+0.52}_{-0.51} $  \\
 $1/q$ & $0.25 $ & ${0.24}^{+0.02}_{-0.02}$ & ${0.24}^{+0.02}_{-0.02}$  \\
 $\chi_{\text{eff}}$ & $0.02$ & ${0.00}^{+0.03}_{-0.03} $ & ${0.00}^{+0.03}_{-0.03}  $ \\
 $e_0$ & $0$ & ${0.01}^{+0.02}_{-0.01} $ & - \\
 $\zeta_0$ & - & ${3.09}^{+2.57}_{-2.48} $ & - \\
$\theta_{\text{JN}}$ & $0.1$ &  ${0.52}^{+0.29}_{-0.3} $ & ${0.51}^{+0.29}_{-0.30} $  \\
 $d_{L}$ & $600$ & ${521.35}^{+64}_{-93} $ &  ${523}^{+63}_{-94} $ \\
$\phi_{\text{ref}}$ & $0.1$ & ${3.15}^{+2.53}_{-2.52} $ &  ${3.16}^{+2.54}_{-2.52} $ \\
 $\rho^{\mathrm{N}}_{\mathrm{mf}}$  & $67.98$ &  ${64.34}^{+0.06}_{-0.06} $ &  ${64.42}^{+0.02}_{-0.03} $  \\  [0.1cm]
 \hline
 \hline
    \end{tabular}
 \caption{Injected and median values of the posterior distributions for the synthetic injection with the \texttt{SEOBNRv4} model, recovered with \texttt{SEOBNRv4E\_opt} and \texttt{SEOBNRv4\_ROM}. The median values also report the $90\%$ credible intervals. The binary parameters correspond to the total mass $M$, chirp mass $\mathcal{M}$, inverse mass ratio $1/q$, effective-spin parameter  $\chi_{\text{eff}}$, initial eccentricity $e_0$, initial relativistic anomaly $\zeta_0$, angle between the total angular momentum and the line of sight $\theta_{\text{JN}}$, luminosity distance $d_{L}$, coalescence phase $\phi_{\text{ref}}$ and the network matched-filtered SNR for LIGO-Hanford/Livingston and Virgo detectors  $\rho^{N}_{\mathrm{mf}}$.
  }
 \label{tab:injection_settings}
\end{table}

Regarding initial eccentricity and relativistic anomaly, \texttt{SEOBNRv4E\_opt} measures an initial eccentricity consistent with zero, $e_0 = {0.01}^{+0.02}_{-0.01}$, while the initial relativistic anomaly becomes an uninformative parameter as for quasi-circular orbits the radial phase provides no information about the position of the binary components. 

In conclusion, the \texttt{SEOBNRv4EHM\_opt} model has an accurate zero eccentricity limit comparable to the underlying quasi-circular \texttt{SEOBNRv4HM} model. This is extremely important as eccentricity is a gauge dependent parameter in General Relativity between two well defined limits (circular and parabolic), and being able to unambiguously recover one of them ensures that the values of eccentricity can be quoted with respect to a uniquely defined physical configuration\footnote{For instance, in the \texttt{TEOBResumS} family of waveform models \cite{Nagar:2018gnk,Nagar:2018zoe}, Ref. \cite{Bonino:2022hkj} found a small bias between the quasi-circular \texttt{TEOBResumS-GIOTTO} \cite{Nagar:2021xnh} and the eccentric \texttt{TEOBResumS-Dali} \cite{Nagar:2021gss} models in the quasi-circular limit.}. 


\subsection{Eccentric case}\label{sec:ModelInj}

Two different initial conditions (ICs) for the \texttt{SEOBNRv4EHM} model were presented in Sec. \ref{sec:eccIC}: 1) one based on the specification of the initial eccentricity and relativistic anomaly, $(e_0, \zeta_0)$ at an instantaneous orbital frequency $\omega_0$, hereafter referred as \textit{instantaneous} ICs, and 2) one based on the specification of $(e_0, \zeta_0)$ at an orbit-averaged orbital frequency $\overline{\omega}_0$, hereafter called \textit{orbit-averaged} ICs.

\setlength{\extrarowheight}{8pt}
\begin{table*}[!]
    \centering 
\begin{tabular*}{\textwidth}{c@{\extracolsep{\fill}} c c c c c c  c  c  c  c c}
 \hline
 \hline
 Parameter & $M/M_\odot$ & $\mathcal{M}/M_\odot$ & $1/q$ &  $\chi_{\text{eff}}$ & $e_0$ &  $\zeta_0$ & $\theta_{\text{JN}}$ & $d_{L}$ &  $\phi_{\text{ref}}$ &  $\rho^{\mathrm{N}}_{\mathrm{mf}}$  \\[0.05cm]
 \hline \hline
 \centering
 \makecell[cc]{Injected \\ value} & ${76.45}$  & ${28.0}$  & ${0.33}$  & ${-0.22}$  & $\textbf{0.1}$  & ${1.2}$  & ${0.1}$  & ${800.0}$  & ${0.1}$  & ${46.3}$   \\\hline
 \makecell[c]{$\overline{\omega}_0$ ICs}  & ${77.93}^{+1.82}_{-1.88} $  & ${28.09}^{+0.57}_{-0.59} $  & ${0.32}^{+0.03}_{-0.02} $  & ${-0.23}^{+0.05}_{-0.05} $  & ${0.1}^{+0.01}_{-0.01} $  & ${1.04}^{+0.46}_{-0.44} $  & ${0.53}^{+0.32}_{-0.32} $  & ${699.63}^{+96.27}_{-140.91} $  & ${3.14}^{+2.52}_{-2.52} $  & ${46.3}^{+0.04}_{-0.06} $  \\
 \makecell[cc]{$\omega_0$ ICs}  & ${77.67}^{+1.84}_{-1.91} $  & ${27.9}^{+0.64}_{-0.68} $  & ${0.31}^{+0.03}_{-0.03} $  & ${-0.24}^{+0.05}_{-0.06} $  & ${0.1}^{+0.01}_{-0.01} $  & ${1.2}^{+0.07}_{-0.09} $  & ${0.53}^{+0.32}_{-0.31} $  & ${691.15}^{+96.25}_{-140.43} $  & ${3.12}^{+2.51}_{-2.53} $  & ${46.24}^{+0.04}_{-0.07} $  \\
 \makecell[cc]{$\overline{\omega}_0$ ICs \\ ($\zeta_0=0$)}  & ${76.01}^{+1.69}_{-1.63} $  & ${26.84}^{+0.4}_{-0.43} $  & ${0.3}^{+0.03}_{-0.02} $  & ${-0.31}^{+0.04}_{-0.05} $  & ${0.09}^{+0.01}_{-0.01} $  & -  & ${0.54}^{+0.32}_{-0.31} $  & ${650.18}^{+89.02}_{-126.32} $  & ${3.15}^{+2.69}_{-2.51} $  & ${46.23}^{+0.04}_{-0.07} $    \\
\makecell[cc]{$\omega_0$ ICs \\ ($\zeta_0=0$) }   & ${77.86}^{+1.9}_{-1.95} $  & ${27.99}^{+0.65}_{-0.67} $  & ${0.31}^{+0.03}_{-0.03} $  & ${-0.24}^{+0.05}_{-0.05} $  & ${0.1}^{+0.01}_{-0.03} $  & - & ${0.53}^{+0.32}_{-0.32} $  & ${697.18}^{+97.41}_{-139.64} $  & ${3.15}^{+2.48}_{-2.56} $  & ${46.08}^{+0.04}_{-0.06} $   \\
\makecell[cc]{$e_0=0$}  & ${77.01}^{+2.07}_{-2.16} $  & ${28.07}^{+0.54}_{-0.55} $  & ${0.33}^{+0.03}_{-0.03} $  & ${-0.23}^{+0.05}_{-0.05} $  & - & -  & ${0.54}^{+0.32}_{-0.32} $  & ${715.39}^{+99.1}_{-145.79} $  & ${3.14}^{+2.49}_{-2.52} $  & ${45.39}^{+0.03}_{-0.05} $   \\  [0.1cm]
 \hline\hline
 \makecell[cc]{Injected \\ value} & ${76.45}$  & ${28.0}$  & ${0.33}$  & ${-0.22}$  & $\textbf{0.2}$  & ${1.2}$  & ${0.1}$  & ${800.0}$  & ${0.1}$  & ${46.75}$  & \\\hline
 \makecell[c]{$\overline{\omega}_0$ ICs}  & ${78.37}^{+2.23}_{-2.18} $  & ${28.49}^{+0.51}_{-0.53} $  & ${0.33}^{+0.03}_{-0.03} $  & ${-0.21}^{+0.05}_{-0.05} $  & ${0.2}^{+0.01}_{-0.01} $  & ${0.82}^{+0.48}_{-0.42} $  & ${0.53}^{+0.32}_{-0.32} $  & ${716.81}^{+96.1}_{-142.65} $  & ${3.17}^{+2.51}_{-2.53} $  & ${46.65}^{+0.04}_{-0.06} $ \\
 \makecell[cc]{$\omega_0$ ICs}  & ${77.31}^{+1.81}_{-1.78} $  & ${28.0}^{+0.65}_{-0.63} $  & ${0.32}^{+0.03}_{-0.02} $  & ${-0.24}^{+0.06}_{-0.06} $  & ${0.2}^{+0.01}_{-0.01} $  & ${1.21}^{+0.04}_{-0.04} $  & ${0.53}^{+0.32}_{-0.31} $  & ${698.53}^{+95.21}_{-141.05} $  & ${3.16}^{+2.49}_{-2.55} $  & ${47.07}^{+0.04}_{-0.07} $  \\
 \makecell[cc]{$\overline{\omega}_0$ ICs \\ ($\zeta_0=0$)}  & ${77.08}^{+2.06}_{-1.93} $  & ${27.49}^{+0.34}_{-0.35} $  & ${0.31}^{+0.02}_{-0.02} $  & ${-0.26}^{+0.04}_{-0.04} $  & ${0.2}^{+0.01}_{-0.01} $  & -  & ${0.54}^{+0.31}_{-0.32} $  & ${676.57}^{+87.97}_{-129.13} $  & ${3.15}^{+2.87}_{-2.16} $  & ${46.61}^{+0.04}_{-0.06} $   \\
\makecell[cc]{$\omega_0$ ICs \\ ($\zeta_0=0$) }   &  ${79.12}^{+1.75}_{-1.73} $  & ${28.76}^{+0.59}_{-0.6} $  & ${0.33}^{+0.03}_{-0.02} $  & ${-0.19}^{+0.05}_{-0.05} $  & ${0.24}^{+0.03}_{-0.02} $  & -  & ${0.53}^{+0.32}_{-0.32} $  & ${734.54}^{+97.61}_{-147.69} $  & ${3.2}^{+2.51}_{-2.52} $  & ${46.61}^{+0.04}_{-0.06} $   \\
\makecell[cc]{$e_0=0$}  & ${82.95}^{+2.39}_{-2.54} $  & ${31.6}^{+1.09}_{-0.94} $  & ${0.39}^{+0.05}_{-0.04} $  & ${-0.17}^{+0.06}_{-0.06} $  & -  & -  & ${0.53}^{+0.32}_{-0.31} $  & ${874.23}^{+125.43}_{-179.57} $  & ${3.11}^{+2.55}_{-2.5} $  & ${43.38}^{+0.04}_{-0.06} $  \\  [0.1cm]
 \hline
 \hline
    \end{tabular*}   
 \caption{Injected and median values of the posterior distributions for two synthetic injections with the \texttt{SEOBNRv4E\_opt} model with initial eccentricities $e_0 = [0.1,0.2]$. The median values also report the $90\%$ credible intervals. The binary parameters correspond to the total mass $M$, chirp mass $\mathcal{M}$, inverse mass ratio $1/q$, effective spin parameter  $\chi_{\text{eff}}$, initial eccentricity $e_0$, initial relativistic anomaly $\zeta_0$, angle between the total angular momentum and the line of sight $\theta_{\text{JN}}$, luminosity distance $d_{L}$, coalescence phase $\phi_{\text{ref}}$ and the network matched-filtered SNR for LIGO-Hanford/Livingston and Virgo detectors  $\rho^{N}_{\mathrm{mf}}$. For each injection the recovery is done with the  \texttt{SEOBNRv4E\_opt} model using orbit-averaged ICs ($\overline{\omega}_0$ ICs), instantaneous ICs ($\omega_0$ ICs), and setting $\zeta_0=0$ for both the orbit-averaged and instantaneous ICs, $ \overline{\omega}_0$ ICs  ($\zeta_0=0$) and $\overline{\omega}_0$ ICs  ($\zeta_0=0$), respectively. Additionally, the values recovered setting the initial eccentricity to zero, $e_0=0$, are also shown.
  }
 \label{tab:Eccinjections}
\end{table*}

Here, we study the implications of these ICs by performing two mock-signal injections in zero detector noise using the \texttt{SEOBNRv4E\_opt} model as a signal with two different initial eccentricities $e_0 = [0.1,0.2]$, and recovering with the \texttt{SEOBNRv4E\_opt} model. We consider a configuration with mass ratio $q=3$, initial relativistic anomaly $\zeta_0 =1.2$, total mass $M=76.4 M_\odot$ and BH's dimensionless spins $\chi_1 = 0.5$ and $\chi_2 = -0.1$ defined at a starting frequency of 20Hz\footnote{The starting frequency of the injected signal is orbit-average or instantaneous in order to be consistent with the ICs used for the recovery of the parameters.}

\begin{figure*}[htpb!]  
\includegraphics[width=2.07\columnwidth]{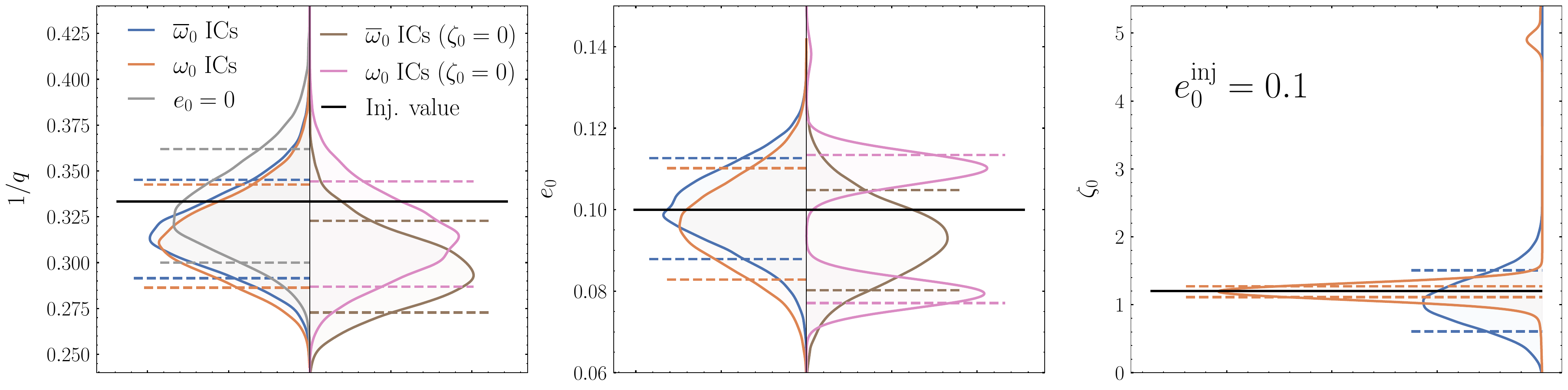} \\
\includegraphics[width=2.07\columnwidth]{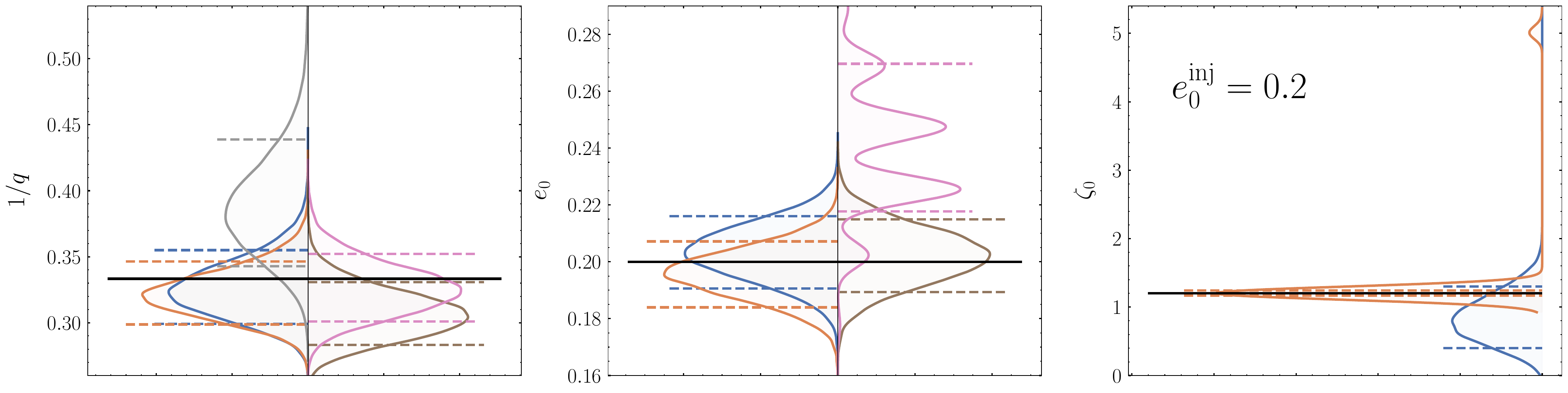}  
 \caption{ \textit{Top row}: Violin plots for the posterior distributions of the inverse mass ratio, initial eccentricity and relativistic anomaly for a synthetic BBH signal injection using the \texttt{SEOBNRv4E\_opt} model with $e_0=0.1$ and $\zeta_0=1.2$ at 20Hz. 
 The posteriors are computed using the \texttt{SEOBNRv4E\_opt} model with orbit-averaged ICs (blue), instantaneous ICs (orange), setting $\zeta_0=0$ for the orbit-averaged (brown) and instantaneous (pink) ICs, and for the zero-eccentricity case $e_0=0$ (gray). The cases where $\zeta$ is fixed during sampling ($\zeta_0 = 0$) have been placed on the right side of the x-axis to ease the visualization of the results.  The dashed vertical lines indicate the $90\%$ credible intervals. The solid vertical lines correspond to the injected parameters, which are also shown in Table \ref{tab:injection_settings}. \textit{Bottom row}: Same as in the upper row but for an injected signal with initial eccentricity $e_0=0.2$ at 20Hz.}
\label{fig:eccInj}
\end{figure*}

The priors on the sampling parameters are chosen as in
Sec. \ref{sec:QClimit}. The injection and recovery of the parameters
are performed using the \texttt{SEOBNRv4E\_opt} model with orbit-averaged
and instantaneous ICs. Furthermore, we also assess the importance of
the relativistic anomaly parameter, $\zeta_0$, by recovering the
parameters by setting the initial starting point of the orbit at
periastron ($\zeta_0=0$), which is different from the injected value
$\zeta_0 =1.2$. For completeness we also measure the binary parameters
in the quasi-circular limit ($e_0=0$). The resulting posterior
distributions of the different cases are shown in
Fig. \ref{fig:eccInj} for the inverse mass-ratio, initial eccentricity
and initial relativistic anomaly. The injected values, the median
values and $90\%$ credible intervals of the posterior distributions of
some parameters are provided in Table \ref{tab:Eccinjections}.

The left side of the violin plots in Fig. \ref{fig:eccInj} shows the comparison of the orbit-averaged and instantaneous ICs as well as the zero-eccentricity case. We find that both ICs are able to accurately recover the injected parameters for all
the parameters of the two injected eccentricities, $e_0=0.1$ and
$e_0=0.2$. In the case of the instantaneous ICs the $\zeta_0$
posterior is sharply peaked at the injected value due to the fact that
for these ICs a change in $\zeta_0$ at a fixed value of $e_0$ provides
a completely different evolution, while for the orbit-averaged ICs
variations of $\zeta_0$ at a fixed value $e_0$ describe different
values of the radial phase in the same orbit as shown in
Fig. \ref{fig:OrbitalFreqExample}.
These different descriptions of the orbits have also implications in
the efficiency of the sampler and the cost of the parameter estimation
runs. For instance, the $e_0=0.2$ injections were performed with the
same sampler settings as described in Sec. \ref{sec:PEsetup} and using
6 nodes of 32 cores each\footnote{The parameter estimation runs were
  performed using computing nodes (dual-socket, 16-cores per socket,
  SMT-enabled AMD EPYC (Milan) 7351 (2.4 GHz), with 8 GB RAM per core)
  of the Hypatia cluster at the Max Planck Institute for Gravitational
  Physics in Potsdam.} with an averaged wall clock time of $11.8$ and
$35.7$ hours for the orbit-averaged and instantaneous ICs,
respectively. This demonstrates that the orbit-averaged ICs lead to a
more computationally efficient sampling of the parameter space than
the instantaneous ICs without a loss of accuracy.

We focus now on the right side of the violin plots in Fig. \ref{fig:eccInj}, where the results of setting the initial relativistic-anomaly parameter 
to a different value ($\zeta_0=0$) from the injected
one ($\zeta_0=1.2$) for both the instantaneous and the orbit-averaged
ICs are displayed.
 For the orbit-averaged ICs, neglecting the relativistic anomaly
parameter can lead to biases in the quasi-circular parameters like the
inverse mass ratio, while the eccentricity parameter is accurately
recovered (see right-side distributions in Fig. \ref{fig:eccInj}). This can be explained by the
fact that during the parameter-estimation run the template waveform
compensates the lack of another eccentric degree of freedom by
modifying the rest of the parameters in order to describe more
accurately the signal. This points out the relevance of taking into
account the radial-phase parameter when using orbit-averaged ICs, and
is in agreement with the findings of Ref. \cite{Islam:2021mha} during
the construction of an eccentric surrogate-waveform model. Regarding
the instantaneous ICs, the quasi-circular parameters are recovered
with no biases, but a multi-modal posterior in the eccentricity
parameter is found in the middle panels of Fig. \ref{fig:eccInj}. 
These multimodalities can be explained by a degeneracy in the waveforms with
fixed $\zeta_0$ at a relatively high total mass ($M\sim 76 M_\odot$),
and the fact that the model assumes circularization at
merger-ringdown. As shown in Sec. \ref{sec:NRInj} these
multimodalities can sometimes be an artifact of the specific
parameterization chosen, and be removed by going to a definition of
eccentricity based on the waveform. However, this is not the case for
the injections shown in this section, and the multimodalities remain
even after a redefinition of the eccentric parameters, indicating that for these particular injections these are true multi-modalities in the posterior distribution.
Therefore, the instantaneous ICs when neglecting $\zeta_0$ can
substantially complicate the sampling as well as the measurement of
the eccentricity parameter due to the multimodalities.

Finally, we also show in Fig. \ref{fig:eccInj} and Table
\ref{tab:Eccinjections} the results of sampling with the initial
eccentricity set to zero. For the injection with small eccentricity at
the injected SNR ($\sim 46$) the quasi-circular model is still able to
recover the parameters accurately within the $90 \%$ credible
intervals. However, for the high eccentricity injection ($e_0=0.2$)
this is no longer the case and the biases in parameters like the chirp
mass can be as large as $8\%$ with respect to the injected value.

In summary, the orbit-averaged ICs provide a more efficient sampling
of the eccentric parameter space than the instantaneous ones without a
loss of accuracy. As a consequence we adopt the orbit-averaged ICs
hereafter for the analysis using the \texttt{SEOBNRv4EHM\_opt} model
in this paper. Furthermore, we have shown that neglecting the radial-phase 
parameter, as currently done in the \texttt{TEOBResumS-Dali} and
the \texttt{SEOBNREHM} \cite{Liu:2021pkr} models can lead to biases in
the recovered parameters for both instantaneous and orbit-averaged
ICs, unless one varies additional parameters. For instance, in
Ref. \cite{Bonino:2022hkj} to avoid biases in the posteriors with
\texttt{TEOBResumS-Dali}, which only employs the eccentricity
parameter to describe elliptical orbits, the starting frequency of the
waveform is also sampled during the parameter estimation
runs. However, we find more natural to keep the starting frequency of
the waveform fixed as done in the LVK analysis of the Gravitational
Wave Transient Catalogs
\cite{LIGOScientific:2020ibl,LIGOScientific:2021usb,LIGOScientific:2021djp},
and thus, use two eccentric parameters, eccentricity and relativistic
anomaly, which can vary freely during the parameter-estimation run.

\subsection{Numerical-relativity injections}\label{sec:NRInj}
In Ref. \cite{Ramos-Buades:2021adz}, the \texttt{SEOBNRv4EHM} model was shown to be accurate to with an unfaithfulness below $1\%$ for a dataset of public eccentric NR waveforms from the SXS catalog \cite{Hinder:2017sxy,Boyle:2019kee}. In this section we further investigate the accuracy of the \texttt{SEOBNRv4EHM} model against NR waveforms by performing zero-noise injections, and recovering the parameters with the \texttt{SEOBNRv4E\_opt} model. 

We consider a set of 3 public eccentric simulations
\texttt{SXS:BBH:1355}, \texttt{SXS:BBH:1359} and
\texttt{SXS:BBH:1363}, which correspond to equal-mass, nonspinning
configurations with initial eccentricities measured from the orbital
frequency at first periastron passage of $0.07$, $0.13$ and $ 0.25$,
respectively (see Table I of Ref. \cite{Ramos-Buades:2021adz} for
details). For these injections we choose a total mass $M=70 M_\odot$,
inclination with respect to the line of sight of the BBH $\iota=0$
rad, coalescence phase $\phi_{\mathrm{ref}}=0$ rad, and luminosity
distance $d_L=$2307 Mpc, which produces a three-detector (LIGO
Hanford, LIGO Livingston and Virgo) network-SNR of
$\rho_{\mathrm{mf}}^N = 20$ when using the LIGO and Virgo PSD at
design sensitivity. The priors are the same as in
Sec. \ref{sec:ModelInj}, with the only exception that for the run
corresponding to the \texttt{SXS:BBH:1363} NR waveform we set a larger
upper bound for the eccentricity prior of $e_0 \in [0,0.5]$ in order
to avoid railing of the posterior against the upper bound of the
prior.

\setlength{\extrarowheight}{8pt}
\begin{table}[h!]
    \centering
    \begin{tabular}{ c  c  c  c c  }
 \hline
 \hline
 Parameter & \makecell[cc]{Injected \\ value} & \makecell[cc]{ \small SXS:1355} & \makecell[cc]{ \small SXS:1359} & \makecell[cc]{ \small SXS:1363}  \\
 \hline
 \centering
$M/M_\odot$ &  ${70.0}$ &  ${70.87}^{+2.47}_{-2.27} $ & ${70.41}^{+2.45}_{-2.45} $ & ${69.81}^{+2.32}_{-2.72} $ \\
 $\mathcal{M}/M_\odot$ &  ${30.47}$ & ${30.41}^{+0.98}_{-0.95} $ & ${30.26}^{+1.04}_{-1.14} $ & ${30.06}^{+0.98}_{-1.21} $ \\
 $1/q$ & ${1.0}$ & ${0.79}^{+0.17}_{-0.19} $ & ${0.8}^{+0.16}_{-0.19} $ & ${0.81}^{+0.15}_{-0.17} $ \\
 $\chi_{\text{eff}}$ & $0.0$ & ${0.02}^{+0.08}_{-0.08} $ & ${0.01}^{+0.09}_{-0.1} $ & ${-0.0}^{+0.08}_{-0.1} $ \\
 $e_0$ & - & ${0.06}^{+0.05}_{-0.05} $ & ${0.14}^{+0.03}_{-0.04} $ & ${0.29}^{+0.09}_{-0.05} $ \\
 $\zeta_0$ & - &  ${2.23}^{+1.37}_{-1.16} $ & ${1.01}^{+4.67}_{-0.75} $ & ${3.28}^{+1.6}_{-0.45} $ \\
$\theta_{\text{JN}}$ & $0.0$ &  ${0.62}^{+0.48}_{-0.38} $ & ${0.61}^{+0.48}_{-0.37} $ & ${0.61}^{+0.47}_{-0.37} $ \\
 $d_{L}$ & ${2307}$  &  ${1831}^{+373}_{-560} $ & ${1818}^{+374}_{-556} $ & ${1859}^{+378}_{-571} $ \\
$\phi_{\text{ref}}$ & $0.0$ & ${3.15}^{+2.5}_{-2.52} $ & ${3.14}^{+2.51}_{-2.5} $ & ${3.16}^{+2.52}_{-2.52} $ \\
 $\rho^{\mathrm{N}}_{\mathrm{mf}}$  & $20.0$ &   ${19.07}^{+0.09}_{-0.14} $ & ${19.05}^{+0.09}_{-0.15} $ & ${19.02}^{+0.17}_{-0.15} $ \\ [0.1cm]
\hline
\multirow{2}{*}{$e_{\rm gw}$} & Injected  & ${0.07}$  &  ${0.13}$  &  ${0.25}$   \\
& Measured  &  ${0.06}^{+0.05}_{-0.05} $ & ${0.14}^{+0.04}_{-0.04} $ & ${0.26}^{+0.02}_{-0.03} $ \\ 
\multirow{2}{*}{$l_{\rm gw}$} & Injected  & ${1.96}$  &  ${0.81}$  &  ${4.27}$   \\
& Measured  &  ${2.25}^{+1.19}_{-1.11} $ & ${1.33}^{+1.7}_{-0.93} $ & ${4.32}^{+0.63}_{-0.54} $ \\  [0.1cm]
 \hline
 \hline
    \end{tabular}
 \caption{Injected and median values of the posterior distributions for three synthetic NR injections with the same quasi-circular parameters and different initial eccentricities, and recovered with \texttt{SEOBNRv4E\_opt}. The median values also report the $90\%$ credible intervals. The binary parameters correspond to the total mass $M$, chirp mass $\mathcal{M}$, inverse mass ratio $1/q$, effective-spin parameter  $\chi_{\text{eff}}$, initial eccentricity $e_0$, initial relativistic anomaly $\zeta_0$, angle between the total angular momentum and the line of sight $\theta_{\text{JN}}$, luminosity distance $d_{L}$, coalescence phase $\phi_{\text{ref}}$ and the network matched-filtered SNR for LIGO-Hanford/Livingston and Virgo detectors  $\rho^{N}_{\mathrm{mf}}$. At the bottom of the table the injected and measured GW eccentricity, $e_{\rm gw}$, and GW mean anomaly, $l_{\rm gw}$, are also reported.}
 \label{tab:nr_injection}
\end{table}

\begin{figure*}[htpb!] 
\includegraphics[width=0.685\columnwidth]{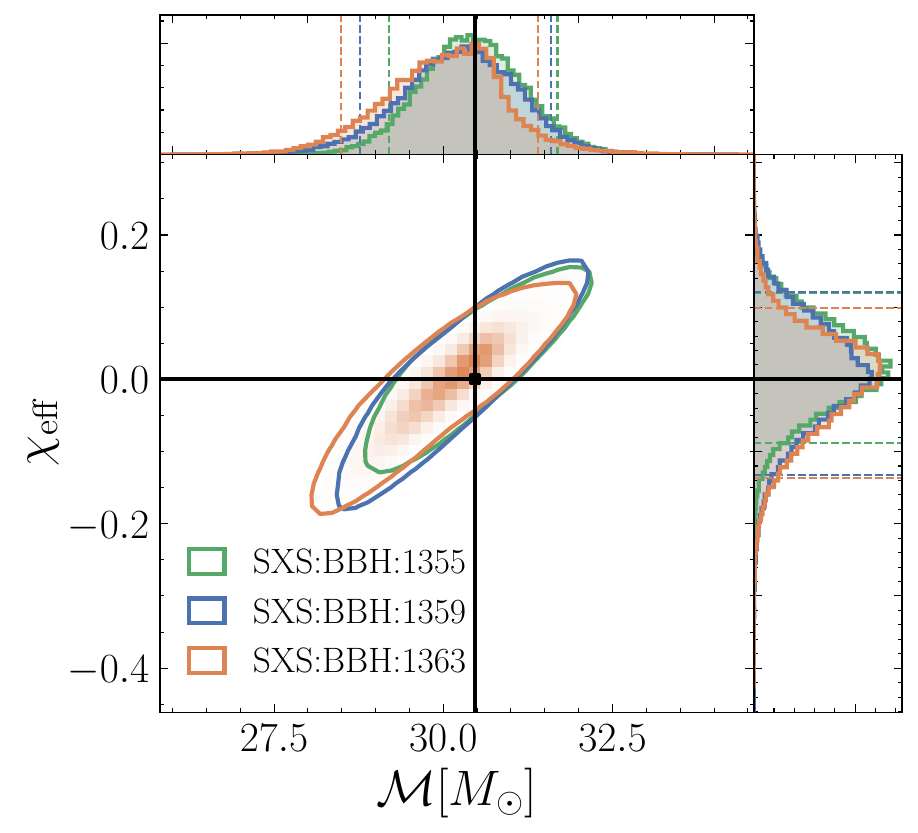} 
\includegraphics[width=0.685\columnwidth]{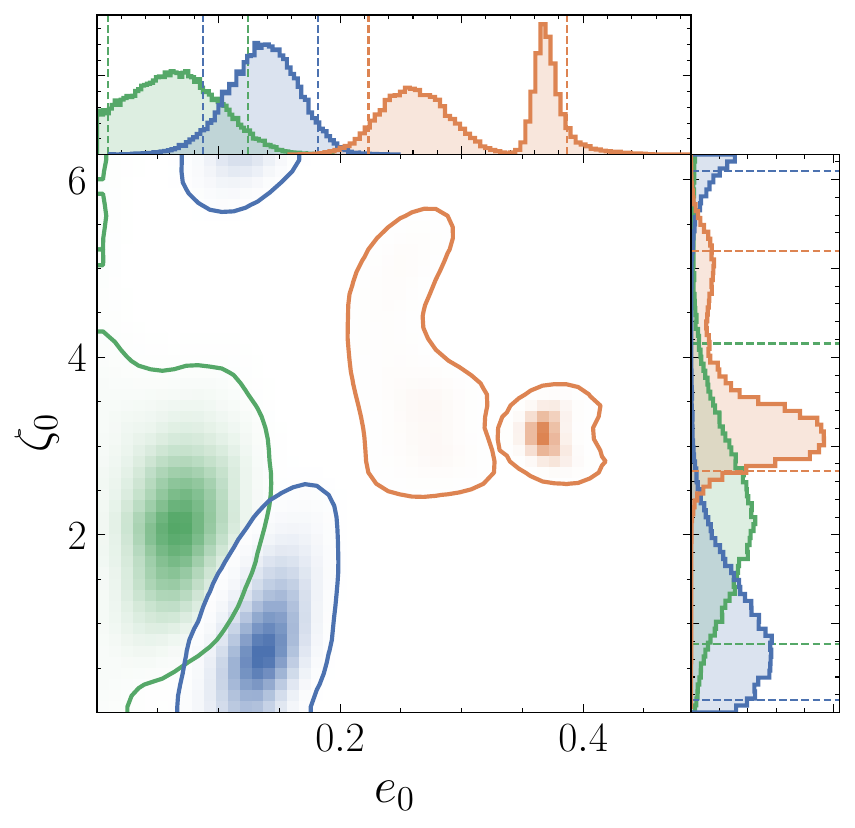} 
\includegraphics[width=0.685\columnwidth]{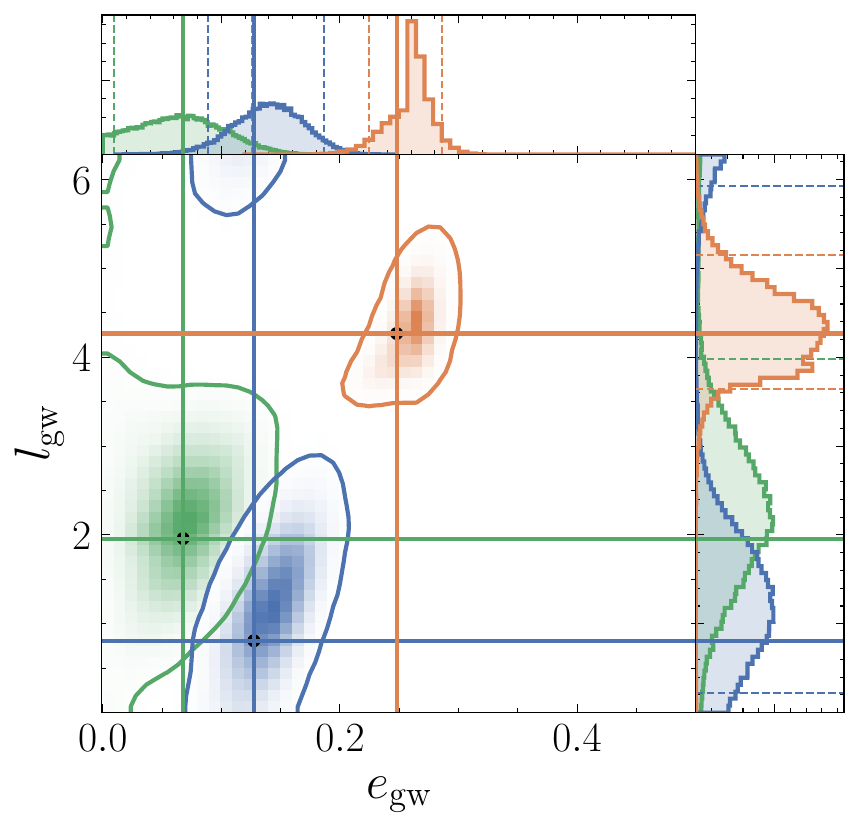}  
 \caption{2D and 1D posterior distributions for some relevant parameters from the equal mass nonspinning synthetic BBH signals with total mass $70 M_\odot$.  The signal waveforms correspond to the NR waveforms from the public SXS catalog {\tt SXS:BBH:1355}, {\tt SXS:BBH:1359} and {\tt SXS:BBH:1363} with GW eccentricities $e_{\rm gw}=0.05,0.1,0.25$ and GW mean anomalies $e_{\rm gw}=0.05,0.1,0.25$ defined at 20Hz, respectively. The other parameters are specified in Table \ref{tab:nr_injection}.  In the 2D posteriors the solid contours represent the $90\%$ credible intervals and black dots show the values of the parameters of the injected signal.  In the 1D posteriors they are represented by dashed and solid vertical lines, respectively.    The parameter estimation is performed with the \texttt{SEOBNRv4E\_opt} model. \textit{Left:} Chirp mass and effective-spin parameter. \textit{Middle:} Initial eccentricity and relativistic anomaly at 20Hz.  \textit{Right:} GW eccentricity and GW mean anomaly at 20Hz.  
 }
\label{fig:nrInj}
\end{figure*}

Before analysing the results of the injections, we consider the problem of mapping the eccentric parameters, eccentricity and radial phase, from the NR waveforms to the \texttt{SEOBNRv4E\_opt} model. This problem stems from the gauge-dependency of the eccentricity parameter in General Relativity, and can be avoided by adopting a common definition of the parameters defining elliptical orbits in the NR simulation and the SEOBNR model. In particular, we adopt a definition of eccentricity, $e_{\rm gw}$, measured from the frequency of the (2,2)-mode with the correct Newtonian limit \cite{Ramos-Buades:2022lgf}
\begin{subequations}
\label{eq:eqEccDef}
\begin{align}
  e_{\rm gw} &= \cos (\psi/3)- \sqrt{3} \sin (\psi/3),
  \intertext{with}
\psi  &= \arctan\left( \frac{1-e_{\omega_{22}}^2}{2 e_{\omega_{22}}} \right),\\
    e_{\omega_{22}} &= \frac{\omega^{1/2}_{22,p}-\omega^{1/2}_{22,a}}{\omega^{1/2}_{22,p} + \omega^{1/2}_{22,a}},
\end{align}
\end{subequations}
where $\omega_{22,a}, \omega_{22,p}$ refer to the values of the (2,2)-mode frequency at apastron and periastron, respectively. 
As the radial-phase parameter describing a binary in an elliptic orbit, we use the mean anomaly, with the following definition \cite{Schmidt:2017btt}
\begin{equation}
    l_{\rm gw} = 2\pi \frac{t-t^p_i}{t^p_{i+1}-t^p_{i}} ,
    \label{eq:eq8}
\end{equation}
where $t^p_i$ is the time of the i-th periastron passage measured from the (2,2)-mode frequency.  

These definitions of eccentricity and mean anomaly can be applied to
the posterior distributions from the parameter-estimation runs as a
post-processing step employing its highly efficient implementation in
the \texttt{gw\_eccentricity} Python package\footnote{We have used
  version \texttt{v1.0.2} from the public git repository
  \url{git@github.com:vijayvarma392/gw_eccentricity.git}.}
\cite{Shaikh:2023ypz}. The procedure consists in evaluating the
waveform for each sample of the posterior distributions, and applying the
\texttt{gw\_eccentricity} package to measure the eccentricity and mean
anomaly at a desired point in the evolution. The process can be parallelized to measure the eccentricity and mean anomaly in a much smaller timescale than an
actual parameter-estimation run\footnote{For the parameter-estimation
  runs in this section, and when running on 4 nodes of 32 cores (their
  description can be found in Sec. \ref{sec:ModelInj}) the
  post-processing step takes less than 30 minutes.}.

In the \texttt{SEOBNRv4EHM} model the eccentricity is specified at an
orbit-averaged orbital frequency, however, the eccentricity definition
introduced in Eq. \eqref{eq:eqEccDef} is based on the (2,2)-mode
frequency. As shown in Ref. \cite{Ramos-Buades:2022lgf}, for eccentric
binaries the instantaneous orbital and (2,2)-mode frequencies are not
related by a simple factor of 2, as in the quasi-circular case, and
this can cause that for some samples in the posterior distribution,
the instantaneous (2,2)-mode frequency does not reach the starting
frequency specified in the initial conditions of the
\texttt{SEOBNRv4EHM} model. In order to avoid that this situation
prevents the eccentricity measurement for some samples we have also
implemented an option to integrate the EOB dynamics backwards in time
from a specific starting frequency. For the rest of the calculations
involving \texttt{gw\_eccentricity} in this paper, we integrate $2000
M$ backwards in time.

In Fig. \ref{fig:nrInj}, we summarize the parameter-estimation results
of the injections. We report the marginalized 1D and 2D posteriors for
the chirp mass $\mathcal{M}$ and the effective-spin parameter
$\chi_{\mathrm{eff}}$, the initial eccentricity and relativistic
anomaly, and the GW eccentricity and mean anomaly measured at 20Hz. In
Table \ref{tab:nr_injection} we provide the values of the injected
parameters and the median of the inferred posterior distribution with
the $90\%$ confidence intervals for both models. The results show that
\texttt{SEOBNRv4E\_opt} is able to recover $\mathcal{M}$ and
$\chi_{\mathrm{eff}}$, as well as the mass ratio and total mass for
all the NR injections within the $90\%$ confidence intervals. The NR
waveforms contain all the multipoles up to $l\leq 8$, while the
\texttt{SEOBNRv4E\_opt} contains only the $(l,|m|)=(2,2)$ modes. This
difference in mode content explains why there are some small biases in
the luminosity distance and $\theta_{\mathrm{JN}}$ parameter.

Regarding the eccentric parameters, the middle panel of
Fig. \ref{fig:nrInj} shows the initial eccentricity and relativistic
anomaly posterior distributions at 20Hz for the three different
injections. For the highest eccentricity injection
(\texttt{SXS:BBH:1363}), the eccentricity posterior is bimodal with two
modes centered at $e_0\sim 0.25$ and $e_0\sim 0.35$. However, when
moving to the definition of eccentricity and mean anomaly based on the
(2,2)-mode frequency introduced in Eq. \eqref{eq:eqEccDef}, we find a
unimodal posterior in $e_{\rm gw}$ and $l_{\rm gw}$, which indicates
that the bimodality in $(e_0,l_0)$ is simply a consequence of the
parametrization of the EOB initial conditions used in the
\texttt{SEOBNRv4E\_opt} model.  Furthermore, the $e_{\rm gw}$ and
$l_{\rm gw}$ parameters at 20Hz are measured from the NR waveforms,
and shown as vertical lines in the right panel of
Fig. \ref{fig:nrInj}. The posterior distributions for $e_{\rm gw}$ and
$l_{\rm gw}$ are consistent within the $90 \%$ credible intervals with
the injected values for the three NR waveforms.

The results of the injections demonstrate that \texttt{SEOBNRv4E\_opt}
is able to accurately recover the eccentric and non-eccentric
parameters of the injected NR waveforms, and they are consistent with
the low unfaithfulness values of \texttt{SEOBNRv4E} against NR
waveforms reported in Ref. \cite{Ramos-Buades:2021adz}. Further
studies of the accuracy of the model will require larger datasets of
eccentric NR waveforms including spins and higher eccentricities, and
we leave for future work investigating the waveform systematics of the
\texttt{SEOBNRv4EHM\_opt} model and its biases against NR waveforms.

\subsection{Analysis of GW events}\label{sec:GWevents}
  
In this section, we analyze 3 GW events recorded by the LIGO and Virgo detectors
\cite{LIGOScientific:2018mvr,LIGOScientific:2021usb,LIGOScientific:2021djp} during the first and third observing runs:
GW150914, GW151226 and GW190521. We employ strain data from the
Gravitational Wave Open Source Catalog (GWOSC)
\cite{LIGOScientific:2019lzm}, and the released PSD and calibration
envelopes included in the Gravitational Wave Transient Catalogs
GWTC-2.1 \cite{LIGOScientific:2021usb}, and their respective
parameter-estimation samples releases\footnote{For GW190521 we employ the samples of the \texttt{SEOBNRv4PHM} model from Refs. \cite{LIGOScientific:2020iuh,LIGOScientific:2020ufj}, which were produced using \texttt{parallel Bilby}.}.

We analyze GW150914, GW151226 and GW190521 using
\texttt{SEOBNRv4EHM\_opt} and \texttt{SEOBNRv4E\_opt} with
\texttt{parallel Bilby} and the settings described in
Sec. \ref{sec:PEsetup}. For \texttt{SEOBNRv4EHM\_opt}, we restrict to a
mode content $l\leq 4$ in order to avoid the increase of computational
cost due to the high sampling rates necessary to resolve the
$(5,5)$-mode, as at the current SNRs the impact of this mode on
accuracy is limited.

\begin{figure*}[htpb!] 
\includegraphics[width=0.75\columnwidth]{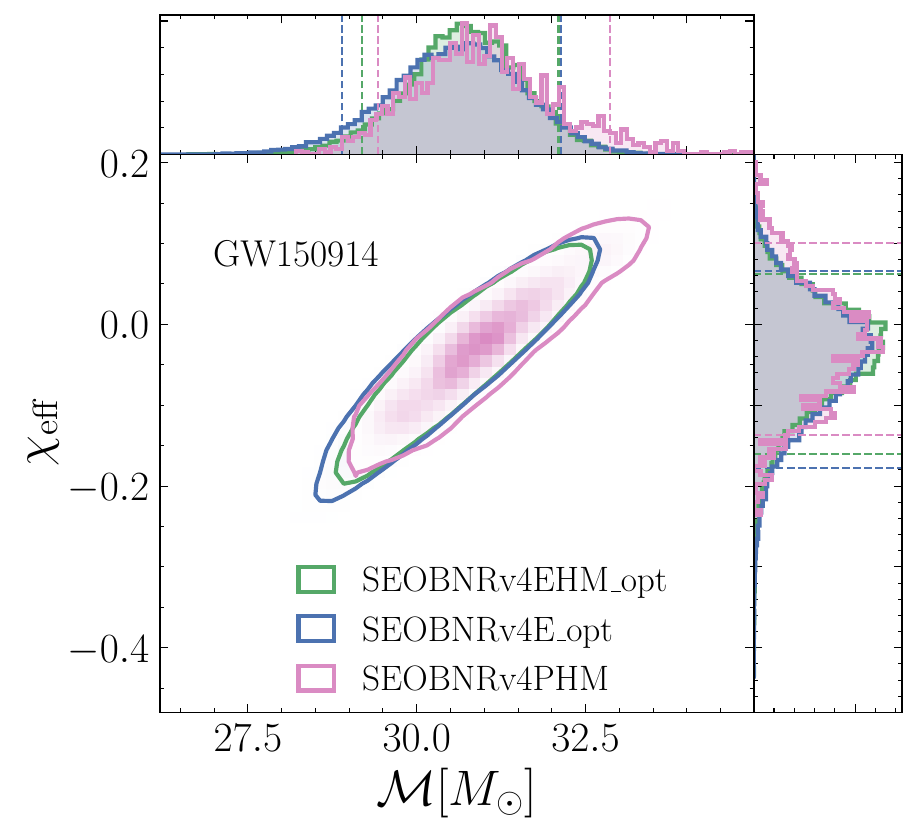} 
\includegraphics[width=0.65\columnwidth]{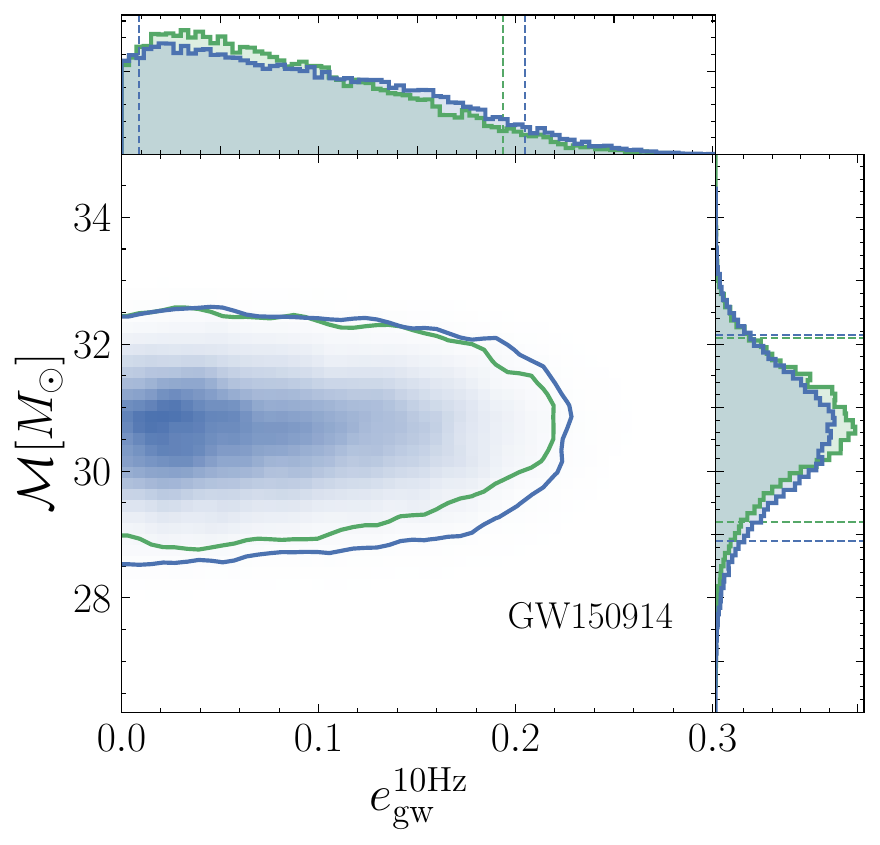}  
\includegraphics[width=0.65\columnwidth]{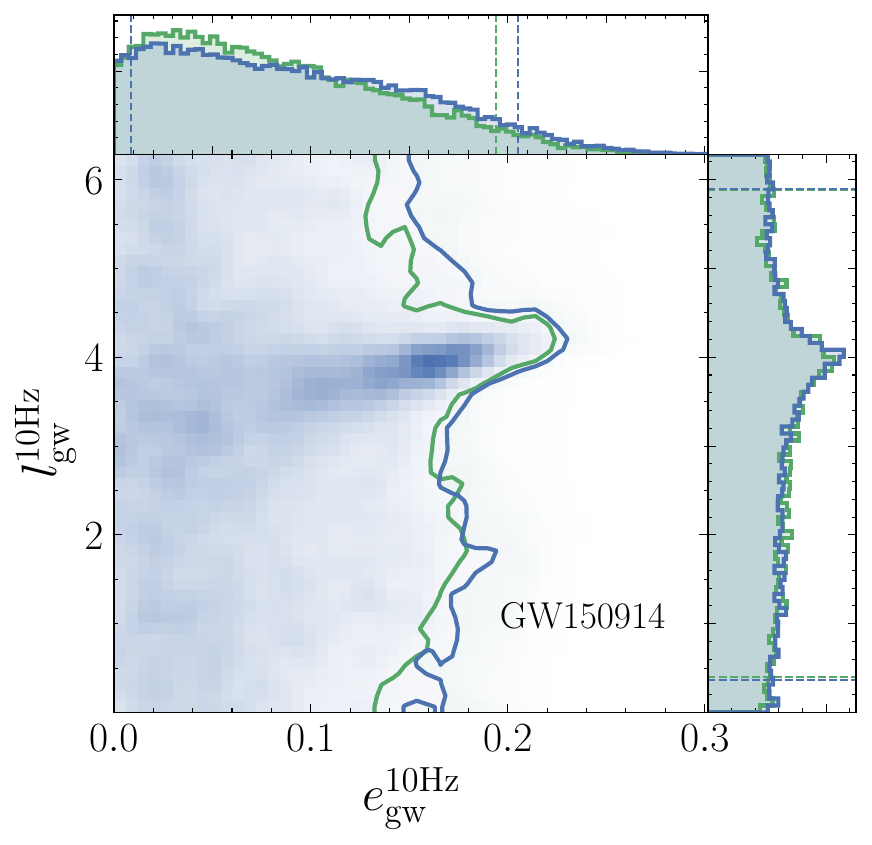} \\
\includegraphics[width=0.75\columnwidth]{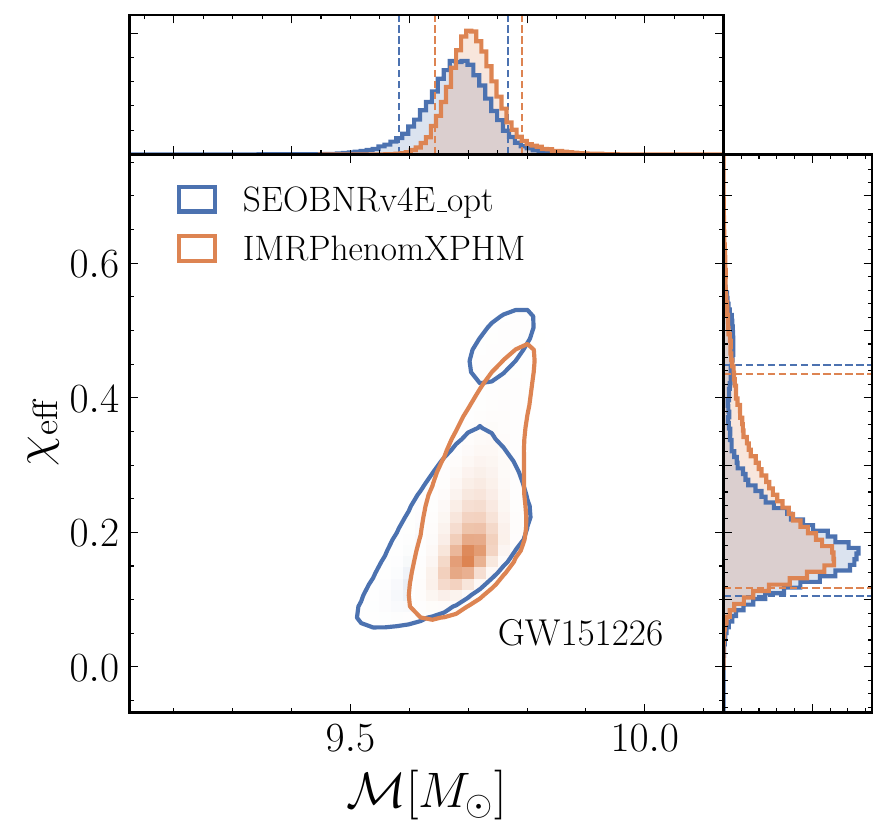} 
\includegraphics[width=0.65\columnwidth]{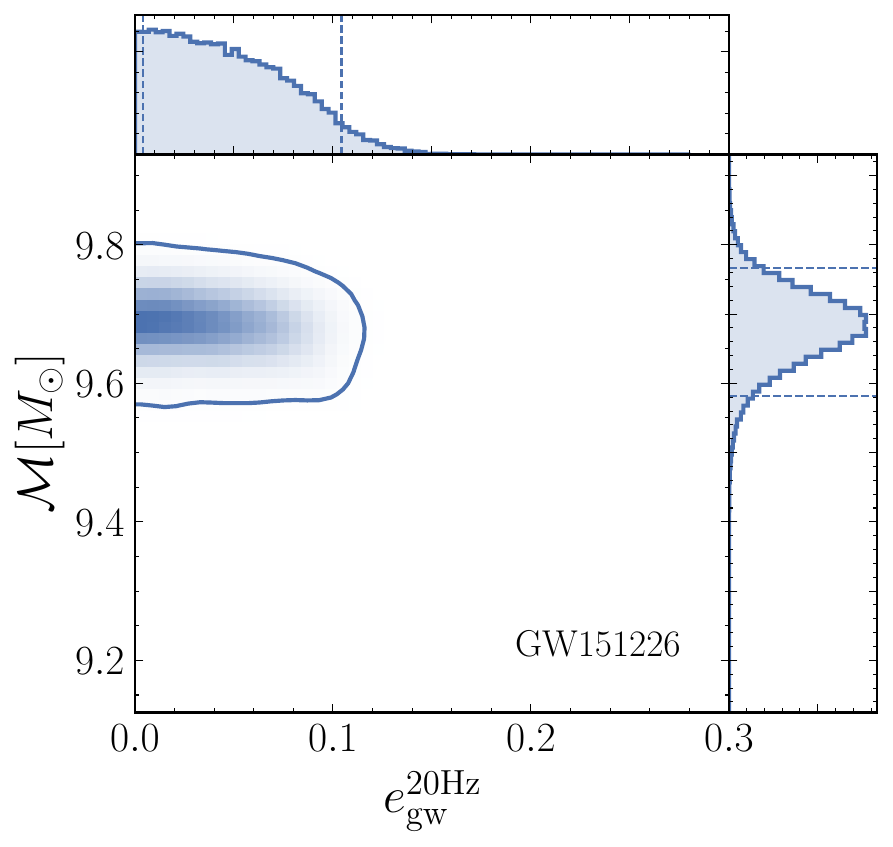}  
\includegraphics[width=0.65\columnwidth]{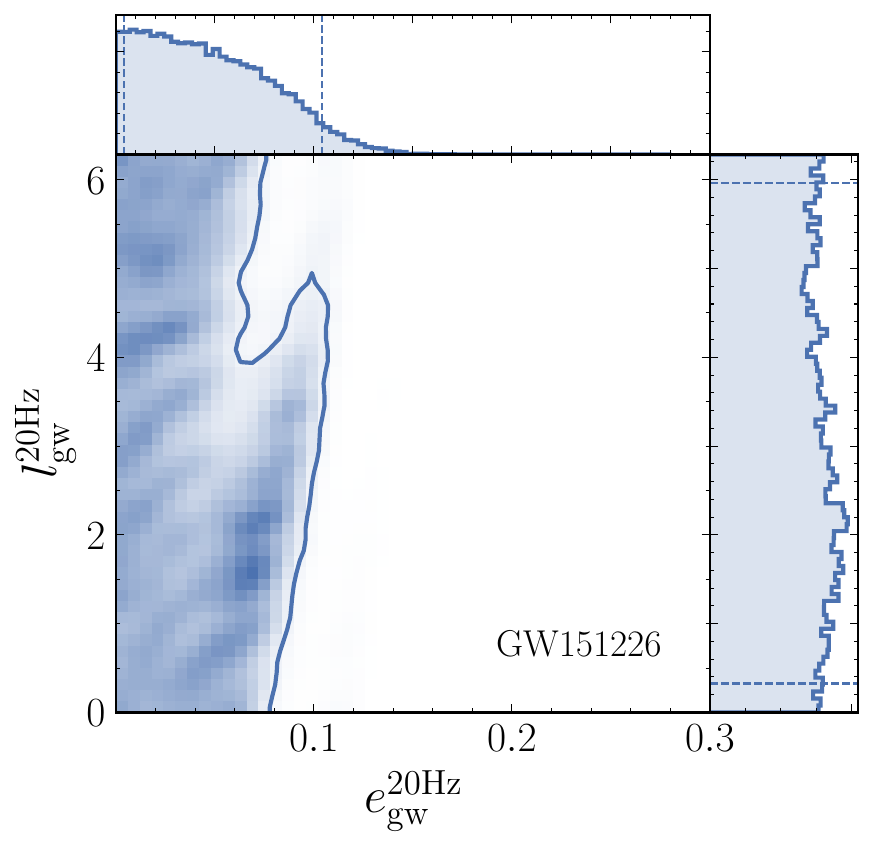} \\
\includegraphics[width=0.75\columnwidth]{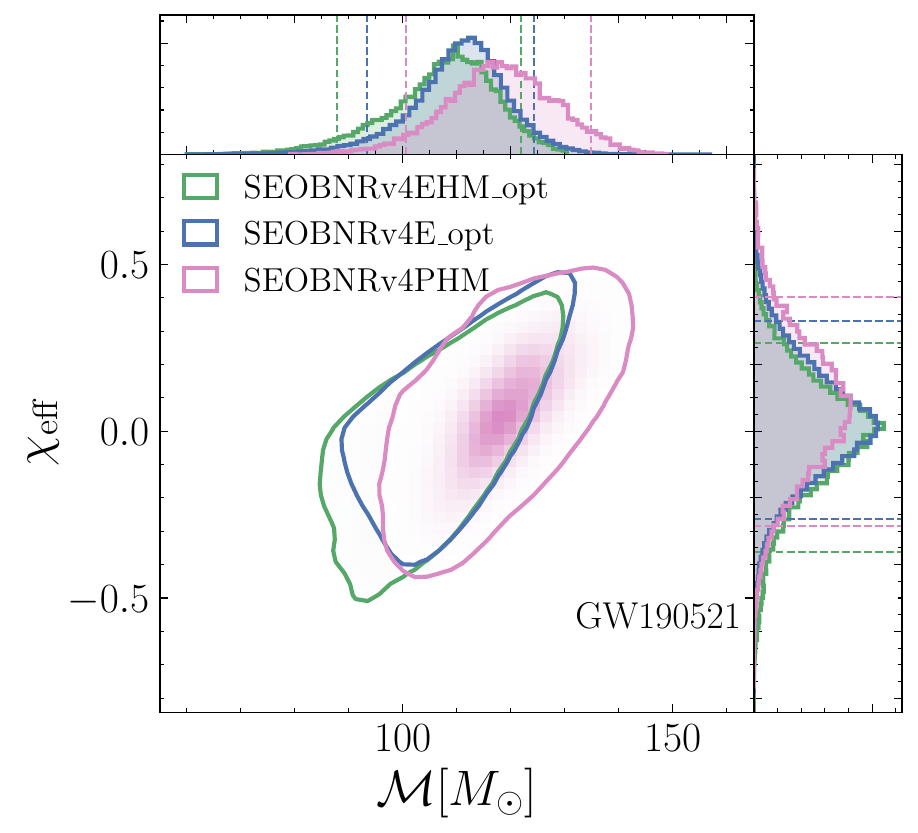} 
\includegraphics[width=0.65\columnwidth]{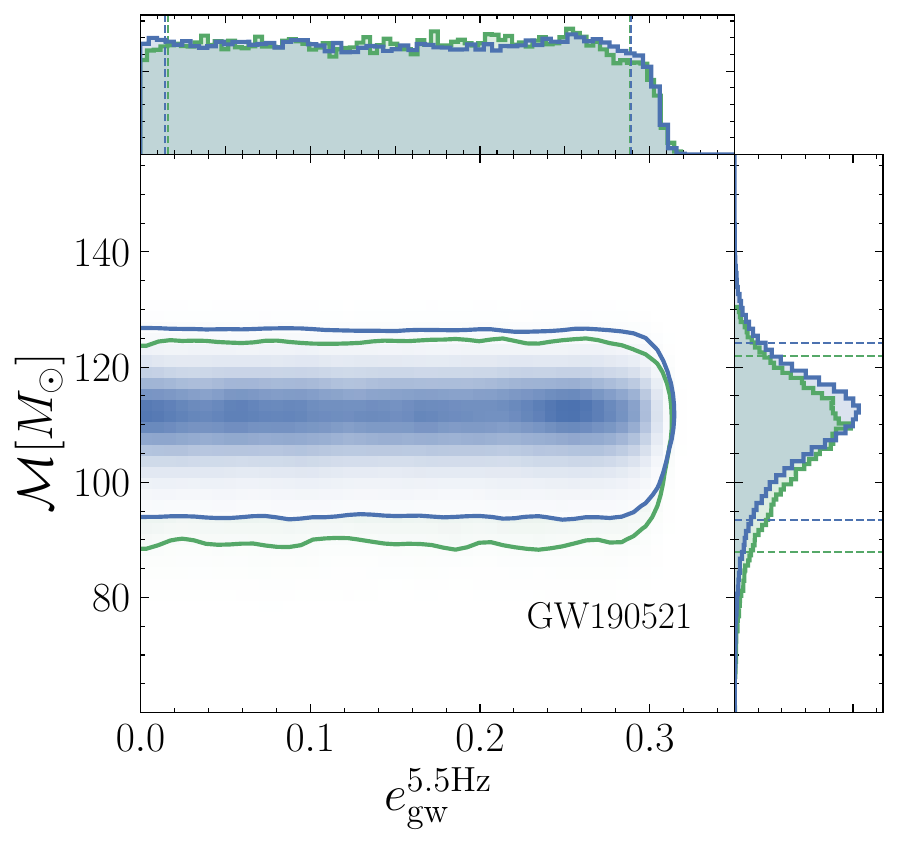} 
\includegraphics[width=0.65\columnwidth]{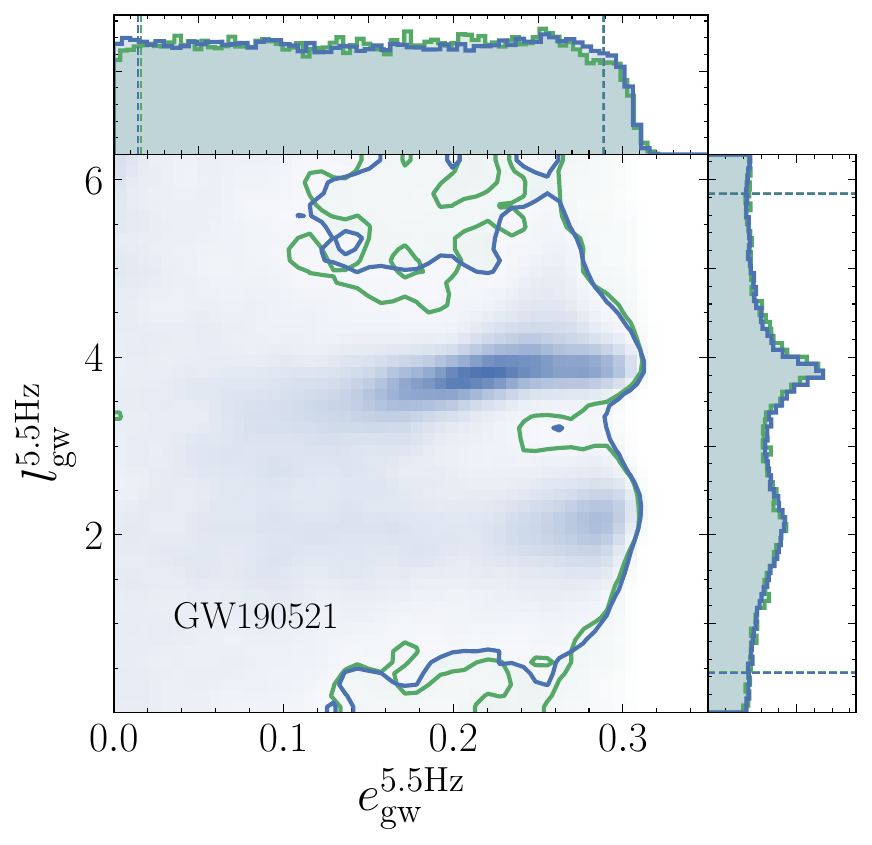}  \\ 
\vspace{-0.3cm}
  \caption{Effective-spin parameter, chirp mass, GW eccentricity and GW mean anomaly parameters inferred for the real GW events analysed with \texttt{SEOBNRv4E\_opt} and \texttt{SEOBNRv4EHM\_opt}. Comparisons are presented with the \texttt{SEOBNRv4PHM} (when available) and the \texttt{IMRPhenomXPHM} models from the GWTC-2.1 catalog \cite{LIGOScientific:2021usb}, except for GW190521 where samples are from Refs.  \cite{LIGOScientific:2020iuh,LIGOScientific:2020ufj}. The waveform starting frequency for GW150914 is $f_{\rm start }= 10$Hz, for the low total mass event GW151226 is $f_{\rm start }= 20$Hz, in order to reduce the computational cost, while for GW190521 $f_{\rm start }= 5.5$Hz in order to have the mode content in band at the likelihood minimum frequency ($f_{\rm min }= 11$Hz).
 }
\label{fig:gwEvents}
\end{figure*}

\setlength{\extrarowheight}{8pt}
\begin{table*}[!]
    \centering 
\begin{tabular*}{\textwidth}{c@{\extracolsep{\fill}} c c c c c c c  c  c  c  c c }
 \hline
 \hline
Event & Waveform & $M/M_\odot$ & $\mathcal{M}/M_\odot$ & $1/q$ &  $\chi_{\text{eff}}$ & $e_{\rm gw}$ &  $l_{\rm gw}$ & $\theta_{\text{JN}}$ & $d_{L}$   & $ \log \mathcal{BF} $\\[0.05cm]
 \hline 
 \centering
 \multirow{3}{*}{ \makecell[cc]{  GW150914	\\ \fontsize{7.5pt}{7pt}\selectfont ($f_{\rm start}=10$Hz) } }  & \fontsize{7.5pt}{7pt}\selectfont  \texttt{SEOBNRv4E\_opt}  & ${70.83}^{+2.7}_{-2.88} $ & ${30.61}^{+1.21}_{-1.32} $ & ${0.85}^{+0.12}_{-0.16} $ & ${-0.04}^{+0.08}_{-0.1} $ &  ${0.08}^{+0.1}_{-0.07} $ & ${3.31}^{+2.2}_{-2.57} $ & ${2.48}^{+0.4}_{-0.55} $ & ${387}^{+145}_{-133} $ & $283.9^{+0.1}_{-0.1} $ \\
                           & \fontsize{7.5pt}{7pt}\selectfont \texttt{ SEOBNRv4EHM\_opt}  &${70.89}^{+2.42}_{-2.59} $ & ${30.72}^{+1.08}_{-1.15} $ & ${0.89}^{+0.09}_{-0.13} $ & ${-0.03}^{+0.08}_{-0.09} $ & ${0.08}^{+0.09}_{-0.06} $ & ${3.27}^{+2.23}_{-2.49} $ & ${2.64}^{+0.3}_{-0.37} $ & ${440}^{+117}_{-120} $ & $284.2^{+0.1}_{-0.1} $  \\
                           & \fontsize{7.5pt}{7pt}\selectfont \texttt{SEOBNRv4PHM}  &   ${71.32}^{+3.39}_{-2.8} $ & ${30.92}^{+1.51}_{-1.22} $ & ${0.9}^{+0.08}_{-0.12} $ & ${-0.02}^{+0.1}_{-0.09} $ & -  & -  & ${2.71}^{+0.26}_{-0.59} $ & ${492}^{+103}_{-146} $  & - \\[0.08cm]
 \hline  
 \multirow{2}{*}{ \makecell[cc]{   GW151226	\\ \fontsize{7.5pt}{7pt}\selectfont ($f_{\rm start}=20$Hz) } }  & \fontsize{7.5pt}{7pt}\selectfont  \texttt{SEOBNRv4E\_opt}  &  ${22.82}^{+3.46}_{-0.59} $ & ${9.68}^{+0.07}_{-0.07} $ & ${0.66}^{+0.26}_{-0.32} $ & ${0.18}^{+0.13}_{-0.06} $ & ${0.04}^{+0.05}_{-0.04} $ & ${2.95}^{+2.67}_{-2.3} $ & ${1.06}^{+1.7}_{-0.73} $ & ${468}^{+170}_{-183} $ & $40.4^{+0.1}_{-0.1} $ \\                           
                           & \fontsize{7.5pt}{7pt}\selectfont \texttt{IMRPhenomXPHM}  &  ${23.71}^{+6.13}_{-1.36} $ & ${9.71}^{+0.06}_{-0.05} $ & ${0.52}^{+0.35}_{-0.29} $ & ${0.2}^{+0.17}_{-0.07} $ & -  & -  & ${0.87}^{+1.94}_{-0.57} $ & ${471}^{+124}_{-154} $ &$ 47.6^{+0.1}_{-0.1}$ \\[0.08cm]
\hline 
  \multirow{3}{*}{ \makecell[cc]{  GW190521	\\ \fontsize{7.5pt}{7pt}\selectfont ($f_{\rm start}=5.5$Hz) } }  & \fontsize{7.5pt}{7pt}\selectfont  \texttt{SEOBNRv4E\_opt}  &  ${259.92}^{+21.63}_{-20.06} $ & ${111.15}^{+9.88}_{-12.69} $ & ${0.74}^{+0.2}_{-0.25} $ & ${0.02}^{+0.23}_{-0.21} $ & ${0.15}^{+0.12}_{-0.12} $ & ${3.14}^{+2.26}_{-2.28} $ & ${1.27}^{+1.43}_{-0.86} $ & ${3924}^{+1434}_{-1460} $ & $77.8^{+0.1}_{-0.1} $ \\
                           & \fontsize{7.5pt}{7pt}\selectfont \texttt{ SEOBNRv4EHM\_opt}  & ${253.97}^{+21.48}_{-24.94} $ & ${108.4}^{+10.56}_{-15.27} $ & ${0.72}^{+0.22}_{-0.24} $ & ${-0.01}^{+0.21}_{-0.26} $ & ${0.15}^{+0.12}_{-0.12} $ & ${3.16}^{+2.25}_{-2.3} $ & ${0.88}^{+1.89}_{-0.55} $ & ${4172}^{+1262}_{-1286} $ & $78.6^{+0.1}_{-0.1} $ \\
                           & \fontsize{7.5pt}{7pt}\selectfont \texttt{SEOBNRv4PHM}  &  ${279.54}^{+36.74}_{-28.65} $ & ${118.13}^{+12.94}_{-13.15} $ & ${0.74}^{+0.2}_{-0.31} $ & ${0.06}^{+0.26}_{-0.27} $ & -  & -  & ${1.39}^{+1.28}_{-0.93} $ & ${3964}^{+1557}_{-1474} $ & -  \\[0.08cm]
 \hline
 \hline
    \end{tabular*} 
 \caption{Median and $90\%$ credible intervals for the total mass, chirp mass, inverse mass ratio, effective-spin parameter, GW eccentricity, GW mean anomaly, $\theta_{\rm JN}$, luminosity distance and signal-versus-noise (natural) log Bayes factor, $\log \mathcal{BF}$, measured with the \texttt{SEOBNRv4E\_opt} and \texttt{SEOBNRv4EHM\_opt} models for GW150914, GW151226 and GW190521. Additionally, we include the results of \texttt{SEOBNRv4PHM} for GW150914 and GW190521 and of \texttt{IMRPhenomXPHM} for GW151226 from the GWTC-2.1 catalog \cite{LIGOScientific:2021usb}, except for GW190521 where samples are from Refs.  \cite{LIGOScientific:2020iuh,LIGOScientific:2020ufj}. The waveform starting frequency for GW150914 is $f_{\rm start }= 10$Hz, for the low total mass event GW151226 is $f_{\rm start }= 20$Hz, in order to reduce the computational cost, while for GW190521 is  $f_{\rm start }= 5.5$Hz in order to have the mode content in band at the likelihood minimum frequency ($f_{\rm min }= 11$Hz).
  }
 \label{tab:gwEvents}
\end{table*}

\begin{table}[h!]
\centering
\begin{tabular}{l c c c c c c c }
 \hline
 \hline
  \makecell[cc]{GW event \\ sampler} & \makecell[cc]{Modes} & \multicolumn{2}{c}{\makecell[cc]{Data \\ settings}} &\multicolumn{2}{c}{\makecell[cc]{Sampler \\ settings}} & \makecell[cc]{Computing \\ resources} & Runtime   \\
 \hline
 & &  \makecell[cc]{srate  \\ (Hz)} & \makecell[cc]{seglen\\(s)} & nact & nlive & cores$\times$nodes &   \\
 \hline
   \multirow{2}{*}{ \makecell[cc]{  GW150914	\\ \fontsize{7.5pt}{7pt}\selectfont ($f_{\rm start}=10$Hz) } }  & \fontsize{7.5pt}{7pt}\selectfont  $(2,|2|)$ & 4096  & 8 & 30 & 2048 & $32\times 10$ & 6.7h\\
                           & \fontsize{7.5pt}{7pt}\selectfont HMs  & 4096 & 8 & 30 & 2048 & $32\times 10$ & 3d 5h  \\[0.08cm]
 \makecell[cc]{  GW151226	\\ \fontsize{7.5pt}{7pt}\selectfont ($f_{\rm start}=20$Hz) }   & \fontsize{7.5pt}{7pt}\selectfont  $(2,|2|)$ & 4096  & 8 & 30 & 2048 & $32\times 16$ & 1d 20h \\[0.08cm]
  \multirow{2}{*}{ \makecell[cc]{  GW190521	\\ \fontsize{7.5pt}{7pt}\selectfont ($f_{\rm start}=5.5$Hz) } }  & \fontsize{7.5pt}{7pt}\selectfont  $(2,|2|)$ & 4096  & 8 & 30 & 8192 & $32\times 10$ & 1d 6h\\
                           & \fontsize{7.5pt}{7pt}\selectfont HMs  & 4096  & 8 & 30 & 2048 & $32\times 16$ & 1d 5h  \\[0.08cm] 
 \hline
\end{tabular}
\caption{Settings and evaluation time for the different parameter estimation runs on real GW events. The mode content indicates the use of \texttt{SEOBNRv4E\_opt} ($(2,|2|)$) or \texttt{SEOBNRv4EHM\_opt} ( HMs $\equiv$ higher modes) to perform the run. Sampling rate ($\text{srate}$) and data segment duration ($\text{seglen}$) are specified in the data settings, while the number of autocorrelaction times ($\text{nact}$) and number of live points ($\text{nlive}$) are specified in the sampler settings. The time reported is walltime, while the total computational cost in CPU hours can be obtained multiplying this time by the reported number of CPU cores employed.}
\label{tab:peruntime}
\end{table}

\subsubsection*{GW150914}
The first observation of GWs from a BBH coalescence, GW150914, had one of the highest SNRs ($\sim 23.7$) of the GW events observed during the first three observing runs  \cite{LIGOScientific:2016aoc,LIGOScientific:2021usb} of the LVK. The binary parameters are consistent with a nonspinning binary  with comparable masses \cite{LIGOScientific:2016ebw}. 

We choose priors in inverse mass ratio, $1/q\in [0.05,1]$, and chirp mass, $\mathcal{M} \in [20,50] M_\odot$, such that the induced priors in component masses are uniform. Uniform priors were also used for initial eccentricity, $e_0 \in [0,0.3]$, and the initial relativistic anomaly, $\zeta_0 \in [0.,2\pi]$. The other priors are chosen as in Sec. \ref{sec:NRInj}. For the analysis we use a starting frequency of 10Hz at which the waveform is generated and, thus at which the initial eccentricity and relativistic anomaly are defined. This choice of starting frequency ensures that the higher order modes in  \texttt{SEOBNRv4EHM\_opt} are in band at the minimum frequency of 20Hz at which the likelihood calculation starts. We also remark the importance that both \texttt{SEOBNRv4EHM\_opt} and \texttt{SEOBNRv4E\_opt} are generated at the same starting frequency, so that both have the same priors in initial eccentricity. 

The posterior distributions of chirp mass, effective-spin parameter,
GW eccentricity and GW mean anomaly are displayed in the top row of
Fig. \ref{fig:gwEvents}. In Table \ref{tab:gwEvents} we also report
the median values and the $90 \%$ credible intervals of the posterior
distributions for other binary parameters. For comparisons we include
in our analysis the samples of \texttt{SEOBNRv4PHM}
\cite{Ossokine:2020kjp} from the GWTC-2.1 catalog
\cite{LIGOScientific:2021usb}. We find that binary parameters like
chirp mass, effective-spin parameter and mass ratio measured by
\texttt{SEOBNRv4E\_opt} and \texttt{SEOBNRv4EHM\_opt} are consistent
with the ones measured with  \texttt{SEOBNRv4PHM}. This is expected as
GW150914 is consistent with a nonspinning binary, and thus the
effects of spin-precession which are accurately described by
\texttt{SEOBNRv4PHM} are negligible. Regarding the eccentric
parameters, although both \texttt{SEOBNRv4E\_opt} and
\texttt{SEOBNRv4EHM\_opt} have median values of eccentricity distinct
from zero, $e^{\rm 10Hz}_{\rm gw}= {0.08}^{+0.1}_{-0.07} $ and $e^{\rm
  10Hz}_{\rm gw}= {0.08}^{+0.09}_{-0.06} $, respectively, the
posterior distributions have a strong support in the zero eccentricity
region, which is in agreement with other analyses of GW150914 with
eccentric waveforms
\cite{LIGOScientific:2016ebw,Romero-Shaw:2019itr,Bonino:2022hkj,Iglesias:2022xfc}. Additionally,
we have produced a run with \texttt{SEOBNRv4E\_opt} setting $e_0=0$,
and obtained a signal-to-noise natural log Bayes factor of
$284.6^{+0.1}_{-0.1} $ which is slightly larger than
$283.9^{+0.1}_{-0.1} $ obtained for \texttt{SEOBNRv4E\_opt} when
sampling in $(e_0,\zeta_0)$ (see Table
\ref{tab:gwEvents}). Unfortunately, this 
information is not available for the \texttt{SEOBNRv4PHM} model 
in the public GWOSC results, but GWTC2-1 does include a log Bayes factor for the similar \texttt{IMRPhenomXPHM} model, of $\log \mathcal{BF} = 303.45^{+0.14}_{-0.14}$. These
results indicate that the non-precessing eccentric hypothesis is
disfavoured against the precessing-spin quasi-circular one, and that
GW150914 is more consistent with a quasi-circular BBH merger.
 
\subsubsection*{GW151226}

GW151226 is one of the GW events with lowest total mass observed in
the first observing run, and it was identified in the GWTC-1 catalog
\cite{LIGOScientific:2018mvr} to exclude support for $\chi_{\rm eff} =
0$ at $> 90\%$ probability. Furthermore, Ref. \cite{OShea:2021ugg}
analyzed this event with the \texttt{TEOBResumS-Dali} model
\cite{Nagar:2021xnh} and \texttt{parallel Bilby}, with a uniform prior
in eccentricity, and constrained the initial eccentricity to be
$e_0<0.15$ at $90 \%$ at a starting frequency of 10Hz.  Moreover, 
Ref. \cite{Romero-Shaw:2019itr} using the \texttt{SEOBNRE} model
\cite{Cao:2017ndf} and the reweighing technique with a log-uniform prior
in eccentricity found a much tighter constraint $e_0<0.04$ at 10Hz.

For our analysis we use a uniform prior in initial eccentricity $e_0
\in [0,0.3]$ and relativistic anomaly $\zeta_0 \in [0,2 \pi]$, and priors
in inverse mass ratio, $1/q\in [0.125,1]$, and chirp mass,
$\mathcal{M} \in [5,100] M_\odot$, such that they are uniform in
component masses. The rest of the priors are chosen as in the analysis of
GW150914. We use a starting frequency of 20Hz at which the waveform is
generated and, at which $e_0$ and $\zeta_0$ are defined. Due to
the low total mass of the event, we restrict to a higher starting
frequency than in the case of GW150914, and we only use the
\texttt{SEOBNRv4E\_opt} model in order to reduce the computational
cost.

The results are shown in the middle row of Fig. \ref{fig:gwEvents},
while in Table \ref{tab:gwEvents} the median values and the $90 \%$
credible intervals of the posterior distributions are reported. For
comparisons we include in our analysis the samples of
\texttt{IMRPhenomXPHM} \cite{Pratten:2020ceb} from the GWTC-2.1
catalog \cite{LIGOScientific:2021usb}. The quasi-circular binary
parameters like chirp mass, effective-spin parameter and mass ratio
measured by \texttt{SEOBNRv4E\_opt} show differences with respect to
the ones measured by \texttt{IMRPhenomXPHM}. This can be explained
because of the different physical content of each model, as
\texttt{IMRPhenomXPHM} includes higher-order modes and describes
precessing-spin quasi-circular binaries, while \texttt{SEOBNRv4E\_opt}
includes only the $(l,|m|)=(2,2)$ modes and describes non-precessing
eccentric binaries. This translates into a preference of
\texttt{IMRPhenomXPHM} for unequal masses and larger effective spin
values than \texttt{SEOBNRv4E\_opt}. Apart from these differences, the
posterior distributions of \texttt{SEOBNRv4E\_opt} and
\texttt{IMRPhenomXPHM} have large overlapping regions as can be
observed in the left plot in the middle row of
Fig. \ref{fig:gwEvents}.

The value of eccentricity measured by \texttt{SEOBNRv4E\_opt} is
$e^{\rm 20Hz}_{\rm gw}= {0.04}^{+0.05}_{-0.04} $, and the posterior
distribution of GW eccentricity (middle and right plots in
Fig. \ref{fig:gwEvents}) have strong support at zero
eccentricity. This combined with an uninformative posterior
distribution for the GW mean anomaly indicate that GW151226 is more
consistent with a quasi-circular binary than a non-precessing
eccentric binary with $e_0\leq 0.3$.
The comparison between the value of eccentricity in Table
\ref{tab:gwEvents} with the ones in
Refs. \cite{Romero-Shaw:2019itr,OShea:2021ugg} is challenging due to
the following issues: the value of eccentricity that we report is
computed at a different starting frequency (20Hz), than the one (10Hz)
used in Refs.  \cite{Romero-Shaw:2019itr,OShea:2021ugg}, and the
definition of eccentricity that we are using is based on the waveforms
generated from the samples, while Refs.
\cite{Romero-Shaw:2019itr,OShea:2021ugg} report the initial
eccentricities considered in the definitions of the initial conditions
in the \texttt{TEOBResumS-Dali} and \texttt{SEOBNRE} models. In order
to estimate the eccentricity value at some other frequency we can
apply the Newtonian relation between the eccentricity and frequency
\cite{Peters:1964zz} $e = e_{\rm ref} (f/f_{\rm ref})^{-19/18}$ to our
eccentricity measurement to obtain $e_{\rm gw}^{\rm 10 Hz} \sim 0.1$,
which is closer to the value reported in
Ref.~\cite{OShea:2021ugg}. This fact can be explained probably due to
the fact that Ref. \cite{Romero-Shaw:2019itr} uses a log-uniform prior
in eccentricity which puts more weight on the low eccentricity region
than a uniform prior.

Finally, we have also produced a run with \texttt{SEOBNRv4E\_opt}
setting $e_0=0$, and obtained a signal-to-noise natural log Bayes factor of
$39.8^{+0.1}_{-0.1} $ which is slightly smaller than
$40.4^{+0.1}_{-0.1} $ obtained for \texttt{SEOBNRv4E\_opt} when
sampling in $(e_0,\zeta_0)$ (see Table \ref{tab:gwEvents}). The difference in log Bayes factors is $\sim 0.6$ which indicates a minor
preference for the eccentric hypothesis. When comparing to the
quasi-circular precessing-spin results of \texttt{IMRPhenomXPHM} from
the GWTC-2.1 catalog we find $\log \mathcal{BF} =
47.59^{+0.14}_{-0.14}$, which indicates that the eccentric
non-precessing spin hypothesis is less preferred than the
precessing-spin quasi-circular one with a log Bayes factors
of $\sim 7.2$ in favour of the latter.

\subsubsection*{GW190521}

GW190521 is particularly intriguing, with only 4 cycles in the band of the
detectors, thus being consistent with a merger-ringdown dominated
signal. It has been attributed to a variety of physical systems
including a head-on collision of exotic compact objects
\cite{CalderonBustillo:2020fyi}, a nonspinning hyperbolic capture
\cite{Gamba:2021gap} and an eccentric binary
\cite{Gayathri:2020coq,Romero-Shaw:2020thy}, although some other recent studies do not find clear evidence for eccentricity  \cite{Iglesias:2022xfc}.

We analyze GW190521 using the \texttt{SEOBNRv4EHM\_opt} and
\texttt{SEOBNRv4E\_opt} models with a prior uniform in initial
eccentricity, $e_0 \in [0,0.3]$, and relativistic anomaly, $\zeta_0\in
[0,2\pi]$. We employ priors in inverse mass ratio $1/q \in [0.05,1]$,
and chirp mass $\mathcal{M}\in [60,200] M_\odot$ such that the induced priors are uniform in component
masses. The rest of the priors are as in the analysis of GW150914,
except for the luminosity distance prior which is chosen to be uniform
in comoving volume instead of $\propto d_L^2$, in order to match
the settings of Ref. \cite{Ramos-Buades:2023ehm}. The starting
frequency of waveform generation is 5.5Hz such that the higher-order
modes are in band at the minimum frequency of the likelihood
evaluation (11Hz).  As discussed in Ref. \cite{Estelles:2021jnz}, the
analysis of GW190521 using different sampling methods can lead to
systematics in recovering some modes in the posterior distribution. In
order to avoid that, we use a higher number of live points of
8192 for the \texttt{parallel Bilby} runs of this event for
\texttt{SEOBNRv4E\_opt}, while for \texttt{SEOBNRv4EHM\_opt} we use a
lower number of live points 2048 to reduce the computational
cost\footnote{We have also produced results for
  \texttt{SEOBNRv4E\_opt} with a number of live points of 2048 and
  4096, and observed only minimal differences in the posteriors with
  respect to the case with nlive=8192. Therefore, for
  \texttt{SEOBNRv4EHM\_opt}, we restricted to nlive=2048 in order to
  reduce the computational cost of the run.}.

The results of our analysis are shown in the bottom row of
Fig. \ref{fig:gwEvents}, while in Table \ref{tab:gwEvents} the median
values and the $90 \%$ credible intervals of the posterior
distributions are reported. For comparisons we include in our analysis
the samples of \texttt{SEOBNRv4PHM} from Refs.  \cite{LIGOScientific:2020iuh,LIGOScientific:2020ufj}. 
The quasi-circular parameters like
chirp mass or total mass show the largest differences with respect to
the \texttt{SEOBNRv4PHM}, while other parameters like the
effective-spin parameter and the mass ratio are quite consistent in
the median. However, as in the case of non-precessing quasi-circular
models (see Ref. \cite{Estelles:2021jnz}), 
the \texttt{SEOBNRv4EHM\_opt} and \texttt{SEOBNRv4E\_opt} models are not
able to reproduce the secondary mode in the inverse mass ratio
posterior. These differences between \texttt{SEOBNRv4PHM} and
\texttt{SEOBNRv4EHM\_opt} are expected due to the GW190521 signal
being merger-ringdown dominated, and these waveform models include
distinct physical
effects. 

Focusing on the eccentric parameters, we find median values of the GW
eccentricity of $e^{\rm 5.5Hz}_{\rm gw}={0.15}^{+0.12}_{-0.12}$ for
both \texttt{SEOBNRv4EHM\_opt} and \texttt{SEOBNRv4E\_opt}. The large
median values of eccentricity contrast with the uniformative posterior
distribution of eccentricity and the large uncertainty in the $90 \%$
credible intervals, which combined with an uninformative posterior
distribution of the GW mean anomaly (see bottom panels of
Fig. \ref{fig:gwEvents}), indicates that the eccentricity parameter is
poorly constrained in GW190521. This situation can be explained by the
fact that the signal is extremely short, and thus merger-ringdown
dominated. However, in \texttt{SEOBNRv4EHM\_opt} the eccentricity
effects are included only through the inspiral-merger EOB modes, while
at merger-ringdown the binary is assumed to have circularized, and the
merger input values and the ringdown model are the same as in
\texttt{SEOBNRv4HM} (see Sec. \ref{sec:EOBmodel} for
details). For moderate eccentricities, as the ones considered here, and non-precessing spins it has been shown in Refs. \cite{Hinder:2007qu,Huerta:2019oxn,Ramos-Buades:2019uvh} that the effects of eccentricity in the final mass and spin of NR waveforms are subdominant. However, non-precessing large eccentricity, as well as precessing-spin eccentric cases are fairly unexplored. Therefore, in order to clearly measure eccentricity in high-mass systems, the binary should have large enough eccentricity at merger.

Apart from this limitation of the \texttt{SEOBNRv4EHM\_opt}
model\footnote{This is also the case of the other state-of-the-art
  inspiral-merger-ringdown eccentric waveform models, like
  \texttt{TEOBResumS-Dali} and \texttt{SEOBNRE}, used in other
  parameter estimation studies.}, we have also attempted to estimate
the validity of the non-precessing eccentric hypothesis versus the
non-precessing quasi-circular one by producing a run with
\texttt{SEOBNRv4E\_opt} setting $e_0=0$, and comparing the
signal-to-noise Bayes factors. For the zero-eccentricity run with
\texttt{SEOBNRv4E\_opt} we find $\log \mathcal{BF} =
77.7^{+0.1}_{-0.1} $, which is consistent within the error bars with
the values obtained for the eccentric run with \texttt{SEOBNRv4E\_opt},
$\log \mathcal{BF} = 77.8^{+0.1}_{-0.1} $. This points out that the
non-precessing eccentric hypothesis is equally favoured as the
non-precessing quasi-circular one. However, when comparing to the
quasi-circular precessing-spin results from the discovery paper of
GW190521\footnote{We do not compare to the \texttt{SEOBNRv4PHM}
  results from GWTC-2.1, as the log Bayes factor is not provided with
  the public data.} \cite{LIGOScientific:2020iuh} we observe that for
the \texttt{NRSur7dq4} model~\cite{Varma:2019csw} the log Bayes factor
is $\log \mathcal{BF} = 84.49$. This produces a log Bayes
factor between the quasi-circular precessing-spin and non-precessing
eccentric hypothesis of $\sim 6.7$, indicating that the
quasi-circular precessing-spin hypothesis is preferred over the
non-precessing eccentric one with a prior in $e_0\in [0,0.3]$. 

The comparison of Bayes factors from different waveform models can be complicated as the result may be affected not only by the different physical effects included in the models, but also by the waveform systematics between the different waveform approximants. Different waveform models including the same physical effects can lead to different log Bayes factors (see for instance Table II of Ref. \cite{Estelles:2021jnz}). As a consequence, we consider the comparison of the eccentric non-precessing against quasi-circular precessing-spin hypothesis using log Bayes factors from other waveform families  as an approximate estimate, and leave for future work a more detailed study \cite{Gupte2023}.

Besides calibration to eccentric NR waveforms one of the main limitations in the analysis of GW190521 with \texttt{SEOBNRv4EHM\_opt} is the lack of inclusion of spin-precession effects, which impact significantly the morphology of the templates at merger-ringdown, and may substantially modify the measured parameters. This points out the necessity to produce waveform models, which include both eccentricity and spin-precession effects, and there is ongoing work  \cite{Gamboa2023} to include both effects in the new generation of \texttt{SEOBNR} models \cite{Pompili:2023tna,Ramos-Buades:2023ehm} built within the new \texttt{pySEOBNR} python infrastructure \cite{Mihaylovv5}. 
 
Finally, the \texttt{SEOBNRv4EHM\_opt} results have been obtained on the order of a few days or a week using \texttt{parallel Bilby} (see Table \ref{tab:peruntime}). This makes \texttt{SEOBNRv4EHM\_opt} a standard tool that can be used with a highly parallelizable nested sampler like \texttt{parallel Bilby}, and we plan to extend the Bayesian inference study presented here, using the machine-learning code \texttt{DINGO}~\cite{Green:2020dnx,Dax:2021tsq,Dax:2022pxd}, to all the GW events observed during the third-observing run \cite{Gupte2023}.
 
\section{Conclusions}
\label{sec:conclusions}

In this paper we have improved and validated the multipolar non-precessing eccentric \texttt{SEOBNRv4EHM} model presented in Ref. \cite{Ramos-Buades:2021adz}, and shown its applicability to Bayesian inference studies.

The \texttt{SEOBNRv4EHM} model is built upon the quasi-circular accurate NR-calibrated multipolar, non-precessing \texttt{SEOBNRv4HM} model \cite{Cotesta:2018fcv}. The eccentricity effects are included in the GW multipoles up to 2PN order including spin-orbit and spin-spin effects \cite{Khalil:2021txt}. The multipolar \texttt{SEOBNRv4EHM} model includes the $(2,2),(2,1),(3,3),(4,4),(5,5)$ modes, and it is shown in Ref. \cite{Ramos-Buades:2021adz} to have an unfaithfulness against public eccentric NR waveforms below $1\%$. 

Within the \texttt{SEOBNRv4EHM} model, elliptical orbits are described
by two parameters, initial eccentricity, $e_0$, and initial
relativistic anomaly, $\zeta_0$, specified at an instantaneous orbital
frequency $\omega_0$. Here, we present a new parametrization of the
initial conditions, where $e_0$ and $\zeta_0$ are specified at an
orbit-averaged orbital frequency $\overline{\omega}_0$. This new
orbit-averaged initial conditions lead to smoother variations of $e_0$
and $\zeta_0$ across parameter space, and as a consequence more
efficient sampling of the parameter space.

The improvement in sampling efficiency due to the new initial
conditions has also being accompanied with the development of
\texttt{SEOBNRv4EHM\_opt}, a faster version of the
\texttt{SEOBNRv4EHM} model. The \texttt{SEOBNRv4EHM\_opt} model combines a reduction of the
absolute and relative tolerances of the Runge Kutta integrator from
$10^{-10}$ and $10^{-9}$, to $10^{-8}$ and $10^{-8}$, with the use of the optimized Hamiltonian and integrator from Refs. \cite{Devine:2016ovp,Knowles:2018hqq}. These modifications lead to a
factor of $\sim 3-7$ speed-up depending on the binary parameters.  The
reduction of the tolerances implies a reduction in accuracy of the
\texttt{SEOBNRv4EHM\_opt} model in the corners of parameter space 
(i.e., high eccentricities and high spins), where the model usage 
is limited, because it is very
sensitive to the attachment point of the inspiral and merger-ringdown
EOB modes. The trade-off between accuracy and efficiency of the new
\texttt{SEOBNRv4EHM\_opt} model, with a waveform evaluation time of
$\mathcal{O}(100)$ms, makes it a competitive model for use in
parameter-estimation studies.

Given the accuracy and computational efficiency of
\texttt{SEOBNRv4EHM\_opt}, we have performed a Bayesian inference
study on mock signals and real GW events detected by the LVK
collaboration. We have first investigated the quasi-circular limit of
the \texttt{SEOBNRv4EHM\_opt} model by computing the unfaithfulness
against the quasi-circular \texttt{SEOBNRv4HM} model
\cite{Cotesta:2018fcv} for 4500 random configurations in the parameter
space $q\in[1,50]$, spins $\chi_{1,2}\in [-0.9,0.9]$ and total masses
$M \in [20,300] M_\odot$, at a dimensionless starting frequency of $M
\omega = 0.023$. The results show that \texttt{SEOBNRv4EHM\_opt} has a
median unfaithfulness of $3.8 \times 10^{-5}$ and no cases with
unfaithfulness $>1\%$, indicating that the quasi-circular limit is
accurately recovered. Furthermore, we have performed a mock-signal 
injection into zero noise using \texttt{SEOBNRv4} \cite{Bohe:2016gbl}
as a signal, and \texttt{SEOBNRv4E\_opt} and \texttt{SEOBNRv4\_ROM}
\cite{Bohe:2016gbl} as templates. We have considered a configuration with
mass ratio $q=4$, total mass $M=90.08 M_\odot$ and BH's dimensionless
spins $\chi_1 = 0.5$ and $\chi_2 = -0.1$ defined at 20Hz. The recovery
of quasi-circular parameters, like mass ratio, chirp mass or the
effective-spin parameter by \texttt{SEOBNRv4E\_opt} agrees remarkably
well with the ones from \texttt{SEOBNRv4\_ROM}. While the initial
eccentricity and relativistic anomaly measured by
\texttt{SEOBNRv4E\_opt} are $e_0=0.01^{+0.02}_{-0.01}$ and $\zeta_0 =
3.09^{+2.57}_{-2.48}$, which indicate that the signal is compatible
with zero eccentricity. Therefore, \texttt{SEOBNRv4EHM\_opt} is able
to correctly reproduce the zero-eccentricity limit, with an accuracy
comparable to the underlying quasi-circular model \texttt{SEOBNRv4HM}.

Moving to the eccentric sector, we have studied with zero-noise
injections, using the \texttt{SEOBNRv4E\_opt} as a signal and template,
the impact of the initial conditions based on $(e_0,\zeta_0)$
specified at $\omega_0$, and the new prescription, where
$(e_0,\zeta_0)$ are specified at $\overline{\omega}_0$. For this study
we have chosen a configuration with two different initial
eccentricities $e_0 = [0.1,0.2]$, mass ratio $q=3$, initial
relativistic anomaly $\zeta_0 =1.2$, total mass $M=76.4 M_\odot$ and
BH's dimensionless spins $\chi_1 = 0.5$ and $\chi_2 = -0.1$ defined at
20Hz.  The results show that both the orbit-averaged and the
instantaneous initial conditions are able to accurately recover the
corresponding injected signal, however, the averaged wall clock-time
of the runs using the instantaneous initial conditions is a factor 3
slower than the orbit-averaged initial conditions, due to the
complicated structure of the likelihood across the $(e_0,\zeta_0)$
parameter space in the case of the instantaneous initial conditions
(see Fig. \ref{fig:logLExamples}). As a consequence we adopt the
orbit-averaged initial conditions as the default ones in the
\texttt{SEOBNRv4EHM} and \texttt{SEOBNRv4EHM\_opt} models.

Moreover, we have investigated the impact of neglecting the radial-phase 
parameter, relativistic anomaly, in the previous model
injections by starting the orbits of the templates at periastron (i.e., $\zeta_0=0$). 
The posterior distributions of the orbit-averaged
ICs show larger biases than the instantaneous ICs in the recovery of
the quasi-circular parameters, for instance, $8\%$ bias in the case of
the chirp mass, while in terms of the eccentricity parameter the
instantaneous ICs may develop multimodalities in the posteriors, as is the case of the 
injections considered here, indicating that the parametrization cannot adequately
reproduce the injected signal. For the orbit-averaged ICs the
eccentricity parameter is compatible with the injected value within
the $90\%$ credible intervals. This indicates that neglecting the
radial-phase parameter when performing parameter estimation of
eccentric signals can induce biases not only in the measurement of the
eccentricity, but also in the estimation of other quasi-circular
parameters like mass ratio or the spins due to the strong correlation
of eccentricity with these parameters.

The accuracy of the \texttt{SEOBNRv4EHM} model in the eccentric case
was investigated in Ref. \cite{Ramos-Buades:2021adz} by computing the
unfaithfulness of the model against a set of public eccentric NR
waveforms from the SXS catalog
\cite{Hinder:2017sxy,Boyle:2019kee}. Here we have also validated the
accuracy of the \texttt{SEOBNRv4EHM\_opt} model by performing a set of
injections of synthetic NR signals into a network of LIGO-Virgo detectors at design
sensitivity. We have injected into zero-detector noise three eccentric
NR waveforms, \texttt{SXS:BBH:1355}, \texttt{SXS:BBH:1359} and
\texttt{SXS:BBH:1363}, corresponding to equal-mass, nonspinning
configurations with initial eccentricities measured from the orbital
frequency at first periastron passage of $0.07$, $0.13$ and $ 0.25$,
respectively. For these injections we choose a total mass $M=70
M_\odot$, inclination $\iota=0$ and SNR$=20$. The results are
summarized in Fig. \ref{fig:nrInj} and Table
\ref{tab:nr_injection}. In order to compare the eccentricity from the
NR waveforms and the \texttt{SEOBNRv4EHM\_opt} model, we have adopted a
common definition of eccentricity and radial phase, based on the
frequency of the (2,2)-mode \cite{Ramos-Buades:2022lgf}, and used its
efficient implementation in the open-source \texttt{gw\_eccentricity}
Python package \cite{Shaikh:2023ypz} to post-process the parameter
estimation runs. We have found that the recovery of the parameters
with \texttt{SEOBNRv4E\_opt} does not produce significant biases, and
that the measurement of the GW eccentricity, $e_{\rm gw}$, and GW mean
anomaly, $l_{\rm gw}$, is consistent with the injected values for the
three injections considered. A more comprehensive Bayesian inference
study will be required to assess the modeling inaccuracies and how
they translate into biases in both eccentric and quasi-circular
parameters. Here, new methods of inference such as machine learning
techniques, like \texttt{DINGO} \cite{Green:2020dnx,Dax:2021tsq,Dax:2022pxd}, may offer an alternative
method to efficiently perform large-scale injections campaigns with
moderate computational cost~\cite{Gupte2023}.  

Besides injection studies, we have demonstrated that
\texttt{SEOBNRv4EHM\_opt} can be used as a standard tool in Bayesian
inference studies of real GW events. We have analyzed three GW events
(GW150914, GW151226 and GW190521) detected by the LVK Collaboration in
the first and third observing runs. The eccentricity measured for the
three events is $e^{\text{GW150914}}_{\text{gw, 10Hz}}=
0.08^{+0.09}_{-0.06}$, $e^{\text{GW151226}}_{\text{gw, 20Hz}}=
{0.04}^{+0.05}_{-0.04} $, and $e^{\text{GW190521}}_{\text{gw, 5.5Hz}}=
0.15^{+0.12}_{-0.12}$. As a consequence, we do not find clear
evidence of orbital eccentricity in any of the GW events considered, when
using a non-precessing eccentric model with initial eccentricities
$e_0 \in [0,0.3]$. For the GW150914 and GW151226 we have compared
with the quasi-circular results from the GWTC-2.1 catalog
\cite{LIGOScientific:2021usb}, while for GW190521 with the results from Refs.  \cite{LIGOScientific:2020iuh,LIGOScientific:2020ufj} (the precessing-spin
\texttt{SEOBNRv4PHM} \cite{Ossokine:2020kjp} and
\texttt{IMRPhenomXPHM} \cite{Pratten:2020ceb} models). For GW150914 we
find good agreement with \texttt{SEOBNRv4PHM} due to the fact that the
event is consistent with a nonspinning binary, while for GW151226 and
GW190521, we find some discrepancies, which are likely due to the
inclusion of spin-precession effects in the quasi-circular models,
which are not included in the \texttt{SEOBNRv4EHM\_opt} model. This is
a clear limitation of the \texttt{SEOBNRv4EHM\_opt} model as well as
all the current existing inspiral-merger-ringdown eccentric waveform models. There is
ongoing work~\cite{Gamboa2023} to include such effects in the new
generation of \texttt{SEOBNR} models
\cite{Pompili:2023tna,Ramos-Buades:2023ehm} developed within the new
\texttt{pySEOBNR} infrastructure \cite{Mihaylovv5}.

Regarding the analysis of real GW events, we plan in the future to extend the Bayesian inference study presented here, using the machine-learning code {\tt DINGO}, to all the GW events detected during the third-observing run \cite{Gupte2023} in order to set constraints on the eccentricity of the observed population of BBHs.

\section*{Acknowledgments}
\label{acknowledgements}

It is a pleasure to thank Mohammed Khalil for providing the expressions of the orbit-averaged orbital frequency used throughout the paper. We also would like to thank Sergei Ossokine and Harald Pfeiffer for helpful discussions about the implementation of the model and the initial conditions, as well as Marta Colleoni for useful comments to improve the manuscript. The computational work for this manuscript was carried out on the computer cluster \texttt{Hypatia} at the Max Planck Institute for Gravitational Physics in Potsdam.

This research has made use of data or software obtained from the Gravitational Wave Open Science Center (gwosc.org), a service of LIGO Laboratory, the LIGO Scientific Collaboration, the Virgo Collaboration, and KAGRA. LIGO Laboratory and Advanced LIGO are funded by the United States National Science Foundation (NSF) as well as the Science and Technology Facilities Council (STFC) of the United Kingdom, the Max-Planck-Society (MPS), and the State of Niedersachsen/Germany for support of the construction of Advanced LIGO and construction and operation of the GEO600 detector. Additional support for Advanced LIGO was provided by the Australian Research Council. This material is based upon work supported by NSF's LIGO Laboratory which is a major facility fully funded by the National Science Foundation. Virgo is funded, through the European Gravitational Observatory (EGO), by the French Centre National de Recherche Scientifique (CNRS), the Italian Istituto Nazionale di Fisica Nucleare (INFN) and the Dutch Nikhef, with contributions by institutions from Belgium, Germany, Greece, Hungary, Ireland, Japan, Monaco, Poland, Portugal, Spain. KAGRA is supported by Ministry of Education, Culture, Sports, Science and Technology (MEXT), Japan Society for the Promotion of Science (JSPS) in Japan; National Research Foundation (NRF) and Ministry of Science and ICT (MSIT) in Korea; Academia Sinica (AS) and National Science and Technology Council (NSTC) in Taiwan.


\appendix
\section{Derivation of the orbit-averaged orbital frequency}\label{sec:AppendixA}

The initial conditions for eccentric orbits were derived in Sec.~II.C of Ref.~\cite{Ramos-Buades:2021adz}, given an   initial (instantaneous) orbital frequency $\omega$, eccentricity $e$, and relativistic anomaly $\zeta$.
In order to perform orbital evolutions starting from different points on (roughly) the same eccentric orbit, we derive an expression for the instantaneous frequency in terms of the orbit-averaged frequency.

We use the Keplerian parametrization
\begin{equation}
r = \frac{1}{u_p (1 + e \cos \zeta)},
\label{eq:eq1A}
\end{equation} 
and perform the following steps, in a PN expansion up to 2PN order:
\begin{enumerate}
\item Calculate the orbit-averaged azimuthal frequency 
\begin{align}
\overline{\omega} &\equiv \frac{1}{T_r} \oint \dot{\phi} \, dt  = \frac{1}{T_r} \oint \frac{\partial H}{\partial p_\phi} \left(\frac{\partial H}{\partial p_r}\right)^{-1} dr 
\nonumber\\
&= \frac{2}{T_r} \int_{0}^{\pi} \frac{\partial H}{\partial p_\phi} \left(\frac{\partial H}{\partial p_r}\right)^{-1} \frac{dr}{d\zeta} \, d\zeta,
\end{align}
where $T_r$ is the radial period, and is given by
\begin{align}
T_r &\equiv \oint dt  = 2\int_{0}^{\pi}  \left(\frac{\partial H}{\partial p_r}\right)^{-1} \frac{dr}{d\zeta} d\zeta,
\end{align}
yielding

\onecolumngrid
\begin{align}
\overline{\omega} &= \left(u_p-e^2 u_p\right)^{3/2} \bigg\lbrace
1 
+ \frac{u_p}{2c^2} \left[\nu -e^2 (\nu -6)\right]  
+\frac{u_p^{3/2}}{2c^3} \left(3 e^2+1\right)  \left[(\nu -2) \chi _S-2 \delta  \chi _A\right]
+ \frac{3u_p^2}{8c^4} \Big[e^4 \left(\nu ^2-5 \nu +24\right)-2 e^2 \left(4 \nu \sqrt{1-e^2}  \right.    \nonumber\\
&\quad \left.  -10 \sqrt{1-e^2}+\nu ^2+\nu +3\right) +8 \sqrt{1-e^2} \nu -20 \sqrt{1-e^2}+\nu ^2-13 \nu +20\Big] 
-\frac{3u_p^2 }{2c^4} \Big[ 2\delta \chi _A \chi _S  \left[e^2 (\nu -1)-\nu \right] \nonumber\\
&\qquad
+ e^2 (4 \nu -1) \chi _A^2
+ \left[2 (\nu -1) \nu -e^2 \left(2 \nu ^2-2 \nu +1\right)\right] \chi _S^2
\Big]
\bigg\rbrace  + \mathcal{O}\left(1/c^5\right),
\end{align}
where the powers of $1/c$ indicate the PN order of the different terms.
\item Invert $\overline{\omega}(u_p,e)$ to obtain $u_p(\overline{\omega},e)$, i.e. the inverse semilatus rectum as a function of the averaged frequency, which at leading order is given by
\onecolumngrid
\begin{equation}
\begin{split}
u_p(\overline{\omega},e) &= \frac{\overline{\omega }^{2/3}}{1-e^2} +\frac{\overline{\omega }^{4/3} \left(e^2 (\nu -6)-\nu
   \right)}{3 c^2 \left(e^2-1\right)^2}  +\frac{\left(3 e^2+1\right) \overline{\omega }^{5/3} \left( 2   \delta  \chi _A-(\nu -2) \chi _S\right)}{3 c^3   \left(1-e^2\right)^{5/2}} \\
   & + \frac{\overline{\omega }^2}{4 c^4
   \left(e^2-1\right)^3}  \bigg\lbrace-2 e^2 \left[\nu  \left(4 \delta  \chi _A \chi _S+8
   \chi _A^2+4 \sqrt{1-e^2}+4 \chi _S^2+7\right)-4 \delta  \chi _A \chi _S-2
   \chi _A^2-10 \sqrt{1-e^2}-4 \nu ^2 \chi _S^2-2 \chi _S^2+3\right]  \\
   &   +\nu  \left(8 \delta  \chi _A \chi _S+8 \sqrt{1-e^2}+8 \chi _S^2-13\right)+e^4 (7
   \nu -12)-20 \left(\sqrt{1-e^2}-1\right)-8 \nu ^2 \chi _S^2 \bigg\rbrace  + \mathcal{O}\left(1/c^5\right).
\end{split}
\end{equation}

\item Calculate the instantaneous orbital frequency (in Keplerian parametrization)
\onecolumngrid
\begin{align}
\begin{split}
\omega(u_p,e,\zeta) &= \frac{\partial H}{\partial p_\phi} = u_p^{3/2} (e \cos \zeta +1)^2  - \frac{u_p^{5/2}}{2
   c^2} (e \cos \zeta+1)^2 \left[\left(e^2-1\right) \nu +4 e \cos \zeta \right] \\
   & -\frac{u_p^3}{2 c^3} \left(e^2-2 e \cos \zeta +1\right) (e \cos \zeta +1)^2
   \left(2 \delta  \chi _A-(\nu -2) \chi _S\right)  \\
   & +\frac{u_p^{7/2}}{8
   c^4} (e \cos \zeta+1)^2 \bigg \lbrace -16 e^2 \nu  \chi _A^2+8 e \cos
   \zeta  \left[\nu  \left(8 \delta  \chi _A \chi _S+8 \chi _A^2+e^2+8 \chi
   _S^2-1\right)-2 \left(2 \delta  \chi _A \chi _S+\chi _A^2+\chi
   _S^2+2\right)-8 \nu ^2 \chi _S^2\right]   \\
   &    +4 e^2 \cos 2 \zeta  \left[(4 \nu -1) \chi _A^2+2 \delta  (2 \nu -1) \chi _A \chi _S-(1-2 \nu )^2 \chi
   _S^2\right]-8 \delta  e^2 \nu  \chi _A \chi _S+8 \delta  e^2 \chi _A \chi
   _S+4 e^2 \chi _A^2+24 \delta  \nu  \chi _A \chi _S+3 e^4 \nu ^2   \\
   &   -3 e^4 \nu -6
   e^2 \nu ^2+2 e^2 \nu +8 e^2 \nu ^2 \chi _S^2-8 e^2 \nu  \chi _S^2+4 e^2 \chi
   _S^2+16 e^2+3 \nu ^2-15 \nu -24 \nu ^2 \chi _S^2+24 \nu  \chi _S^2 \bigg\rbrace  + \mathcal{O}\left(1/c^5\right) .
\end{split}
\end{align}

\item Plug $u_p(\overline{\omega},e)$ in $\omega(u_p,e,\zeta)$  to obtain $\omega(\overline{\omega},e,\zeta)$, which reads
\onecolumngrid
\begin{align}
\label{omegaInstAvg}
\omega &= \frac{\overline{\omega } (e \cos \zeta +1)^2}{\left(1-e^2\right)^{3/2}}
-\frac{e \overline{\omega }^{5/3} (3 e+2 \cos \zeta ) (e \cos \zeta +1)^2}{c^2\left(1-e^2\right)^{5/2}}  -\frac{e \overline{\omega }^2(e+\cos \zeta ) }{c^3\left(e^2-1\right)^3}\left(1 + e \cos \zeta \right)^2 \left[2 \delta  \chi _A-(\nu -2) \chi _S\right]\nonumber\\
&\quad
-\frac{\overline{\omega }^{7/3} (e \cos \zeta +1)^2}{12c^4 \left(1-e^2\right)^{7/2}}
\bigg\lbrace
12 e^4 (\nu -6)+8 e \left(e^2 (\nu -15)-\nu +6\right) \cos \zeta -3 e^2 \left[2 \left(6 \sqrt{1-e^2}+7\right) \nu -30 \sqrt{1-e^2}+17\right]\nonumber\\
&\qquad
+18 \left(\sqrt{1-e^2}-1\right) (2 \nu -5)
\bigg\rbrace  + \frac{e \overline{\omega }^{7/3} (e \cos \zeta +1)^2}{2 c^4\left(1-e^2\right)^{7/2}}
\bigg\lbrace
2 \delta  \chi _A \chi _S \left[e (2 \nu -1) \cos (2 \zeta )+2 e (\nu -1)+(8 \nu -4) \cos \zeta \right] \nonumber\\
&\qquad
+\chi _S^2 \left[-e \left(4 \nu ^2+(1-2 \nu )^2 \cos (2 \zeta )-4 \nu +2\right)-4 (1-2 \nu )^2 \cos \zeta \right]  
+(4 \nu -1) \chi _A^2 \left[e (\cos (2 \zeta )+2)+4 \cos \zeta \right]
\bigg\rbrace + \mathcal{O}\left(1/c^5\right).
\end{align}
\end{enumerate}

\twocolumngrid
Thus, we start with a given initial orbit-averaged frequency $\overline{\omega}$, eccentricity $e$, and relativistic anomaly $\zeta$.
Then, use Eq.~\eqref{omegaInstAvg} to compute the instantaneous frequency $\omega(\overline{\omega},e,\zeta)$, and follow the same procedure as in Ref.~\cite{Ramos-Buades:2021adz} to obtain the initial conditions for the dynamical variables.


\bibliography{biblio}

\end{document}